\documentclass[twocolumn,showpacs,preprintnumbers,amsmath,amssymb,nofootinbib]{revtex4}

\usepackage{graphicx}
\usepackage{dcolumn}
\usepackage{bm}

\usepackage{epstopdf}

\def\bea{\begin{eqnarray}}
\def\eea{\end{eqnarray}}

\def\pp{\mbox{$p$-$p$}}
\def\auau{\mbox{Au-Au}}

\def\pbpb{\mbox{Pb-Pb}}
\def\aa{\mbox{$A$-$A$}}
\def\pa{\mbox{$p$-$A$}}
\def\da{\mbox{$d$-$A$}}
\def\ha{\mbox{$h$-$A$}}
\def\na{\mbox{$N$-$A$}}
\def\nn{\mbox{$N$-$N$}}

\def\ee{\mbox{$e^+$-$e^-$}}

\def\qqbar{\mbox{$q$-$\bar q$}}
\def\pt{$p_t$}
\def\v2{$v_2$}
\def\yt{$y_t$}
\def\mt{$m_t$}

\begin{document} 

\preprint{Version 2.9}

\title{Distinguishing the nonjet azimuth quadrupole from QCD jets and hydrodynamic flows\\ via 2D angular correlations and quadrupole spectrum analysis
}

\author{Thomas A.\ Trainor}\affiliation{CENPA 354290, University of Washington, Seattle, Washington 98195}


\date{\today}

\begin{abstract}
According to the flow narrative commonly applied to high-energy nuclear collisions a 1D cylindrical-quadrupole component of 2D angular correlations conventionally denoted by quantity $v_2$ is interpreted to represent elliptic flow: azimuth modulation of transverse or radial flow in noncentral nucleus-nucleus (A-A) collisions. Jet angular correlations may also contribute to $v_2$ data as ``nonflow'' depending on the method used to calculate $v_2$, but 2D graphical methods can achieve accurate separation. The nonjet (NJ) quadrupole component exhibits various properties inconsistent with a flow or hydro interpretation, including the observation that NJ-quadrupole centrality variation in $A$-$A$ collisions has no relation to strongly-varying jet modification (``jet quenching'') in those collisions commonly attributed to jet interaction with a dense flowing medium. In the present study I report isolation of quadrupole spectra from $p_t$-differential $v_2(p_t)$ data obtained at the relativistic heavy ion collider (RHIC) and large hadron collider (LHCr). I demonstrate that NJ quadrupole spectra have characteristics very different from the single-particle spectra for most hadrons, that quadrupole spectra indicate a common boosted hadron source for a small minority of hadrons that ``carry'' the  quadrupole structure, that the narrow source-boost distribution is characteristic of an expanding thin cylindrical shell (also strongly contradicting a hydro interpretation), and that in the boost frame a single universal quadrupole spectrum (L\'evy distribution) on transverse mass $m_t$ accurately describes data for several hadron species scaled according to their statistical-model abundances. The quadrupole spectrum shape changes very little from RHIC to LHC energies. Taken in combination those characteristics strongly suggest a unique {\em nonflow} (and nonjet) QCD mechanism for the NJ quadrupole conventionally represented by $v_2$.
\end{abstract}


\maketitle

 \section{Introduction} \label{intro}

 In high-energy nucleus-nucleus (\aa) collisions elliptic flow, as the physical interpretation of an azimuth-quadrupole component of angular correlations denoted by symbol $v_2$, has played a central role in supporting arguments claiming quark-gluon plasma (QGP) formation~\cite{perfect,qgp1,qgp2,keystone}. According to the conventional flow narrative elliptic flow should be sensitive to the early stage of high-energy \aa\ collisions where quarks and gluons are believed to be the more-likely degrees of freedom. Correlation data demonstrating the presence of elliptic flow might confirm large energy and matter densities and copious parton rescattering to achieve a thermalized QGP~\cite{hydro}. 

The RHIC experimental program has seemed to provide strong evidence confirming what may be termed  a flow-QGP narrative based on data obtained with certain preferred measures and techniques. It was therefore concluded in 2005 that a ``strongly-coupled QGP'' or ``perfect liquid'' is formed in central \auau\ collisions at RHIC energies~\cite{perfect}. However, one should distinguish between (a) the physical mechanism of elliptic flow and (b) the observed phenomenon of a cylindrical quadrupole on azimuth near midrapidity. The existence of (a) might imply (b) but observation of (b) does not {\em require} (a), and other observations may falsify (a). The present study applies novel analysis methods to recent LHC $v_2(p_t,b)$ data for identified hadrons. Some analysis results appear to contradict essential elements of the flow-QGP narrative.

Extensive studies of two-dimensional (2D) angular correlations~\cite{axialci,anomalous,multipoles,ppquad} have established that there are two main contributions to an observed azimuth quadrupole: (a) a nonjet (NJ) quadrupole component and (b) a jet-related quadrupole contribution derived from a same-side (on azimuth) 2D jet peak representing {\em intra}\,jet angular correlations. Contribution (b) is often referred to as ``nonflow'' without acknowledging the dominant jet mechanism. The NJ quadrupole can be isolated accurately from jet-related and other contributions by model fits to 2D angular correlations~\cite{anomalous,davidhq}. $v_2$ data trends obtained with that method are inconsistent with an elliptic-flow interpretation in several ways~\cite{davidhq,noelliptic,v2ptb}. Jet-related bias of  published $v_2$ data has been estimated for cases where the preferred analysis method is simply related to 2D angular correlations~\cite{davidhq,v2ptb}. Physical interpretation of the NJ quadrupole remains a central issue: does it represent elliptic flow or some alternative mechanism~\cite{gluequad,nohydro,noelliptic,v2ptb}?

Although quadrupole $v_2$ data have played a central role in claims of QGP formation, hadron production near midrapidity appears to be dominated by two other mechanisms: (a) longitudinal projectile-nucleon dissociation (soft) and (b) transverse large-angle parton scattering and jet formation (hard). The two mechanisms form the basis for the two-component (soft + hard) model (TCM) of hadron yields, spectra and correlations~\cite{ppquad,hardspec,anomalous,aliceptfluct,alicetomspec}. The TCM then provides an essential context for interpretation of NJ quadrupole data.

To better understand the relation between the NJ quadrupole and hydrodynamic (hydro) theory expectations for flows, $v_2(p_t,b)$ data for three species of identified hadrons from 200 GeV \auau\ collisions were transformed to obtain {\em quadrupole spectra}~\cite{quadspec}. The three quadrupole spectra were found to be consistent with emission from a common boosted hadron source. The inferred narrow boost distribution suggested emission from an expanding thin cylindrical shell. The three spectra were found to be equivalent modulo rescaling by factors consistent with a statistical model of hadron abundances~\cite{statmodel}. Quadrupole-spectrum parameters were very different from those for single-particle (SP) spectra for most produced hadrons from the same collisions. The study concluded that the NJ quadrupole may be independent of most hadrons and may represent a unique mechanism unrelated to a flowing dense medium or QGP.

In the present study several $v_2$ analysis methods and associated data trends for \pt-integral data are  compared, with emphasis on the extent to which jet-related correlations contribute to $v_2$ data from various methods. The quadrupole spectrum concept is introduced and methods are presented for obtaining quadrupole spectra by combining $v_2(p_t)$ data and SP \pt\ spectra $\bar \rho_0(p_t)$. Quadrupole spectra from the cited analysis of 200 GeV \auau\ data are reviewed to illustrate the method. Recent $v_2(p_t,b)$ data for four hadron species from seven centrality classes of 2.76 TeV \pbpb\ collisions are then analyzed similarly to obtain corresponding quadrupole spectra. The main goal of the present study is determination of the collision-energy and \aa\ centrality dependence of  NJ quadrupole spectra so as to further test  the relation (if any) between the NJ quadrupole and hydro mechanisms, and possibly to elucidate alternative (QCD) mechanisms.

This article is arranged as follows:
Section~\ref{narrative} compares a ``flow-QGP'' narrative inspired by nucleus-nucleus experiments at lower collision energies with ``nuclear transparency'' inferred from hadron-nucleus experiments at higher energies.
Section~\ref{methods} reviews general analysis methods for high-energy nuclear collisions.
Section~\ref{v2meths}  describes analysis methods applied to azimuth quadrupole correlations.
Section~\ref{quadspecmeth}  defines new methods specific to inference of azimuth quadrupole spectra.
Section~\ref{200gevquad} reviews a previous quadrupole spectrum analysis of 200 GeV \auau\ data.
Section~\ref{quadspec} extends quadrupole spectrum analysis to data from 2.76 TeV \pbpb\ collisions.
Section~\ref{edep} reviews quadrupole-spectrum centrality and collision-energy trends and possible factorizations.
Section~\ref{syserr} discusses systematic uncertainties.
Sections~\ref{disc} and~\ref{summ}  present discussion and summary.
Appendix~\ref{boost} reviews the relativistic kinematics of boosted hadron sources, and 
Appendix~\ref{spspec} discusses single-particle identified-hadron spectra used in the present study.

\section{a-a flow-QGP $\bf vs$ $\bf h$-a transparency} \label{narrative}

The flow-QGP narrative describes a combination of high-energy nuclear collision phenomena in terms of  ``...dense, thermally equilibrated, strongly interacting matter, the quarkgluon plasma (QGP)''~\cite{hydro}. The flow narrative is based on certain {\em a priori} assumptions, preferred analysis methods, physical interpretations and hydro theory that have evolved over time with increasing collision energies at a succession of accelerators. The conjectured phenomena include ``jet quenching'' (modification or absorption of jets in the dense medium or QGP), various flows as manifested in spectra and correlations, and especially elliptic flow interpreted to provide the basis for claims of ``perfect liquid'' formed in high-energy \aa\ collisions~\cite{perfect}. The flow-QGP narrative provides one limiting case. The other limit is {\em transparent} \aa\ collisions described as linear superpositions of elementary \nn\ collisions (e.g.\ \pp\ collisions in isolation). To provide a context for the present study of NJ quadrupole spectra the two limiting cases are summarized briefly.


First demonstration of flowing {\em nucleonic} matter in relativistic \aa\ collisions~\cite{bevalac} represented a major achievement of the Bevalac program. Similar flows persisted at the Brookhaven alternating gradient synchrotron (AGS) up to nucleon-nucleon (\nn) center-of-momentum (CM) energy $\sqrt{s_{NN}} \approx 5$ GeV~\cite{agsv2}, the basic degrees of freedom progressing with increasing collision energy from nucleon clusters to nucleons to hadronic resonances. It was concluded that a flow description is essential for nuclear collisions within that energy interval, that a thermodynamic state might be established and that a QCD phase transition might be relevant at higher energies.

Progressing approximately in parallel with the Bevalac \aa\ program were fixed-target studies of \ha\ collisions within the CM energy interval $\sqrt{s_{NN}} \approx 10$-30 GeV (50-400 GeV beam energy). Whereas Bevalac-AGS \aa\ results below 5 GeV were consistent with projectile stopping and nearly-opaque nuclei the \ha\ results were interpreted to indicate 
``...the apparent near transparency of nuclear matter to coherently produced multiparticle states...''~\cite{witpa2} attributed to relativistic and quantum effects.  
In effect, time dilation prevents interaction of virtual projectile-hadron {\em fragments} within the target-nucleus volume. Most hadron fragments are formed only later outside the collision volume: 
``The lack of significant nuclear cascading has led to the conclusion that the high energy secondaries [``shower'' particles with $\beta > 0.7$, mainly pions] produced in the fundamental hadron-nucleon collision take a long time to form compared to nuclear [target nucleus A] dimensions''~\cite{witpa3}.
``The limited space-time development of hadronic matter inside a hit nucleus is a probable explanation of the low pion multiplicities in [\ha] reactions ....the [shower] pions do not transfer any significant energy to the target nucleus''~\cite{otterlundpa}.
``...the bulk of the [hadron] production in central rapidity region takes place long after the constituent [of a projectile hadron] passed through the target [nucleus]...''~\cite{bialas}.

The produced hadrons (shower particles, projectile fragments) from \ha\ collisions can be identified with the soft component of the TCM. The quantity $\bar \nu$ as defined for instance in Ref.~\cite{witpa3} measures the number of target-nucleus participants (average number of inelastic collisions for projectile hadron $h$). A linear relation was observed between produced (shower) hadrons (in the target hemisphere) and $\bar \nu$~\cite{otterlundpa}. In the more-modern language of \aa\ collisions ($\nu$ is defined below in an \aa\ context) the \ha\ results imply a linear relation between the soft component and the number of projectile participants $N_{part}$ (defined below). If \aa\ collision are linear superpositions of \na\ collisions (as a limiting case) then according to \ha\ results the \aa\ soft component (the majority of produced hadrons) is formed {\em mainly outside the \aa\ collision volume} and does not interact significantly within that volume: produced-hadron rescattering is minimal.


Despite evidence from \ha\ collisions, including implications for nuclear transparency, a narrative extrapolated from Bevalac-AGS observation of flowing nucleonic matter was preferred in anticipation of RHIC startup. The basic argument for a dense, flowing medium relies on inferred particle and energy densities assuming that {\em almost all detected hadrons} emerge from the dense medium~\cite{perfect}. In central 200 GeV \auau\ collisions $\eta$ density $dn_{ch}/d\eta \approx 700$, $\langle p_t \rangle \approx$ 0.5 GeV/c and collision volume $\tau \pi R_A^2$ with $\tau \approx 1$ fm/c lead to estimated energy density $\approx 5$ GeV/fm$^3$ and particle density $\approx 10$/fm$^3$~\cite{hydro}. According to lattice QCD such densities should provide conditions required for a deconfinement phase transition to a QGP.  A cross section of 1-2 mb with mean free paths less than 1 fm would imply copious rescattering,  thermalization and flows responding to large density gradients.

A key element in the conflict between \ha\ transparency and the claimed \aa\ QGP is interpretation of the NJ quadrupole component of azimuth correlations.  If the NJ quadrupole is indeed a flow manifestation the flow-QGP narrative is supported, but if the quadrupole component is demonstrated to be a novel {\em nonflow} QCD phenomenon the conjectured flowing dense medium or QGP is unlikely. In the present study $v_2(b)$ (\pt-integral) and $v_2(p_t,b)$ (\pt-differential) data as inferred from several sources are reconsidered in the context of jets and flows. Contributions to $v_2$ data from jets are identified and removed, and the remaining NJ $v_2$ component is compared with hydro expectations to challenge the claim that $v_2$ data represent a flow phenomenon. The main result of this study is extraction of quadrupole spectra from $v_2(p_t)$ data that appear to be inconsistent with the flow-QGP narrative.

 \section{General analysis methods} \label{methods}

This section reviews basic analysis methods for yields, spectra and correlations applied to high-energy \pp, \pa, \da\ and \aa\ collisions within a TCM context. 

\subsection{\aa\ collision geometry} \label{aageom}
 
Minimum-bias distributions of \aa\ cross section $\sigma$ on nucleon participant number $N_{part}$ and \nn\ binary collision number $N_{bin}$ are accurately described by {\em power-law trends} leading to simple parametrizations in terms of fractional cross section $\sigma/\sigma_0$~\cite{powerlaw}
\bea \label{glauber}
(N_{part}/2)^{1/4} &=& 0.5^{1/4}\frac{\sigma }{\sigma_0} + (N_{part,max}/2)^{1/4} \left(1-\frac{\sigma }{\sigma_0}\right)
\nonumber \\ 
N_{bin}^{1/6} &=&  0.5^{1/6}\frac{\sigma }{\sigma_0} + N_{bin,max}^{1/6}\left(1-\frac{\sigma }{\sigma_0}\right),
\eea
with $N_{part,max} = 382$ and $N_{bin,max} = 1136$ for 200 GeV \auau\ collisions (with $\sigma_{NN} = 42$ mb). Those parametrizations describe Glauber simulations at the percent level. The same trends may be used for all energies above $\sqrt{s_{NN}} \approx 30$ GeV as purely geometric centrality measures. Another useful centrality measure is participant path length $\nu \equiv 2 N_{bin} / N_{part}$ which provides good visual access to the more-peripheral data required to test a \nn\ linear-superposition hypothesis. Because A for Pb is 5\% larger than for Au $N_{part}$ increases accordingly. $N_{bin} \propto N_{part}^{4/3}$ should then increase by 7\% and ratio $\nu \propto N_{part}^{1/3}$  by less than 2\% (omitted as negligible). 

Parameter $\nu$ defined above should correspond to mean path length $\nu_{hA}$ defined previously for $h$-$A$ collisions. The mean number of nucleons encountered in target  $A$ by incident hadron $h$ ($h$-$N$ ``collisions'') is defined by~\cite{witpa}
\bea
\nu_{hA} &=& A \frac{\sigma_{hN}}{\sigma_{hA}}
\approx \frac{N_{bin}}{N_{part}},
\eea
where for a given integrated luminosity $\sigma_{hA}$ measures the mean number of participant (interacting) incident beam hadrons per target nucleus  and $A \sigma_{hN}$ measures the corresponding mean number of  binary $h$-$N$ encounters. Since $N_{part}$ for \aa\ collisions usually represents the total number of participants in two nuclei the corresponding \aa\ expression is $\nu_{AA} = 2 N_{bin} / N_{part} \rightarrow \nu$ as defined above.

Eccentricity $\epsilon(b)$ measures the shape of the transverse overlap region for noncentral \aa\ collisions ($b$ is the \aa\ impact parameter). Optical eccentricity $\epsilon_{opt}$ is accurately parametrized by~\cite{davidhq}
\bea
\epsilon_{opt}(N_{bin}) \hspace{-.00in} &=& \hspace{-.00in} 0.2\,\log_{10}(N_{bin}/1)\log_{10}(1136/N_{bin})^{0.8}~~~
\\ \nonumber
 &\rightarrow& 1.5 x(1-x)^{0.8},
\eea
essentially a beta distribution on logarithm ratio $x = \log(N_{bin}) /  \log(N_{bin,max}) \approx  \log(N_{part}) /  \log(N_{part,max})$ with $N_{bin,max} \approx 1136$ and $N_{part,max} \approx 382$ for 200 GeV \auau\ collisions.  Because $\epsilon_{opt}$ depends on log ratios it is essentially constant from Au to Pb. Optical eccentricity $\epsilon_{opt}$  based on a continuum optical-model nuclear density is distinguished from Monte Carlo eccentricity $\epsilon_{MC}$ derived from a  model based on discrete nucleons~\cite{alver}. 

\subsection{TCM for $\bf p_t$ and $\bf y_t$ hadron spectra and yields} \label{tcmspectra}

Particle spectra may be presented as densities on transverse momentum \pt, mass $m_t = \sqrt{p_t^2 + m_h^2}$ or rapidity $y_t = \ln[(p_t + m_t)/m_h]$ where $m_h$ is a hadron mass.
The TCM for \aa\ $y_t$ spectra conditional on uncorrected $n_{ch}'$ (representing impact parameter $b$) averaged over $2\pi$ azimuth and some $\eta$ acceptance $\Delta \eta$ is represented by~\cite{hardspec}
\bea \label{ppspec}
\bar \rho_{0}(y_t,b) &=&  \frac{N_{part}}{2}S_{NN}(y_t,b) + N_{bin }H_{AA}(y_t,b)
\\ \nonumber
\frac{2}{N_{part}}\bar \rho_{0}(y_t,b) &=& \bar \rho_{s,NN} \hat S_0(y_t)  + \nu\,  \bar \rho_{h}(b) \hat r_{AA}(y_t,b) \hat H_0(y_t)
\\ \nonumber
\frac{2}{N_{part}}\bar \rho_{0}(b)&=& \bar \rho_{NN} [1 + x(b) (\nu - 1)],
\eea
where  $\bar \rho_s = n_s / 2\pi \Delta \eta$ and $\bar \rho_h = n_h / 2\pi \Delta \eta$ are $\eta$- and $\phi$-averaged soft and hard hadron angular densities corresponding to single \nn\ ($\approx$ \pp) encounters within \aa\ collisions with $H_{AA} \equiv r_{AA} H_{NN} = \bar \rho_h \hat r_{AA} \hat H_0$. Inferred soft and hard {\em reference} $y_t$ spectrum shapes [unit normal $\hat S_0(y_t)$ and $\hat H_0(y_t)$ for \pp\ collisions] are assumed independent of $n_{ch}'$ or $b$~\cite{ppprd}, with parametrized forms defined in Refs.~\cite{hardspec,fragevo}. Conversion from densities on \pt\ or $m_t$ to densities on \yt\ is via Jacobian factor $p_t m_t / y_t$.  

The fixed soft component is a L\'evy distribution on $m_t$
\bea \label{s0}
\hat S_0(m_t) &\equiv& \frac{A(n_0,T_0)}{[1 + (m_t - m_h) / (n_0 T_0)]^{n_0}},
\eea
with slope parameter $T_0$ and L\'evy exponent $n_0$, that goes to a Maxwell-Boltzmann (M-B) exponential on \mt\ in the limit $1/n_0 \rightarrow 0$. The fixed \pp\ hard-component model $\hat H_0(y_t)$ (Gaussian plus exponential tail) is determined by Gaussian centroid $\bar y_t$, Gaussian width $\sigma_{y_t}$ and ``power-law'' parameter $q$. The slope at the transition point from Gaussian to exponential is required to be continuous. An algorithm for computing $\hat H_0(y_t)$ is provided in Ref.~\cite{hardspec} (App.~A). Quantity $r_{AA}(y_t,b)$ includes all information on modification of jet contributions to \aa\ spectra relative to \pp\ over the full \pt\  acceptance, in contrast to ratio $R_{AA}$ that provides no information at lower \pt~\cite{hardspec}.  Product $\hat H_{AA}(y_t,b) =\hat r_{AA}(y_t,b) \hat H_0(y_t)$ is then a unit-normal function describing the  \aa\ spectrum hard component. The {\em Glauber linear superposition} (GLS) model for \aa\ spectra corresponds to $r_{AA}(y_t) \rightarrow 1$ and $H_{AA}(y_t,b) \rightarrow \bar \rho_{h,NN} \hat H_0(y_t)$, the expression describing minimum-bias (MB) \nn\ ($\approx$ \pp) collisions.

The TCM for hadron production near midrapidity is consistent with the assumption that  production proceeds exclusively via low-x gluons, either directly through projectile-nucleon dissociation or indirectly through large-angle gluon scattering and jet formation. As collision energy falls well below 100 GeV valence quarks play an increasing role near midrapidity, the low-x gluon contribution falls to zero near 10 GeV and the TCM is not relevant at  lower energies. See the discussion in Sec.~V-B of Ref.~\cite{alicetomspec} relating to its Fig.~7.

\subsection{2D angular correlations}

Two-particle correlations are structures observed within a particle-pair density defined on 6D momentum space $(p_{t1},\eta_1,\phi_1,p_{t2},\eta_2,\phi_2)$ where $\eta$ is pseudorapidity and $\phi$ is azimuth angle. Given the observed  invariance of angular correlations on sum variables $x_\Sigma = x_1 + x_2$ near midrapidity 2D pair densities can be {\em projected by averaging} onto difference variables $x_\Delta = x_1 - x_2$ to obtain a {\em joint angular autocorrelation} on the reduced space $(p_{t1},p_{t2},\eta_\Delta,\phi_\Delta)$~\cite{inverse}. 
That distribution can be further reduced by integration over \pt\ bins or the entire \pt\ acceptance (e.g.\ to form a {\em marginal projection} onto a single \pt\ variable). The present study emphasizes conventional projection of 2D angular correlations onto 1D azimuth by averaging over $\eta_\Delta$ within some acceptance $\Delta \eta$, for \pt-integral correlations and for marginals on \pt.

A  measured (uncorrected) pair density denoted by $\rho'(x_1,x_2)$ is corrected to $\rho(x_1,x_2)$ through mixed-pair density $\rho_{mix}$ constructed with particle pairs sampled from different but similar events~\cite{reid}. The corrected pair density can then be compared (assuming factorization) with reference $\rho_{ref}(x_1,x_2) = \bar \rho_0(x_1)\bar \rho_0(x_2) \approx \rho_{mix}(x_1,x_2)$ to define the {\em correlated-pair density}
\bea
\Delta \rho(p_{t1},p_{t2},\eta_\Delta,\phi_\Delta) \equiv \rho_{ref} \,(\rho' / \rho_{mix} - 1).
\eea
The corrected correlated-pair density is an {\em extensive} correlation measure that may contain {\em cylindrical multipole} structures on 1D $\phi_\Delta$ (e.g.\ dipole, quadrupole, sextupole, etc.) represented by power-spectrum amplitudes $V^2_m$ (defined below) as well as distinct 2D structures on $(\eta_\Delta,\phi_\Delta)$. In previous studies {\em per-particle} measure $\Delta \rho / \sqrt{\rho_{ref}} \equiv \sqrt{\rho_{ref}}(\rho' / \rho_{mix} - 1) \rightarrow \bar \rho_0(\rho' / \rho_{mix} - 1)$ was defined with multipole amplitudes $V^2_m / \bar \rho_0 =\bar \rho_0 v_m^2$~\cite{anomalous,davidhq}. For direct comparisons with conventional $v_m$ measures {\em per-pair} measure $\Delta \rho / \rho_{ref} =(\rho' / \rho_{mix} - 1)$ was also defined with multipole amplitudes $v_m^2 = V^2_m / \bar \rho_0^2$~\cite{davidhq2,v2ptb}.

Measured 2D angular correlations have a simple structure that can be modeled by a few 1D and 2D functions on $(\eta_\Delta,\phi_\Delta)$~\cite{anomalous,v2ptb}. In discussing 2D models it is convenient to separate azimuth difference $\phi_\Delta$ into two regions: {\em same-side} (SS, $|\phi_\Delta| < \pi / 2$) and {\em away-side} (AS, $|\phi_\Delta - \pi| < \pi / 2$). The six-element fit model of Ref.~\cite{anomalous} includes eleven model parameters but describes more than 150 data degrees of freedom for typical $25\times 25$ data histograms on $(\eta_\Delta,\phi_\Delta)$. The model parameters are thus strongly constrained. A NJ quadrupole component of 2D angular correlations can be extracted accurately via such model fits. For per-particle 2D angular correlations in the form $\Delta \rho / \sqrt{\rho_{ref}}$ the fitted NJ quadrupole amplitude is represented by symbol $A_Q\{2D\} \equiv \bar \rho_0 v_2^2\{2D\}$.

 \section{$\bf v_2$ analysis methods} \label{v2meths}

The form of quantity $v_2 = \langle \cos(2\phi) \rangle$ as a measure of elliptic flow is motivated by expectations that the hadronic final state of ultrarelativistic \aa\ collisions is dominated at lower \pt\ by ``anisotropic [on azimuth] flow''~\cite{ollitrault}. As demonstrated below $v_2$ is effectively (the square root of) a per-pair measure derived from a number of correlated pairs in ratio to a number of reference (mixed) pairs.

$v_2$ can be estimated by several techniques (methods) including {\em nongraphical numerical methods} (NGNM) based on 1D  fits to pair distributions on azimuth-difference $\phi_\Delta$ or a method based on 2D fits to angular correlations on $(\eta_\Delta,\phi_\Delta)$~\cite{davidhq,v2ptb}. The suffix ``$\{\text{method}\}$'' appended to $v_2$-related measures denotes the analysis method: $\{2\}$ or $\{4\}$ signifies inference from two- or four-particle azimuth correlations (integrated over some $\Delta \eta$ acceptance), $\{\text{EP}\}$ signifies inference from an {\em event-plane} method (discussed below), $\{\text{2D}\}$ signifies inference from model fits to 2D angular correlations, and so forth.

Several physical mechanisms may contribute to inferred $v_2$ data depending on the collision system, conventionally separated into ``flow'' (elliptic flow) and ``nonflow'' [e.g.\ jets, Bose-Einstein correlations (BEC), conversion-electron pairs, hadron-resonance correlations]. Various $v_2$ methods are more or less able to differentiate among the several correlation mechanisms (i.e.\ to distinguish flow from nonflow). Whereas model fits to 2D angular correlations arguably retain all information conveyed in those correlations, NGNM analysis applied to projections onto 1D $\phi_\Delta$ may discard crucial information required for such distinctions~\cite{v2ptb}.

In this section several $v_2$ analysis methods are defined and compared. Methods for \pt-integral $v_2(b)$ analysis are reviewed first and then generalized to \pt-differential methods leading to $v_2(p_t,b)$ data and  quadrupole spectra.

\subsection{$\bf v_2(b)$ analysis on 1D azimuth}

Conventional $v_2$ methods project some or all 2D angular correlations onto 1D azimuth and {\em in effect} fit the 1D projection with a Fourier series in which any term may be interpreted to represent a flow~\cite{poskvol,luzum}. It is assumed that some terms (``harmonics'') may relate to an \aa\ reaction plane. 
Whatever the production mechanisms any 2D angular correlations projected onto 1D azimuth difference $\phi_\Delta$ can be described exactly by a Fourier series
\bea \label{power}
\rho(\phi_\Delta) \hspace{-.05in} &=&\hspace{-.05in} \frac{\bar \rho_0(b)}{\Delta \eta} \delta(\phi_\Delta)\hspace{-.02in} +\hspace{-.02in} V_0^2(b) \hspace{-.02in}+\hspace{-.02in} 2 \hspace{-.02in} \sum_{m=1}^\infty \hspace{-.02in} V_m^2(b) \cos(m\phi_\Delta),~~
\eea
where $V_m^2(b) = \bar \rho_0^2(b)\, v_m^2(b)$ and $v_0^2 \equiv 1$. The first term represents self pairs.  The coefficients denoted by $V_m^2\{2\}$ represent {\em all} two-particle azimuth correlations directly.  Since $V_0^2(b) = \bar \rho_0^2 = \rho_{ref}$, correlated-pair density $\Delta \rho(\phi_\Delta) = \rho(\phi_\Delta) - \rho_{ref}$ (excluding self pairs) is represented by  Fourier terms with $m > 0$. 
In the present study I emphasize the $m = 2$ quadrupole term and invert Eq.~(\ref{power}) (excluding self pairs $i = j$) to obtain
\bea \label{v22}
V_2^2\{2\}(b) 
&=& \frac{1}{(2\pi \Delta \eta)^2} \sum_{i \neq j=1 \in \Delta \eta}^{n,n-1} \vec u(2\phi_i) \cdot \vec u(2\phi_j) 
 \nonumber  \\
&=& \rho_0^2(b) \langle \cos(2\phi_\Delta) \rangle \equiv \rho_0^2(b) v_2^2\{2\}(b),
\eea
where $\vec u(\phi)$ is a unit vector in the $\phi$ direction, $\vec u(2\phi_i) \cdot \vec u(2\phi_j) = \cos[2(\phi_i - \phi_j)]$ and $V_2^2\{2\}$ is the basic observable representing the quadrupole amplitude for all projected 1D azimuth correlations, including jet-related correlations as well as what might be associated with flows.

\subsection{$\bf v_2(b)$ analysis by the event-plane method}

In the conventional flow narrative $v_2$ analysis near midrapidity is based on a 1D Fourier decomposition of the $\eta$-averaged particle density $\bar \rho_0(\phi,b)$. The Fourier series is defined relative to a  {\em reaction-plane angle} $\Psi_r$~\cite{poskvol}
\bea \label{fourier1}
 \bar  \rho(\phi,b) &=&\bar \rho_0(b) \left\{ 1 + 2\sum_{m=1}^\infty   v_m(b) \cos[m(\phi - \Psi_r)] \right\},~~
 \eea
where $\bar \rho_0(b) \rightarrow V_0(b)$ is the single-particle 2D angular charge density averaged over acceptance $(\Delta \eta,2\pi)$, and the Fourier amplitudes appear as ratios $ v_m(b) =  V_m(b) /  V_0(b)$. In the defining Ref.~\cite{poskvol} and other sources the Fourier series is  assumed to be dominated by ``anisotropic flows'' relative to a reaction plane but could represent several other physical mechanisms including MB jets (mainly {\em minijets} with $E_{jet} \approx 3$ GeV)~\cite{anomalous}. The equation is nonphysical since $\Psi_r$ is not observable, and the $v_m$ are not accessible by simple 1D series inversion. 

Within the flow narrative $\Psi_r$ must be estimated from some subset of collision products, the estimates described as {\em event-plane angles} $\Psi_m$. Equation~(\ref{fourier1}) is rewritten  as
\bea \label{fourier2}
 \bar \rho(\phi,b) 
&=&\bar \rho_0(b) \left\{ 1 + 2\sum_{m=1}^\infty \hspace{-.05in}   v_m'(b) \cos\left[m(\phi - \Psi_m)\right] \right\},~~~~
 \eea
with the $\Psi_m$ defined by $ \vec Q_m \equiv ({1}/{2\pi \Delta \eta})\sum_{j=1}^n \vec u(m \phi_j) = Q_m\, \vec u(m\, \Psi_m)$ ($Q$ vectors). Coefficients $Q_m$ are commonly interpreted in a context where most azimuth structure is hydrodynamic in origin (flows) and relates to a reaction plane. However, the $\vec Q_m$ may contain substantial ``nonflow'' contributions dominated by Fourier coefficients of a jet-related same-side 2D jet peak~\cite{gluequad,tzyam}. The inferred event-plane angles $\Psi_m$ may even be  unrelated to a physical  \aa\ reaction plane.

For $v_2$ analysis the $m=2$ (quadrupole) $Q$ vector
\bea \label{qvec}
\vec Q_2 &=& \frac{1}{2\pi \Delta \eta} \sum_{j=1}^n \vec u(2\phi_j) \equiv Q_2 \vec u(2 \Psi_2)
\eea
determines quadrupole event-plane angle $\Psi_2$. $\vec Q_2$ thus defined is a 2D angular density, not just a sum over $n_{ch} \rightarrow n$ particles. The $Q$-vector magnitude is obtained from
\bea \label{q22}
Q_2^2\{2\} &=& \vec Q_2 \cdot \vec Q_2 =  \frac{\bar \rho_0}{2\pi \Delta \eta} + V_2^2\{2\}
\eea
including the {\em self-pair term} ${\bar \rho_0}/{2\pi \Delta \eta}$.

In an attempt to avoid ``autocorrelations''\footnote{Within the flow narrative self pairs are denoted by this term which in conventional mathematics refers to a cross-correlation measure initially relating to statistical analysis of time series~\cite{auto}. An autocorrelation on azimuth is defined by Eq.~(\ref{power}) for example.} Eq.~(\ref{qvec}) is modified to exclude the $i^{th}$ particle in the sum over $j$ so that $\vec Q_2 \rightarrow \vec Q_{2i}$~\cite{poskvol}. Although self pairs are then excluded from the implicit double sum in Eq.~(\ref{v2obs}) below they persist as the self-pair term ${\bar \rho_0} / {2\pi \Delta \eta}$ in Eq.~(\ref{q22}) that defines magnitude $Q_2 = \sqrt{Q_2^2} \rightarrow Q_{2i}$, thus requiring an ``event-plane resolution'' correction defined by~\cite{quadspec}
\bea
\langle \cos[2(\Psi_2 - \Psi_r)] \rangle &\approx& \frac{V_2\{2\}}{Q_2\{2\}} =  \sqrt{\frac{v_2^2\{2\} }{1/n_{ch} + v_2^2\{2\}}}.~~~
\eea

The observed (uncorrected) event-plane $v_2' \rightarrow v_{2,obs}$ is obtained by inverting Eq.~(\ref{fourier2}) for $m = 2$ in the form
\bea \label{v2obs}
v_{2,obs} \hspace{-.05in} &\equiv& \hspace{-.05in} \frac{1}{n} \sum_{i=1}^n \vec u(2\phi_i) \cdot \vec u(2\Psi_{2i}) = \langle \cos[2 (\phi - \Psi_2 )] \rangle~~
\eea
which still includes self pairs implicitly through $Q_{2i} \approx Q_2$ and must then be corrected by event-plane resolution $\langle \cos[2(\Psi_2 - \Psi_r)]\rangle$ to obtain final event-plane $v_2$ measure
\bea
v_2\{EP\} &\equiv& \frac{v_{2obs}}{\langle \cos[2(\Psi_2 - \Psi_r)]\rangle} \approx  \frac{v_{2obs}}{{V_2\{2\}}/{Q_2\{2\}}}.
\eea
But Eq.~(\ref{fourier2}) that seems to be equivalent to  Eq.~(\ref{fourier1}) (a SP density) is directly related to a two-particle autocorrelation function. Equation (\ref{v22}) derived from Eq.~(\ref{power}) (a true autocorrelation~\cite{inverse}) can be reexpressed as
\bea
V_2^2\{2\} 
&=& \frac{1}{2\pi \Delta \eta} \sum_{i=1}^n \vec u(2\phi_i) \cdot Q_{2i}\{2\} \vec u(2\Psi_{2i})~~~ \\ \nonumber
&\approx& \bar \rho_0 \frac{v_{2obs}}{{V_2\{2\}}/{Q_2\{2\}}} V_2\{2\}~~~\text{or}
\\ \nonumber
\bar \rho_0^2 v_2^2\{2\}&\approx& \bar \rho_0v_2\{EP\} \bar \rho_0 v_2\{2\},
\eea
implying that $v_2\{EP\} \approx v_2\{2\}$ (the approximation arising only from $Q_{2i} \approx Q_2$). Possible small differences between published $v_2\{EP\}$ and $v_2\{2\}$ ($< 5$\%~\cite{2004}) may be caused by minor deviations from ideal factorization (covariance contributions). ``Subevents'' defined in various manifestations of the EP method~\cite{subevt} are equivalent to conventional binnings on $\eta$ (e.g.\ to form 2D angular autocorrelations~\cite{anomalous}), particle charge or hadron species.

Reference~\cite{poskvol} acknowledges the possibility of $v_2\{2\}$ obtained from a pair density as in Eq.~(\ref{power}) but warns that $v_2^2\{2\}$ is a small quantity and the determination of data errors might be problematic. However, that same pair (autocorrelation) density is actually the source of $v_2\{EP\}$: the established autocorrelation technique is in effect {\em notionally reinvented} in the context of the flow narrative. The event- or reaction-plane concept is not required to analyze azimuth correlations whose relation to a reaction plane remains an open question.  Below I refer to $v_2\{2\}$ unless introducing published $v_2\{EP\}$ data.

\subsection{$\bf v_2(p_t,b)$ analysis on 1D azimuth}

Equation~(\ref{v22}) describing the \pt-integral case can be generalized to a joint distribution on $(p_{t1},p_{t2})$
\bea
 V_2^2\{2\}(p_{t1},p_{t2},b)  \hspace{-.05in}
&=&  \hspace{-.15in}  \sum_{i \in p_{t1} \neq j \in p_{t2}}^{n_{p_{t1}},n_{p_{t2}}} \hspace{-.05in} \frac{\cos[2(\phi_i - \phi_j)]}{(2\pi \Delta \eta)^2 p_{t1}dp_{t1}p_{t2}dp_{t2}}~~~~
\\ \nonumber
&\equiv&V_2(p_{t1},b) V_2(p_{t2},b)  \\ \nonumber
&=& \bar \rho_0(p_{t1}) v_2\{2\}(p_{t1}) \bar \rho_0(p_{t2}) v_2\{2\}(p_{t2}),
\eea
where $n_{p_t}$ is the particle number within a \pt\ bin at \pt\ integrated over acceptance $(\Delta \eta,2\pi)$. The marginal projection onto a single \pt\ variable is
\bea \label{v2pt2}
 V_2^2\{2\}(p_{t},b) 
&=&\frac{1}{(2\pi \Delta \eta)^2} \sum_{i \in p_{t}\neq j=1 }^{n_{p_t},n-1} \frac{\cos[2(\phi_i - \phi_j)]}{p_{t}dp_{t}}~~~
\\ \nonumber
&\equiv&V_2\{2\}(b) V_2\{2\}(p_{t},b)  \\ \nonumber
&=& \bar \rho_0(b) v_2\{2\}(b) \bar \rho_0(p_t,b) v_2\{2\}(p_t,b)
\eea
where $\bar \rho_0(p_t,b)$ is a SP \pt\ spectrum and $V_2(p_t,b) =  \bar \rho_0(p_t,b) v_2(p_t,b)$ includes the NJ {\em quadrupole spectrum} $\bar \rho_2(p_t,b)$ as a factor~\cite{quadspec} (see Sec.~\ref{quadspecmeth}). Inference of quadrupole spectra from $v_2(p_t,b)$ data is a main focus of the present study. Note that the integral relation
\bea \label{v2int}
V_2^2\{2\}(b) &=& \int_0^\infty dp_t\, p_t\, V_2^2\{2\}(p_t,b)~~\text{or}
\\ \nonumber
\bar \rho_0(b) v_2\{2\}(b) &=& \int_0^\infty dp_t\, p_t\, \bar \rho_0(p_t,b) v_2\{2\}(p_t,b)
\eea
 combined with Eq.~(\ref{v2pt2}) is equivalent to Eq.~(\ref{v22}). Whereas $V_2^2\{2\}(b)$ or $V_2\{2\}(b)$ is a basic correlation measure derived from a pair distribution, $\bar \rho_0(b) v_2\{2\}(p_t,b)$ is an arbitrary factorization with questionable interpretation.

\subsection{$\bf v_2\{2D\}$ inferred from 2D angular correlations} \label{v22dpp}

\begin{figure}[h]
    \includegraphics[width=1.49in,height=1.5in]{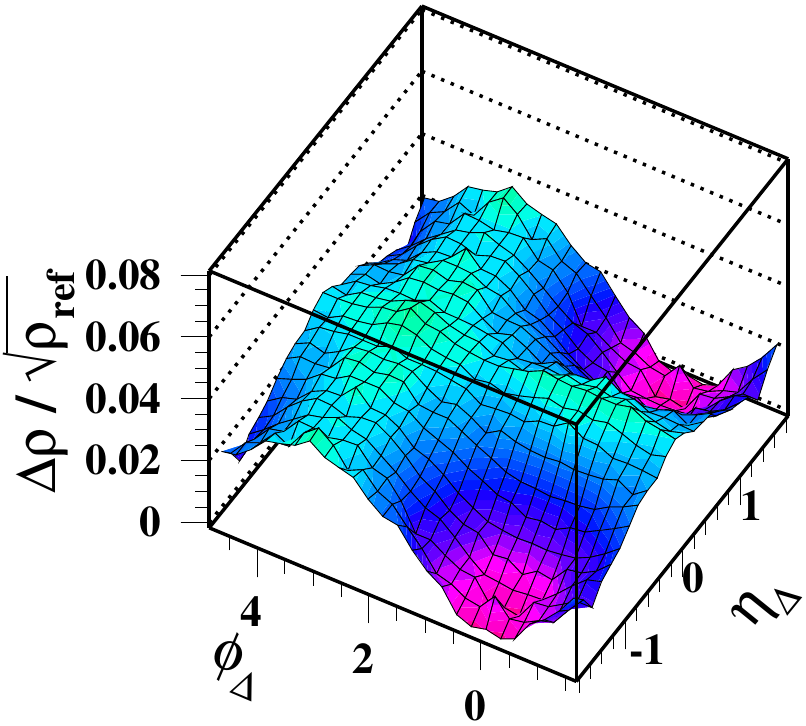}
     \includegraphics[width=1.49in,height=1.48in]{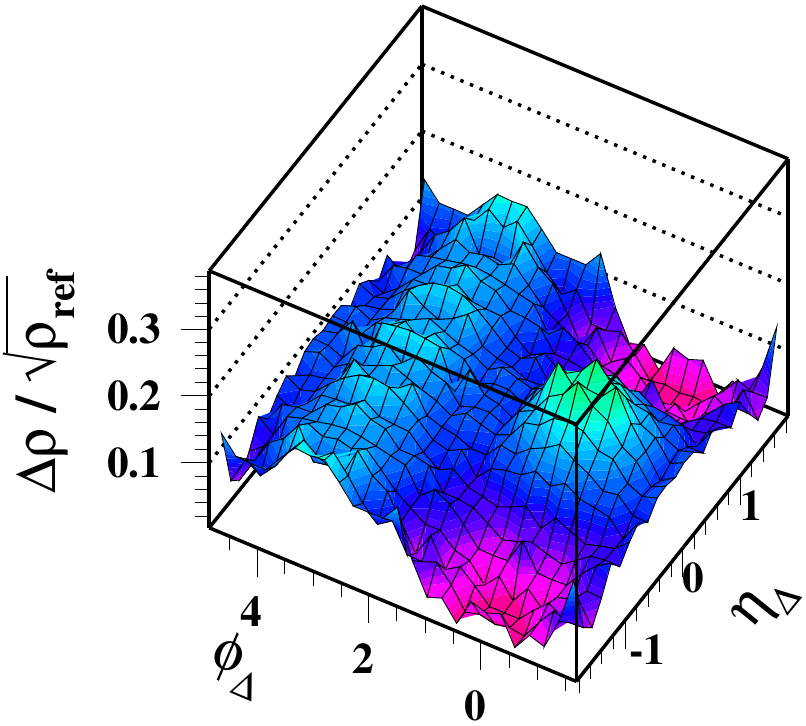}
 \caption{ \label{ppcorr}
2D angular correlations from 200 GeV \pp\ collisions for two charge multiplicity classes ($\bar \rho_0 = 1.8$, 13.7) from Ref.~\cite{ppquad}. Fitted model elements corresponding to a soft component (projectile-proton dissociation), Bose-Einstein correlations and gamma-conversion electrons have been subtracted leaving jet-related and NJ quadrupole contributions.
}
\end{figure}

Figure~\ref{ppcorr} shows angular correlations for two multiplicity classes (index $n = 1$, 6) from a high-statistics study of the charge-multiplicity dependence of 200 GeV \pp\ collisions~\cite{ppquad}. Fitted 2D model elements representing a soft component (projectile dissociation) and BEC/electrons have been subtracted from data histograms to isolate jet-related and NJ-quadrupole components.
The jet-related structures are a SS 2D peak at the angular origin ({\em intra}\,jet correlations) and an AS 1D peak at $\pi$ on azimuth ({\em inter}\,jet correlations). Substantial elongation of the SS 2D peak on azimuth is evident in the left panel. The AS 1D peak is broad enough to be described well by a single azimuth-dipole element~\cite{multipoles}.

\begin{figure}[h]
     \includegraphics[width=1.49in,height=1.47in]{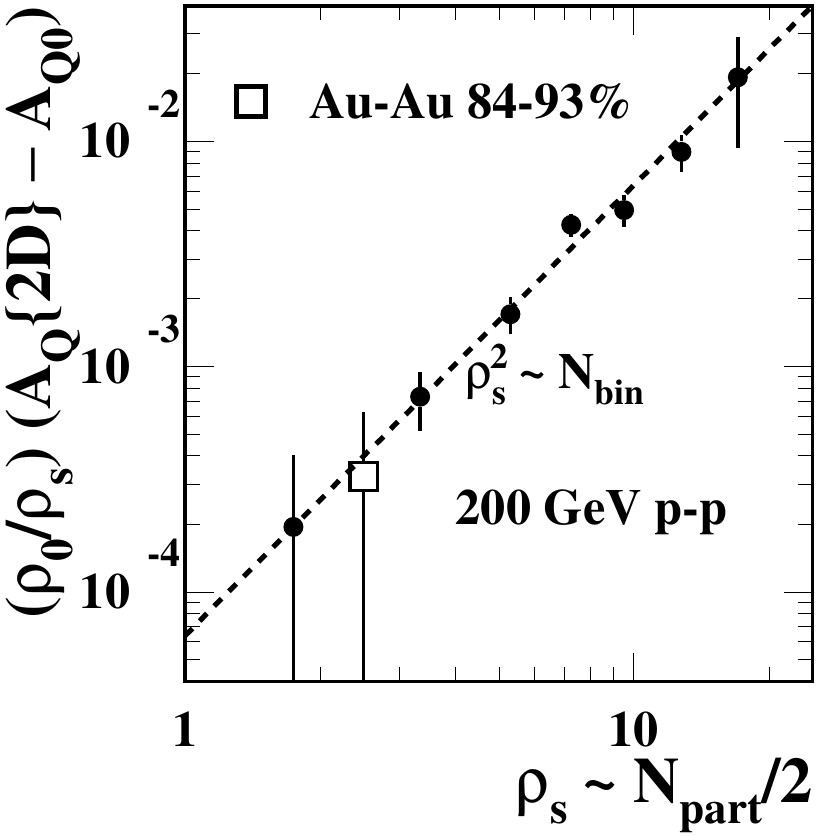}
    \includegraphics[width=1.49in,height=1.5in]{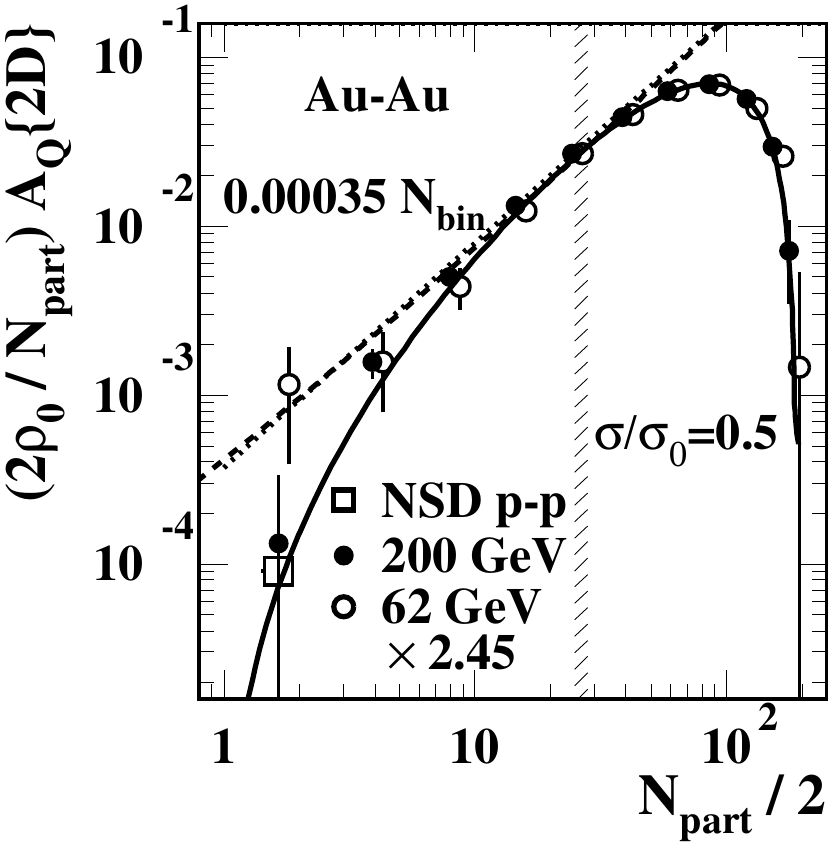}
 \caption{ \label{quadx} 
Left: NJ quadrupole correlations from 200 GeV \pp\ collisions (quantity $\propto$ number of correlated pairs per soft hadron) vs soft-component hadron density $\bar \rho_s$ illustrating a trend $\propto \bar \rho_s^2 \sim N_{bin}$ (dashed line)~\cite{ppquad}.
Right: NJ quadrupole correlations from 62 (open circles) and 200 (solid points) GeV \auau\ collisions  (quantity $\propto$ number of correlated pairs per nucleon participant) vs number of participant pairs illustrating a trend $\propto N_{bin} \epsilon_{opt}^2$ [solid curve, Eq.~(\ref{magic})]~\cite{davidhq,v2ptb}. The 62 GeV data are plotted at $1.1 N_{part}/2$ values for clarity. The 62 and 200 GeV data are consistent with factorization of $V_2(b,\sqrt{s_{NN}})$ trends in that energy interval (see Sec.~\ref{quadenergy}).
} 
\end{figure}

Figure~\ref{quadx} (left) shows the fitted NJ quadrupole amplitude in the form $(\bar \rho_0 / \bar \rho_s) A_Q\{2D\}$ vs soft charge density $\bar \rho_s$ with the trend $\bar \rho_0 A_Q\{2D\} = V_2^2\{2D\} \propto \bar \rho_s^3$ (dashed line)~\cite{ppquad}. A small offset $A_{Q0} = 0.0007$ independent of $\bar \rho_s$ and representing global transverse-momentum conservation has been subtracted from the $A_Q$ data. In previous \pp\ collision studies  the trend dijets $\propto \bar \rho_s^2$ was inferred, suggesting that $\bar \rho_s$ plays the role of participant (low-$x$ gluon) number $N_{part}$ and $\bar \rho_s^2$ plays the role of binary-collision number $N_{bin}$~\cite{tomalicempt,alicetomspec,ppquad}. The \pp\ quadrupole data are then consistent with $V_2^2\{2D\} \propto N_{part} \times N_{bin}$. 

The NJ quadrupole amplitude thus increases very rapidly with increasing charge multiplicity, as is evident by comparing the two panels of Fig.~\ref{ppcorr}. In the right panel the two positive lobes of the NJ quadrupole have doubled the negative curvature of the AS 1D peak at $\pi$ and greatly reduced the positive curvature near $\phi_\Delta = 0$ (for $|\eta_\Delta| > 1$). Quadrupole superposition onto the SS peak gives the  impression that it is much narrower on azimuth, but 2D fits indicate little change in the SS peak shape.

\begin{figure}[h]
    \includegraphics[width=1.49in,height=1.5in]{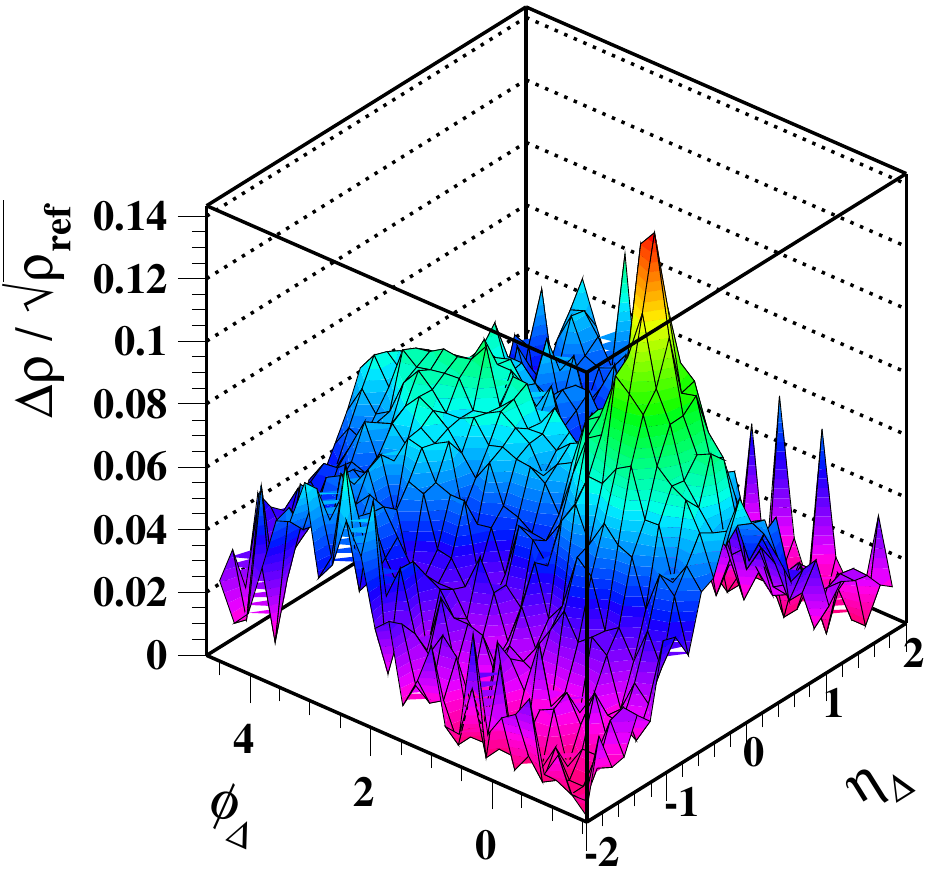}
     \includegraphics[width=1.49in,height=1.48in]{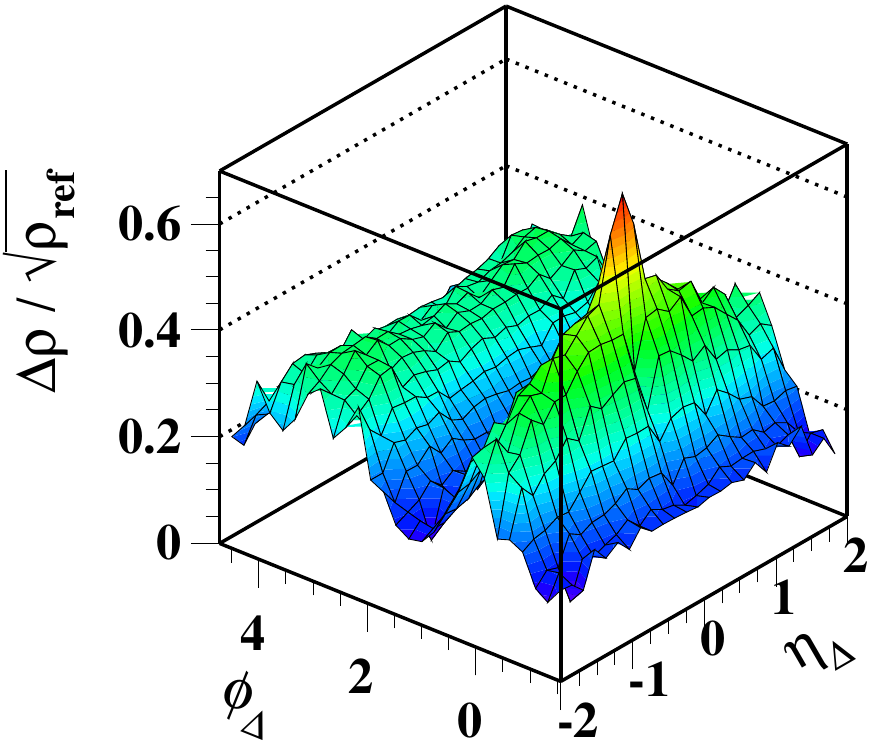}
 \caption{ \label{auau}  
2D angular correlations from most-peripheral (left) and most-central (right) 200 GeV \auau\ collisions~\cite{anomalous}. The prominent structures are jet-related (same-side 2D peak dominated by unlike-sign pairs and away-side ridge), soft (narrow 1D peak on pseudorapidity difference $\eta_\Delta$) and Bose-Einstein correlations (same-side 2D peak dominated by like-sign pairs). The NJ quadrupole is not visible for these cases. The same-side 2D peak is strongly elongated on $\eta_\Delta$ in central collisions.
} 
\end{figure}

Figure~\ref{auau} shows 2D angular correlations from the most peripheral (left) and most central (right) 200 GeV \auau\ collisions. The statistical errors for those histograms are about 4.5 times larger than for the high-statistics \pp\ data in Fig.~\ref{ppcorr}. The same six-element 2D fit model was applied to the \auau\ data~\cite{anomalous}. The peripheral \auau\ data are approximately equivalent to NSD \pp\ data similar to Figure~\ref{ppcorr} (left) (but before subtraction of two model elements). The NJ quadrupole  in those panels is very small compared to both the jet-related structure (in both panels) and the soft component (1D peak on $\eta_\Delta$ at the origin in the left panel).

Figure~\ref{quadx} (right) shows \auau\  NJ quadrupole amplitudes (points) in the form $(2 \bar \rho_0 / N_{part}) A_Q\{2D\}$ vs participant-pair number $N_{part} / 2$ for 62 and 200 GeV data~\cite{davidhq,noelliptic}. The same trend $\bar \rho_0(b)A_Q\{2D\}(b) = V_2^2\{2D\}(b) \propto N_{part}(b) N_{bin}(b) \epsilon_{opt}^2(b)$ (solid curve) describes both energies given a factor $1.4 \times 1.75 =$ 2.45 (with 1.75 from $A_Q\{2D\}$ in Fig.~\ref{v2ep}). The $\epsilon_{opt}$, $N_{part}$ and $N_{bin}$ functions are as defined in Sec.~\ref{aageom}. The 200 GeV solid curve is then defined by (given $\bar \rho_0 = n_{ch} / 2\pi \Delta \eta$)~\cite{davidhq} 
\bea \label{magic}
[2\bar \rho_0(b) /N_{part}] A_Q\{2D\}(b) &=& 0.0022 N_{bin} \epsilon_{opt}^2(b).
\eea
The dashed curve is $(2\bar \rho_0/N_{part}) A_Q\{2D\} = 0.00035 N_{bin}$ consistent with the maximum value of $\epsilon_{opt}$ being near 0.4. The dotted curve is that expression with $N_{bin} \rightarrow (N_{part}/2)^{4/3}$ which is equivalent within a few percent. For \pp\ collisions $N_{bin} \sim  (N_{part}/2)^2 \sim \bar \rho_s^2$ is inferred~\cite{ppquad}. The hatched band marks the position on \auau\ centrality ($\sigma / \sigma_0 \approx 0.5$) of a {\em sharp transition} in the systematics of jet-related 2D angular correlations~\cite{anomalous} coinciding with major changes in the \pt-spectrum hard component~\cite{hardspec}. There is no apparent correspondence between ``jet quenching'' and the NJ azimuth quadrupole~\cite{noelliptic}.

Based on model fits to 2D angular correlations unbiased correlation components can be isolated, including a NJ quadrupole component {\em common to both \pp\ and \aa\ collisions} with equivalent dependence on geometry parameters $N_{part}$ and $N_{bin}$ in either case. Differences (in $N_{part}$ vs $N_{bin}$ and variability of $\epsilon$ or not) imply that the semiclassical eikonal approximation is valid for \aa\ collisions but not for \pp\ collisions. The quadrupole trend is distinct from the dijet $\bar \rho_0 A_X \propto N_{bin}$ trend common to SS 2D peak and AS 1D peak in  \pp\ and \aa~\cite{anomalous,ppquad}.
In contrast to NGNM applied to 1D projections onto azimuth, model fits to 2D angular correlations permit accurate separation of data components (and underlying physical mechanisms) represented by the fit-model elements. Differences are illustrated in the next subsection.

\subsection{Comparison of $\bf v_2$ methods} \label{compare}

Published $v_2$ data are commonly interpreted  to represent some combination of elliptic flow and nonflow~\cite{poskvol}, with several proposed sources for the latter such as hadronic resonances, BEC and jets~\cite{2004}. Various strategies have been proposed to reduce nonflow, including cuts on $\eta$ to exclude an interval on $\eta_\Delta$ near the origin where {\em for \pp\ collisions} the SS 2D peak and contributions from BEC are localized on $\eta$. Here we compare results from model fits to 2D angular correlations with common NGNM based on two- and four-particle cumulants.

All 2D angular correlations include a SS 2D peak~\cite{anomalous}. 
The Fourier amplitudes for given SS-peak azimuth width $\sigma_{\phi_\Delta}$ are represented by factor $F_m(\sigma_{\phi_\Delta})$ for the $m^{th}$ Fourier term (cylindrical multipole)~\cite{tzyam}. Projection of the SS peak onto 1D azimuth depends on its $\eta$ width relative to detector $\eta$ acceptance $\Delta \eta$ as represented by factor  $G(\sigma_{\eta_\Delta},\Delta \eta)$~\cite{jetspec}. A detailed expression for  $G(\sigma_{\eta_\Delta},\Delta \eta)$ with  $\eta$-exclusion cuts is presented in Ref.~\cite{multipoles}. Both expressions assume a unit-amplitude 2D peak. The {\em jet-related} quadrupole amplitude derived from measured SS 2D peak properties is then
\bea \label{aqss}
2 A_Q\{SS\}(b)  &=&  F_m(\sigma_{\phi_\Delta},b)G(\sigma_{\eta_\Delta},\Delta \eta,b) A_{2D}(b),~~ 
\eea
where $A_{2D}(b)$ is the SS peak amplitude. NJ quadrupole amplitude $A_Q\{2D\}$ and jet-related amplitude $A_Q\{SS\}$ represent ``flow'' and ``nonflow'' in more-central \aa\ collisions (where BEC projected on azimuth are negligible).

\begin{figure}[h]
 \includegraphics[width=1.65in,height=1.635in]{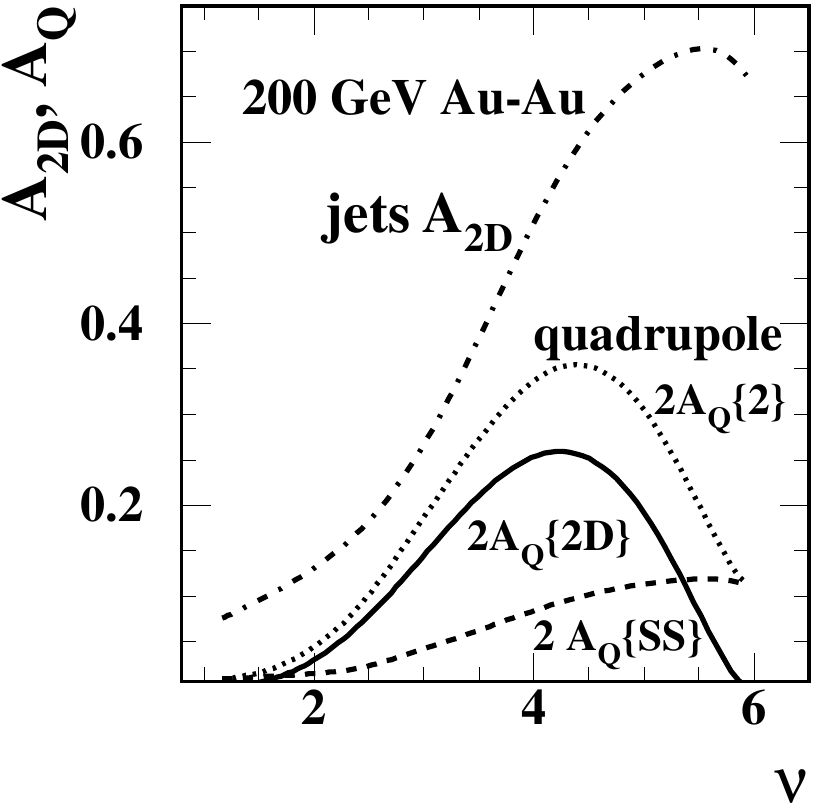}
  \includegraphics[width=1.65in,height=1.635in]{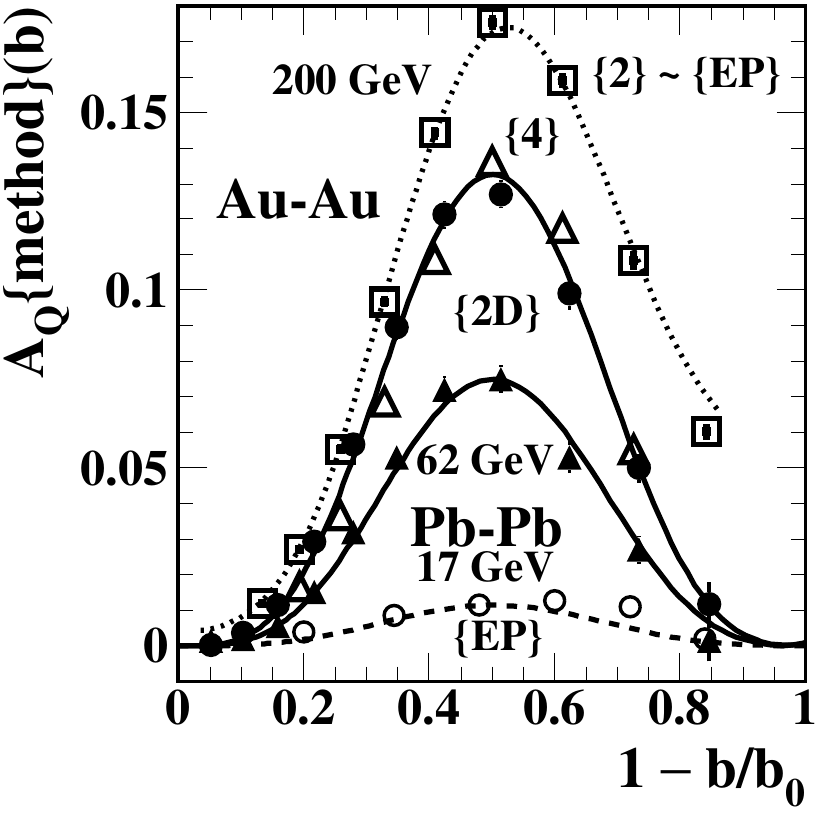}
\caption{\label{v2ep}
Left: SS 2D (jet) peak amplitude $A_{2D}$ (parametrization of data from Ref.~\cite{anomalous}), SS 2D peak quadrupole component $A_Q\{SS\}$ (``nonflow'') inferred from Eq.~(\ref{aqss}) and Ref.~\cite{anomalous} data and NJ quadrupole amplitude $A_Q\{2D\}$ from Eq.~(\ref{magic}), with $A_Q\{2\} = A_Q\{2D\} + A_Q\{SS\}$~\cite{gluequad}. 
Right: Quadrupole results from 2D model fits presented in Fig.~\ref{quadx} (solid points~\cite{davidhq,noelliptic}) compared to published $v_2$ measurements obtained with NGNM methods \{2\},  \{4\}, \{EP\} (open points~\cite{na49v2,2004}). The ratio of 200 to 62 GeV $A_Q\{2D\}$  is approximately 1.75. The dotted curve includes jet-related contribution $A_Q\{SS\}$ obtained from SS 2D peak systematics reported in Ref.~\cite{anomalous}.
 } 
 \end{figure}

Figure~\ref{v2ep} (left) shows a parametrization of the centrality dependence of SS 2D peak amplitude $A_{2D}(b)$ (dash-dotted curve) and three quadrupole amplitudes with~\cite{gluequad}  
\bea
A_Q\{2\}(b) = A_Q\{2D\}(b) + A_Q\{SS\}(b).
\eea
$A_Q\{2D\}$ (solid curve) is defined by Eq.~(\ref{magic})~\cite{davidhq} and $A_Q\{SS\}$ (dashed curve) by Eq.~(\ref{aqss}) using measured SS peak parameters (amplitude and two widths) from Ref.~\cite{anomalous}. The sum $A_Q\{2\}$ (dotted curve) then establishes a {\em prediction} for published $v_2\{2\} \approx v_2\{EP\}$ measurements derived from NGNM cosine fits to 1D projections onto azimuth of all 2D angular correlation structure~\cite{2004}. The dotted curve in the left panel appears (transformed) in the right panel. Centrality measure $\nu$ (Sec.~\ref{aageom}) represents the mean number of \nn\ binary collisions per participant-nucleon pair. Fractional impact parameter $b/b_0$ is derived from fractional cross section $\sigma/\sigma_0$ as $b/b_0 = \sqrt{\sigma/\sigma_0}$.

Figure~\ref{v2ep} (right) shows two-particle cumulant data $A_Q\{2\}$ (open squares) from 200 GeV \auau\ collisions~\cite{2004} and event-plane measurements $A_Q\{EP\}$ (open circles) from 17 GeV \pbpb\ collisions~\cite{na49v2} compared with Eq.~(\ref{magic}) (solid and dashed curves) with energy dependence given by $\log(\sqrt{s_{NN}} / \text{13 GeV})$~\cite{davidhq,noelliptic}. The solid points  are 62 and 200 GeV $A_Q\{2D\}$ data from Fig.~\ref{quadx} (right). The published uncertainties for $A_Q\{2\}$ data have been {\em increased 5-fold} to make them visible (shown as bars within the open squares).
The  NA49 $A_Q\{\text{EP}\}$ data (open circles)~\cite{na49v2} provide a reference for energy scaling. 

The precise agreement between $v_2\{2\} \approx v_2\{EP\}$ data and the prediction based on jet-related structure (dotted curve) is evident.  From a universal NJ quadrupole trend and measured jet-related correlation structure published $v_2\{2\}$ measurements are accurately predicted.   For statistically well-defined $v_2$ methods (e.g.\ $v_2\{2\} \approx v_2\{EP\}$) the dijet (``nonflow'') bias in $v_2$ data can be determined precisely. For NGNM based on complex strategies for nonflow reduction the corresponding $v_2$ data may lie somewhere between $v_2\{2D\}$ and $v_2\{2\}$ limiting cases. According to conventional arguments $v_2\{4\}$ data should fall {\em below} $v_2\{2D\}$ data due to a negative contribution from $v_2$ fluctuations~\cite{v24nojets}, but detailed comparisons show that the statement is not generally true for published $v_2\{4\}$ data~\cite{davidhq,noelliptic}. The open triangles show 200 GeV \auau\ $v_2\{4\}$ data from Ref.~\cite{2004} plotted as $A_Q\{4\}$. Relative to the \{2D\} trend the \{4\} data are systematically lower in more-peripheral collisions and higher in more-central collisions, consistent with significant jet bias increasing with \aa\ centrality (also see Sec.~\ref{quadenergy}).

\section{Quadrupole spectrum methods} \label{quadspecmeth}

\pt-differential $v_2(p_t,b)$ as conventionally defined is a ratio with the general form (for given centrality)
\bea \label{v2struct}
v_2(p_t) \hspace{-.05in} &=& \hspace{-.05in} \frac{V_2\{2D\}(p_t) + \text{jet contribution}}{(N_{part}/2)S_{NN}(p_t) + N_{bin}r_{AA}(p_t) H_{NN}(p_t)},~~~~
\eea
where the denominator is \aa\ SP spectrum $\bar \rho_0(p_t,b)$ represented by its TCM form in Eq.~(\ref{ppspec}), $S_{NN}$ and $H_{NN}$ are soft and hard (jet) spectrum components for \nn\ ($\approx$ \pp) collisions, $r_{AA} \neq 1$ describes modified jet formation (``jet quenching'') in more-central \aa\ collisions, and ``jet contribution'' in the numerator represents a ``nonflow'' contribution that depends on the $v_2$ analysis method. NJ quadrupole physics is confined to $V_2\{2D\}(p_t,b) = \bar \rho_0(p_t,b) v_2\{2D\}(p_t,b)$ whereas conventional $v_2(p_t)$ data may include jet contributions via two aspects of NGNM analysis. 
Eq.~(\ref{v2struct}) is more applicable above 50 GeV where the TCM accurately describes hadron production in \nn\ collisions dominated by low-$x$ gluons.

\subsection{Single-particle spectrum models}

Inference of quadrupole spectra from $v_2(p_t,b)$ data requires matching SP spectra $\bar \rho_0(p_t,b)$ and factorization of $V_2\{2D\}(p_t)$ according to the Cooper-Frye formalism~\cite{quadspec}. 
In the context of a flow narrative it is commonly assumed that most hadrons emerge from a flowing bulk medium and share a common SP spectrum that also exhibits radial (monopole) flow. The NJ quadrupole then represents a simple modulation of the radial-flow component {\em relative to an eventwise reference angle}, and the concept of a unique quadrupole spectrum is not relevant to that context. However,  differential analysis of spectra for identified hadrons from 200 GeV \auau\ collisions failed to detect such a radial-flow component but did establish that a substantial jet-related (hard) spectrum component persists for all \auau\ centralities~\cite{hardspec}. 

Those and other recent results have motivated reconsideration of spectrum models to obtain a more-general spectrum description relying on fewer {\em a priori} assumptions. Would an NJ quadrupole spectrum (inferred from $v_2$ data) represent {\em modulation} of an existing spectrum component  (e.g.\ the TCM soft component) or a distinct radially-boosted hadron source modulated on azimuth? 

A more-general {\em three}-component spectrum model on $(x_t,\phi_r)$ for $x_t = m_t$ or $y_t$ can be expressed as
\bea \label{specspec}
\bar \rho_0(y_t,\phi_r) &= & \bar \rho_1(y_t;\mu_0) + \bar \rho_2[y_t;\mu_2,\Delta y_{t}(\phi_r)],
\\ \nonumber
\bar \rho_0(m_t,\phi_r) &= &\bar  \rho_1(m_t;T_0) + \bar \rho_2[m_t;T_2,\beta_{t}(\phi_r)],
\eea
where $\phi_r \equiv \phi - \Psi_r$ measures azimuth relative to reference angle $\Psi_r$, $\mu_0 = m_h / T_0$, $\mu_2 = m_h / T_2$ and $\bar \rho_2$ is a possible quadrupole (third) component from a radially-boosted source. Parameter $\beta_{t}(\phi_r)$ or $\Delta y_{t}(\phi_r)$ represents a conjectured azimuth-dependent radial boost of the third component. The first term $\bar \rho_1(x_t;\mu_0)$ is the SP spectrum TCM from Refs.~\cite{ppprd,hardspec}. Quadrupole term $\bar \rho_2$ may represent a new particle source or a modification of the SP spectrum soft component (or  hard component). The azimuth-averaged spectrum $\bar \rho_0(y_t,b)$ then includes $\bar \rho_2(y_t,b)$ that may comprise a small fraction of the total.
To clarify the relation the shape and absolute magnitude of azimuth-averaged quadrupole spectra $\bar \rho_2(x_t)$ inferred from $v_2(p_t)$ data should be compared with measured azimuth-averaged SP spectra $\bar \rho_0(x_t)$~\cite{quadspec}. 

\subsection{Quadrupole spectrum definition}

A spectrum for the NJ azimuth quadrupole component may be simply derived from experimental $v_2(p_t)$ data assuming that  
(a) the quadrupole component arises from a hadron source with eventwise azimuth-dependent radial boost distribution $\Delta y_t(\phi_r)$, 
(b) the quadrupole spectrum may appear nearly thermal {\em in the boost frame} 
and 
(c) the quadrupole source {\em may} produce only a fraction of the hadrons in a collision, independent of SP-spectrum TCM soft and hard components.
Given those possibilities the $\eta$-averaged $\phi_r$-dependent spectrum at midrapidity for those hadrons associated with the NJ quadrupole component is then modeled by
\bea \label{fullquad}
\bar \rho_2(y_t,\phi_r) \hspace{-.04in} &\propto& \hspace{-.04in} \exp\{ -\mu_2[\cosh(y_t - \Delta y_{t}(\phi_r)) - 1]  \},
\\ \nonumber
\bar \rho_2(m_t,\phi_r) \hspace{-.04in} &\propto& \hspace{-.04in}  \exp\{ -\left(\gamma_t(\phi_r)[m_t - \beta_t(\phi_r) p_{t} ] - m_h\right)  /T_2\},
\eea
where a M-B distribution for a locally-thermal source is assumed for simplicity. The procedure below may be applied to a more general spectrum model such as a L\'evy distribution~\cite{wilk}.  $\Delta y_{t}(\phi_r)$ defined in Eq.~(\ref{quadboost}) includes fixed monopole and quadrupole boost components.

Based on relativistic kinematics reviewed in App.~\ref{boost} and the assumed boost model expressed by  Eq.~(\ref{quadboost}) the spectrum defined by Eq.~(\ref{fullquad}) can be factored as
\bea \label{factor}
\bar \rho_2(y_t,\phi_r) \hspace{-.0in} &=& A_{2,y_t} \exp\{ -\mu_2[\cosh(y_t - \Delta y_{t0}) - 1]  \}\times 
 \nonumber \\
&& \hspace{-.2in} \exp[m'_t \, \{\cosh[ \Delta y_{t2}\, \cos(2  \phi_r)] - 1\}/T_2] \times 
 \nonumber \\
&& \hspace{-.2in} \exp\{p'_t \, \sinh[\Delta y_{t2}\, \cos(2  \phi_r)] /T_2\} 
 \nonumber \\
&\equiv& \bar \rho_2(y_t;\Delta y_{t0})\times F_1(y_t,\phi_r; y_{t0} ,\Delta y_{t2})\times 
 \nonumber \\
&&F_2(y_t,\phi_r;\Delta y_{t0} ,\Delta y_{t2}).
\eea  
The last line defines azimuth-dependent factors $F_1(y_t,\phi_r)$ and $F_2(y_t,\phi_r)$ in terms of monopole and quadrupole components of the radial boost. The objective is azimuth-averaged quadrupole spectrum  $\bar \rho_2(y_t;\Delta y_{t0})$ emitted from a conjectured boosted hadron source as one factor of Fourier amplitude $V_2(y_t)$ inferred from $v_2(p_t)$ measurements. 


Assuming (for the purpose of derivation) an azimuth-dependent spectrum component $\bar \rho_2(y_t,\phi_r)$ in Eq.~(\ref{factor}) relative to a reaction plane the $m = 2$ Fourier amplitude is
\bea
V_2(y_t)
&=&\frac{1}{2\pi} \int_{-\pi}^{\pi} d\phi\,  \bar \rho_2(y_t,\phi_r) \cos(2 \phi_r).
\eea
The full integral over factors $F_1$ and $F_2$ in Eq.~(\ref{factor}) is
\bea \label{fint}
 \frac{1}{2\pi} \int_{-\pi}^{\pi} \hspace{-.14in} d\phi F_1(y_t,\phi_r) F_2(y_t,\phi_r)\cos(2  \phi_r)  
\hspace{-.05in } &=& \hspace{-.05in}p'_t \frac{\Delta y_{t2}}{2T_2}f(y_t),~~~~
\eea
where $f(y_t;\Delta y_{t0},\Delta y_{t2})$ is an $O(1)$ correction factor determined by ratio $\Delta y_{t0}/\Delta y_{t2}$: $f(y_t)$ remains closer to 1 the smaller is $\Delta y_{t2}/\Delta y_{t0}$~\cite{quadspec}. Combining factors gives
\bea \label{combfac}
V_2(y_t,b;\Delta y_{t0},\Delta y_{t2}) \hspace{-.04in} &=&\bar \rho_0(y_t,b) v_2(y_t,b) 
\\ \nonumber  
&\approx& 
 \hspace{-.04in}p'_t\, \frac{ \Delta y_{t2}(b)}{2T_2} \, \bar \rho_2[y_t,b;\Delta y_{t0}(b)].
\eea
establishing a direct relation between $v_2(y_t)$ data and quadrupole spectrum $\bar  \rho_2(y_t;\Delta y_{t0})$. $p'_t$ is \pt\ in the boost frame, $T_2$ is the quadrupole-spectrum slope parameter, $\Delta y_{t0}$ is the (single fixed value) monopole source boost, and $\Delta y_{t2}$ is the amplitude of the source-boost quadrupole modulation. $v_2(y_t)$ data might conflict with that simple model to reveal a source-boost {\em distribution} on $\Delta y_{t0}$ corresponding to Hubble expansion of a dense medium. 

\subsection{Inferring $\bf \bar \rho_2(y_t;\Delta y_{t0})$ from measured $\bf v_2(y_t)$ data}

A quadrupole spectrum can be inferred from measured quantities by the relation (for fixed monopole boost)
\bea \label{stuff}
\bar \rho_0(y_t)\, \frac{v_2(y_t)}{p_t} \hspace{-.05in} &=& \hspace{-.05in} \left\{\frac{p'_t}{p_t\, \gamma_t(1 - \beta_t)}\right\}\,  \left\{\frac{\gamma_t(1 - \beta_t)}{2T_2}\right\} \times \\ \nonumber
&&f(y_t;\Delta y_{t0},\Delta y_{t2})\,\Delta y_{t2}\,  \bar \rho_2(y_t;\Delta y_{t0}).
\eea
The quantities on the left are measured experimentally. $\bar \rho_2(y_t;\Delta y_{t0})$ on the right is the sought-after quadrupole spectrum. The common monopole boost $\Delta y_{t0}$ and $T_2$ for each hadron species can be estimated accurately from the $\bar \rho_2(y_t;\Delta y_{t0})$ spectrum shape inferred from data, as illustrated below. The first factor on the right, shown in Fig.~\ref{boost3} (right), is determined only by $\Delta y_{t0}$ and deviates from unity only near that intercept on \yt. The numerator of the second factor ($\approx 0.55$) is also determined by $\Delta y_{t0}$. Thus, all factors in the first line on the right and the shape of $\bar \rho_2(y_t;\Delta y_{t0})$ are determined by data on the left.

In the second line on the right there is an ambiguity. The absolute quadrupole yield $\bar \rho_2(b)$ is not accessible from this procedure, only the product of quadrupole boost amplitude $\Delta y_{t2}(b)$ and quadrupole yield $\bar \rho_2(b)$.  Comparison of the inferred quadrupole $\bar \rho_2(y_t,b;\Delta y_{t0})$ spectrum shape (especially the lower edge of the boosted spectrum) with measured azimuth-averaged spectrum $\bar \rho_0(y_t,b)$ for each hadron species might place a lower limit on $\Delta y_{t2}$~\cite{quadspec}. The upper limit $\Delta y_{t2} \leq \Delta y_{t0}$ assumes positive-definite transverse boosts. The two limits could establish an allowed range for quadrupole spectrum integral $\bar \rho_{2}(b)$. $\Delta y_{t2}(b)$ should be common to all  hadron species emitted from a boosted hadron source for a given \aa\ centrality, possibly reducing systematic uncertainty. Without absolute determination of the quadrupole yield one can define a unit-normal spectrum shape from experimental data
\bea \label{shat}
\hat S_2(y_t,b) &\equiv& \frac{V_2(y_t,b) / p_t}{V_2(b) \langle 1/p_t \rangle} \\ \nonumber
&\approx& \frac{\bar \rho_2(y_t,b)}{\bar \rho_2(b)}
\eea
to obtain a quadrupole spectrum shape {\em in the lab frame} from measured quantities. To illustrate those results relations between hydro models and $v_2(p_t)$ data are explored.

\subsection{Predicting $\bf v_2(p_t)$ data from a hydro model}

If the NJ quadrupole spectrum were equivalent to the SP spectrum as commonly assumed Eq.~(\ref{stuff}) reduces to
\bea \label{v2simple}
v_2(p_t;\Delta y_{t0}) &\approx& p'_t(p_t;\Delta y_{t0})  \frac{\Delta y_{t2}}{2T_2} 
\eea
given $f(y_t;\Delta y_{t0},\Delta y_{t2}) \approx 1$ over a relevant \pt\ interval. That ``ideal hydro'' trend is shown below in a conventional $v_2(p_t)$ vs \pt\ plot format and in a modified  format.

\begin{figure}[h]
 \includegraphics[width=1.65in,height=1.65in]{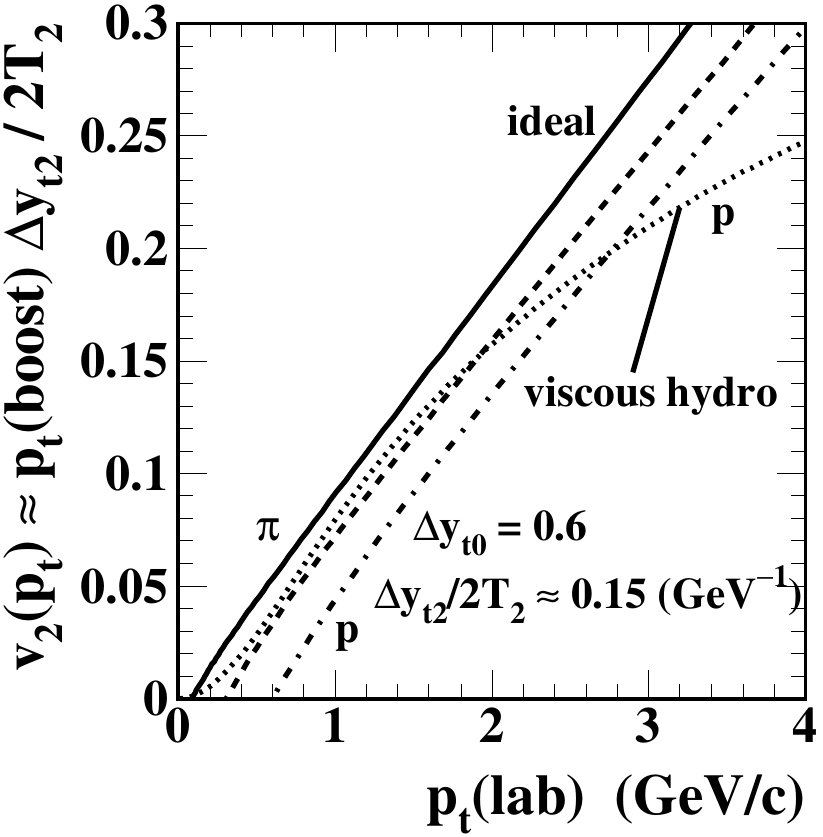}
  \includegraphics[width=1.65in,height=1.625in]{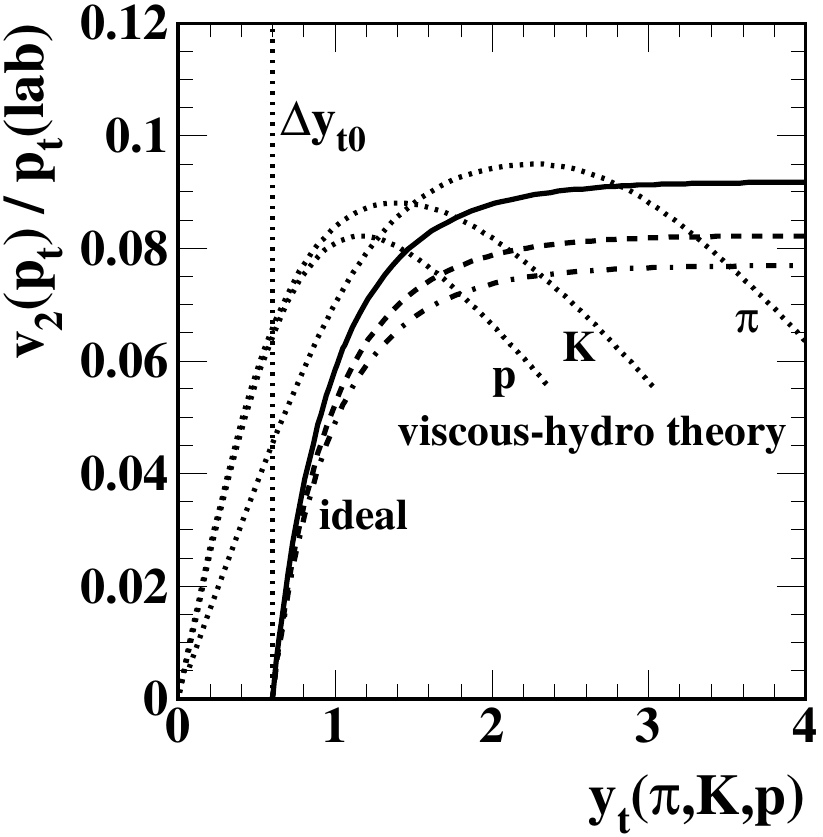}
\caption{\label{schema1}
Left: $v_2(p_t)$ trends for three hadron species vs $p_t(\text{lab})$ (solid, dashed, dash-dotted) assuming ``ideal hydro'' conditions and a monolithic flowing-bulk-medium hadron source. The dotted curve is a viscous-hydro theory prediction for protons~\cite{rom}.
Right: Ratio $v_2(p_t) / p_t(\text{lab})$ vs transverse rapidity $y_t$ defined for each hadron species.  ``Ideal-hydro'' curves have a common form given by Eq.~(\ref{kine}). Viscous-hydro predictions for three hadron species are shown as the dotted curves~\cite{rom}.
 } 
 \end{figure}

Figure~\ref{schema1} (left) shows Eq.~(\ref{v2simple}) for three hadron species ($\pi$, K, p) and fixed $\Delta y_{t0} = 0.6$ (based on results from Ref.~\cite{quadspec}). Expression ${\Delta y_{t2}}/{2T_2} \approx 0.15$/GeV is adjusted so that the ``ideal hydro'' trends (solid, dashed, dash-dotted) correspond approximately to $v_2(p_t)$ data at lower \pt\ in Fig.~\ref{x1} (left) (actual values are 0.17, 0.15 and 0.14 for pions, kaons and protons). $v_2(p_t)$ data suggest that $T_2 \approx 90$ MeV, so ratio $\Delta y_{t2}/\Delta y_{t0} \approx 0.05$ implies that $f(y_t;\Delta y_{t0},\Delta y_{t2})$ from Eq.~(\ref{stuff}) deviates from unity by only a few percent over a relevant \yt\ interval and can be neglected. However, that ratio value is only a lower limit corresponding to assumed $\bar \rho_2(b) = \bar \rho_0(b)$. The dotted curve is a viscous-hydro result for protons from Ref.~\cite{rom}.

Figure~\ref{schema1} (right) shows ratio $v_2(p_t) / p_t$(lab) vs the proper transverse rapidity for each hadron species. The ``ideal'' curves have a universal form that intercepts zero at $\Delta y_{t0}$ and corresponds to Fig.~\ref{boost3} (right) of App.~\ref{boost}. Note that the viscous-hydro predictions for three hadron species (dotted) have very different behavior from the ideal-hydro curves of Eq.~(\ref{v2simple}) over the entire \yt\ interval. 

\section{200 $\bf GeV$ $\bf Au$-$\bf Au$ quadrupole spectra} \label{200gevquad}

Quadrupole spectra $\bar \rho_2(y_t,b)$ can be inferred directly from $v_2$ data. Starting with published $v_2(p_t,b)$ data a procedure is developed to infer corresponding quadrupole spectra and applied to $v_2$ data for three hadron species.

\subsection{NJ quadrupole $\bf v_2(p_t)$ data in two formats}

\begin{figure}[h]
    \includegraphics[width=1.49in,height=1.5in]{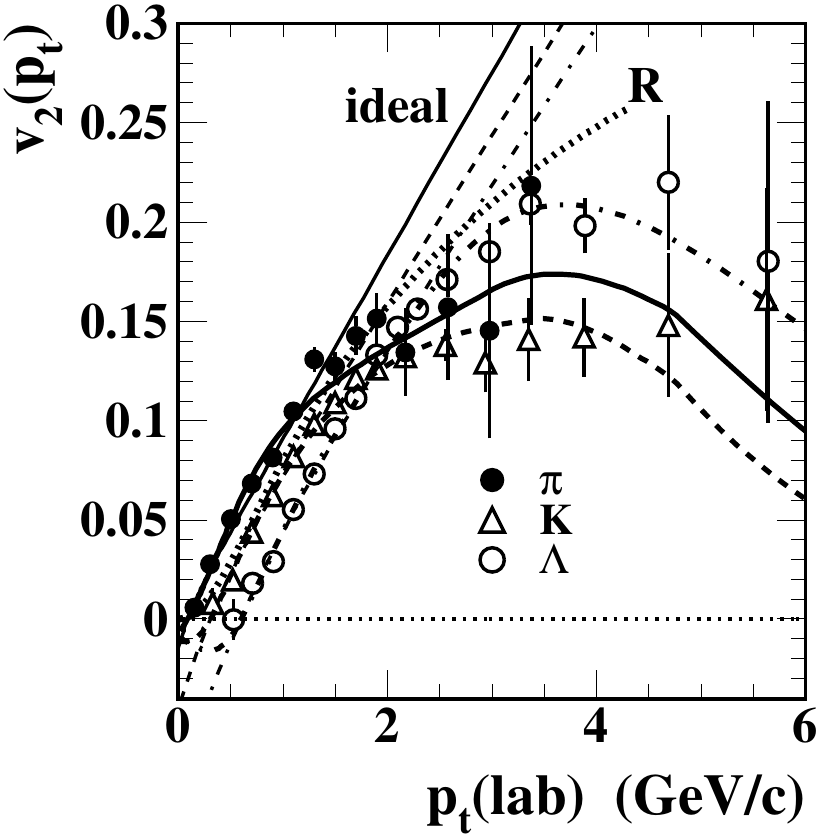}
     \includegraphics[width=1.49in,height=1.48in]{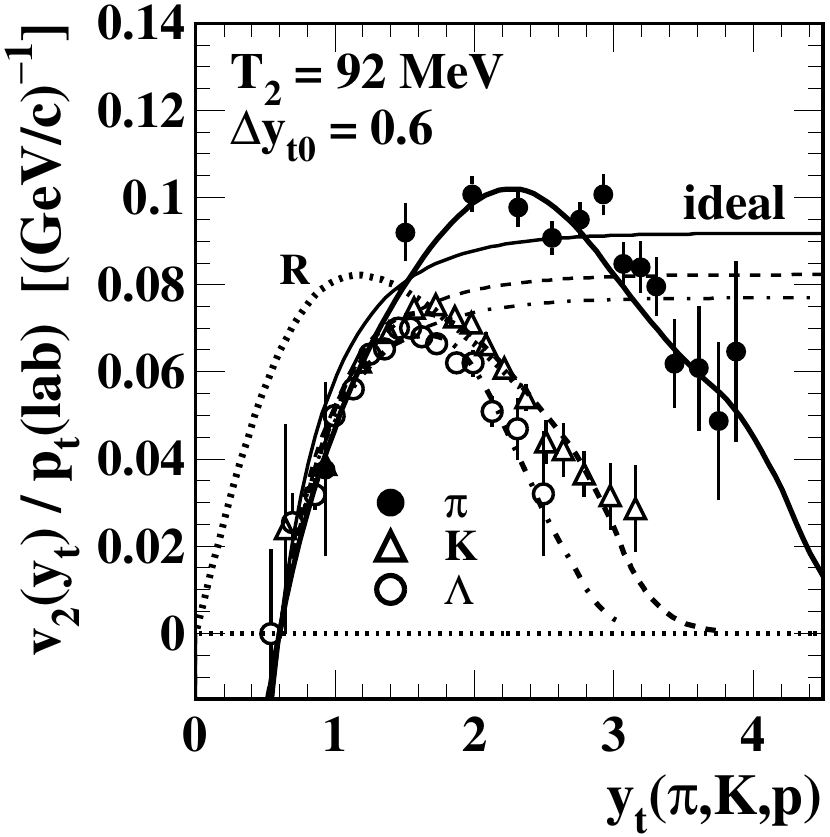}
\caption{ \label{x1}  
Left: $v_2(p_t)$ data for three hadron species plotted in a conventional format. The kaon and Lambda data are from 0-80\% central 200 GeV \auau\ collisions~\cite{v2strange}. The data representing pions are hadron data for 16-24\% 130 GeV \auau\ collisions scaled up by 1.2~\cite{v2pions}.
Right: The same data plotted as ratio $v_2(p_t) / p_t(\text{lab})$ on transverse rapidity. The curves labeled R represent a viscous-hydro prediction for identified protons. The solid, dashed and dash-dotted curves through data are described below.
} 
\end{figure}

Fig.~\ref{x1} (left) shows 200 GeV $v_2(p_t)$ data for three hadron species vs $p_t$ in the conventional plotting format averaged over 0-80\% \auau\ centrality~\cite{v2pions,v2strange}.  The curves extending off the top edge of the panel are $v_2 \propto p_t'$ as in Eq.~(\ref{v2simple}) and Fig.~\ref{schema1} (left) reflecting expected ideal-hydro trends for a single boost value $\Delta y_{t0} = 0.6$ that describes $v_2$ data for $p_t < 1.5$  GeV/c. For Hubble expansion of a bulk medium the source boost distribution should be broad. The solid, dashed and dash-dotted curves passing through data at higher \pt\ are described below.

Fig.~\ref{x1} (right) shows the same data divided by $p_t(\text{lab})$ and plotted on transverse rapidity $y_t$ with proper mass for each hadron species. It is notable that the data for three hadron species pass through a common zero intercept at $y_t = 0.6$ ($\Delta y_{t0}$) consistent with emission from an expanding thin cylindrical shell. The curves approaching a constant value at larger $y_t$ represent the ideal-hydro trends from the left panel. That the $v_2(p_t)$ data drop sharply away from the ideal trends toward zero has been attributed to viscosity of a bulk medium assuming that almost all hadrons emerge from that common medium, but the fall-off could also be explained by quadrupole spectra quite different from SP spectra describing most hadrons. The curves through data are described below.

\begin{figure}[h] 
    \includegraphics[width=1.49in,height=1.5in]{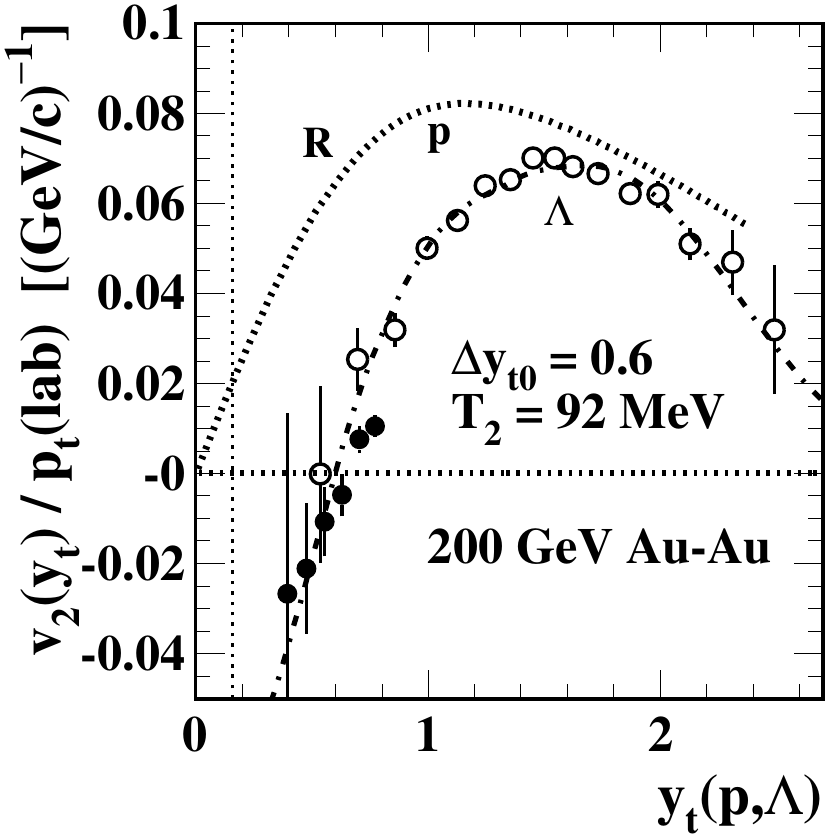}
   \includegraphics[width=1.49in,height=1.47in]{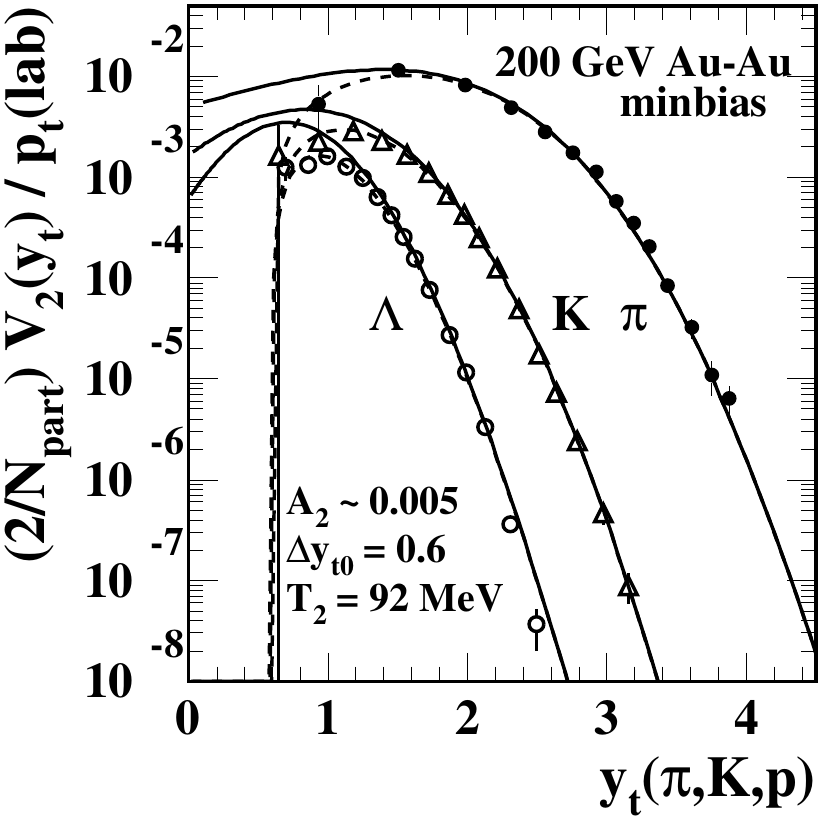}
\caption{\label{x2}  
Left: Lambda data from Fig.~\ref{x1} (right, open points) on an expanded \yt\ scale compared to viscous-hydro theory curve R for protons~\cite{rom}.
The solid points are more-recent Lambda data for 0-10\% central \auau\ collisions~\cite{starnewlambda}. The dotted line marks a detector acceptance limit for charged hadrons at $p_t \approx 0.15$ GeV/c.
Right: Data from Fig.~\ref{x1} (right) multiplied by SP spectra in the form $(2/N_{part}) \bar \rho_0(y_t)$ derived from hadron spectra in Ref.~\cite{hardspec}. The curves are described in the text.
} 
\end{figure}

Fig.~\ref{x2} (left) shows an expanded view of Fig.~\ref{x1} (right) for Lambda hadrons compared to a viscous-hydro theory curve for protons (dotted curves R in several panels). 
The quadrupole source-boost distribution is best determined in  this case by protons or Lambdas for two reasons: (a) For a given detector \pt\ acceptance (lower bound) the data distribution on \yt\ extends to a lower value for more-massive hadrons. The vertical dotted line marks a lower limit for protons or Lambdas whereas the corresponding limit for pions is near $y_t = 1$. (b) Given that $\Delta y_{t0} \approx 0.6$ any data from heavier hadrons with more-limited statistics would provide little additional information.

Viscous-hydro curve R, representing a broad boost distribution consistent with a hydro assumption of \aa\ Hubble expansion, is dramatically falsified by the Lambda $v_2(p_t)$ data.  The solid points are recent Lambda data for 0-10\% \auau\ collisions~\cite{starnewlambda} that follow a data trend with significant negative values below the common intercept near $y_t = 0.6$ and confirm the dash-dotted trend predicted by Ref.~\cite{quadspec}.

\subsection{Quadrupole spectra inferred from $\bf v_2(p_t)$ data}

\begin{figure}[h] 
   \includegraphics[width=3.3in]{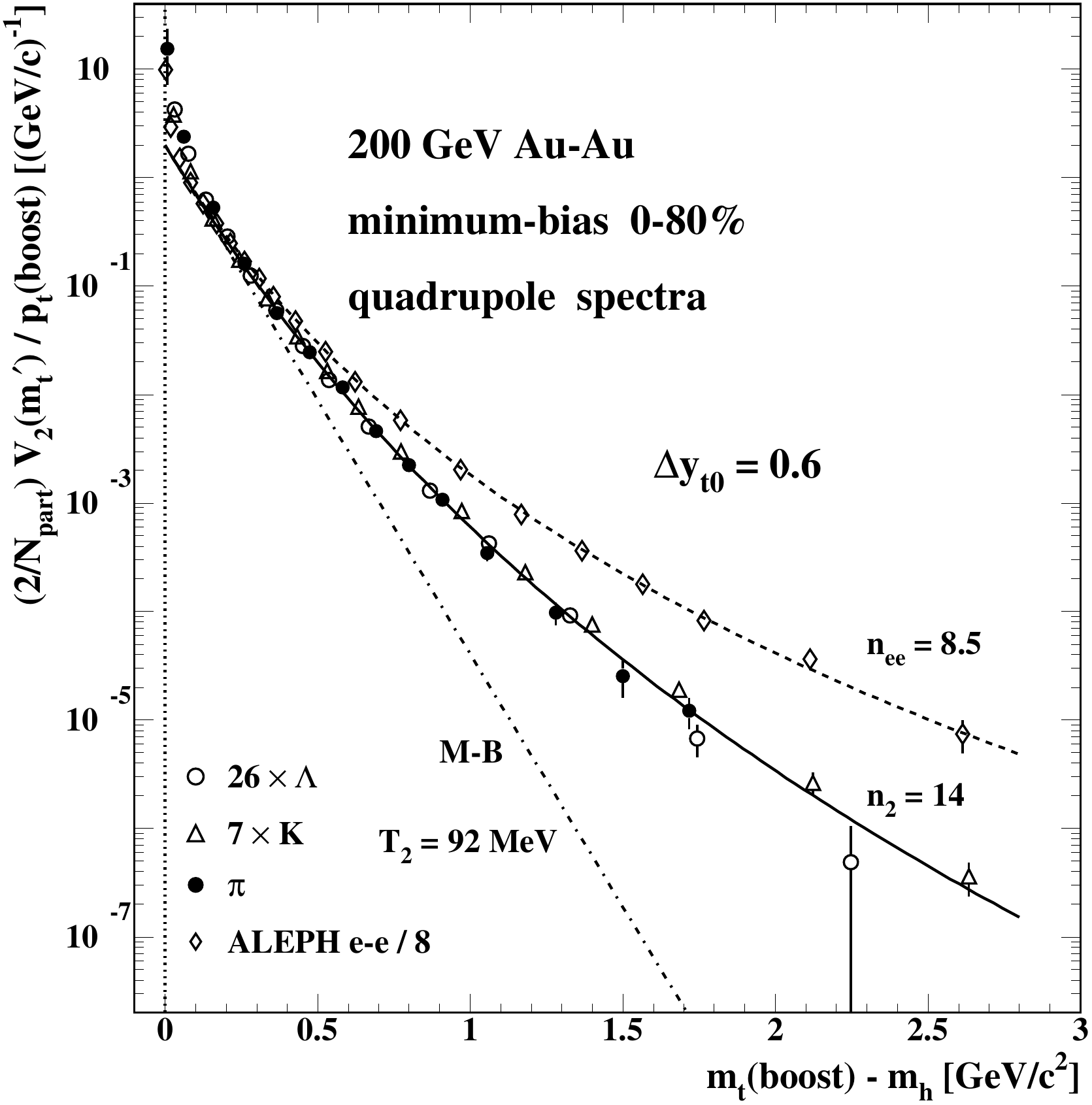}
\caption{\label{xbig}  
Data from Fig.~\ref{x2} (right) divided by the kinematic factor $p_t' / p_t$ defined in Eq.~(\ref{kine}) and transformed to $m_t(\text{boost}) - m_h$. Those data rescaled by the factors indicated in the plot then coincide on a single locus
modeled by a L\'evy distribution (solid curve). The spectrum model parameters are very different from those for SP hadron spectra~\cite{ppprd,hardspec}. The dashed curve is a \pt\ spectrum (perpendicular to thrust axis) for in-vacuum dijets from \ee\ collisions~\cite{eeptspec}.
} 
\end{figure}

Fig.~\ref{x2} (right) shows data from Fig.~\ref{x1} (right) multiplied by per-participant-pair SP spectra in the form $(2/N_{part})\bar \rho_0(y_t)$ for each identified hadron species to obtain $(2/N_{part})V_2(y_t) / p_t(\text{lab})$ (points). The kaon SP spectrum was generated by interpolation of TCM parametrizations of measured spectra for pions and protons~\cite{hardspec}. The curves are back-transformed from a universal quadrupole spectrum on $m_t' - m_h$ from Ref.~\cite{quadspec} (solid curve in Fig.~\ref{xbig}) with (dashed) and without (solid) kinematic factor $p_t' / p_t$ derived from Eq.~(\ref{kine}). The solid curves include an extra factor $\gamma_t (1 - \beta_t) \approx 0.55$.

Fig.~\ref{xbig} shows quadrupole spectra on  $m_t' - m_h$ {\em in the boost frame} for three hadron species as defined by the $y$-axis label. The lab-frame quadrupole spectra in Fig.~\ref{x2} (right) are multiplied by $p_t / p_t' = p_t(\text{lab}) / p_t(\text{boost})$ (since $\Delta y_{t0}$ is precisely known from the common data zero intercept), transformed to $y_t'$ in the boost frame by shifting the data to the left on \yt\ by $\Delta y_{t0}$ and transformed to densities on $m_t' - m_h$ by the Jacobian $y_t' / (m_t' - m_h) p_t'$. The $v_2/p_t$ data errors have been similarly transformed assuming that SP spectrum errors are negligible. The resulting spectra, rescaled with the statistical-model factors indicated in the plot, are found to coincide precisely over the entire $m_t'$ acceptance. Note that the proton data extend to $p_t = 5.6$ GeV/c in the lab frame but only 3 GeV/c in the boost frame (or $m_t' - m_h = 2.25$ GeV/$c^2$).

The solid curve is a L\'evy distribution with $(n_2,T_2)$ parameters indicated. That function is back-transformed to generate the curves through data in previous figures. Up to an overall constant three numbers -- $\Delta y_{t0} = 0.6$, $T_2 = 92$ MeV and $n_2 = 14$ -- accurately describe all MB $v_2(p_t)$ data for three hadron species. Those hadrons associated with the NJ  quadrupole follow a unique spectrum representing not a Hubble-expanding bulk medium but rather a thin shell expanding with fixed radial speed. The quadrupole spectrum is quite different from the SP \pt\ spectrum describing most hadrons. These data include factor $f(y_t;\Delta y_{t0},\Delta y_{t2})$ from Eq.~(\ref{stuff}) that raises the apparent spectrum tail at larger \mt. The L\'evy exponent $n_2 = 14$ should be considered a lower limit -- the actual spectrum may be significantly softer. Scaling factors 7 and 26 for kaons and Lambdas relative to pions are consistent with a statistical model of hadron emission~\cite{statmodel}.

A transverse-mass spectrum for in-vacuum dijets (for \qqbar\ pairs from the large electron-positron collider) describing \pt\ relative to the dijet axis (open diamonds) is included for comparison~\cite{eeptspec}. \ee\ slope parameter $T_{ee} \approx 90$ MeV is essentially the same as for the quadrupole spectrum and substantially lower than $T_0 \approx 145$ MeV for SP hadron spectra from hadron-hadron (e.g.\ \nn) collisions. Note that  the \ee\ \qqbar\ pairs have negligible $k_t$ in the lab frame whereas low-$x$ gluons from projectile protons have substantial initial $k_t$ in the lab as inferred from the acoplanarity of dijet pairs.

\section{2.76 $\bf TeV$ $\bf Pb$-$\bf Pb$ quadrupole spectra} \label{quadspec}

The 200 GeV quadrupole spectrum analysis in Ref.~\cite{quadspec} based on $v_2(p_t)$ MB data for identified hadrons with limited statistics established a novel analysis method. Recent $v_2(p_t,b)$ data from the LHC offer the possibility of high-statistics analysis including collision-energy and \aa\ centrality dependence of quadrupole spectra.

\subsection{Reduction of $\bf v_2(p_t,b)$ data to common loci}

\begin{figure}[h]
     \includegraphics[width=3.3in]{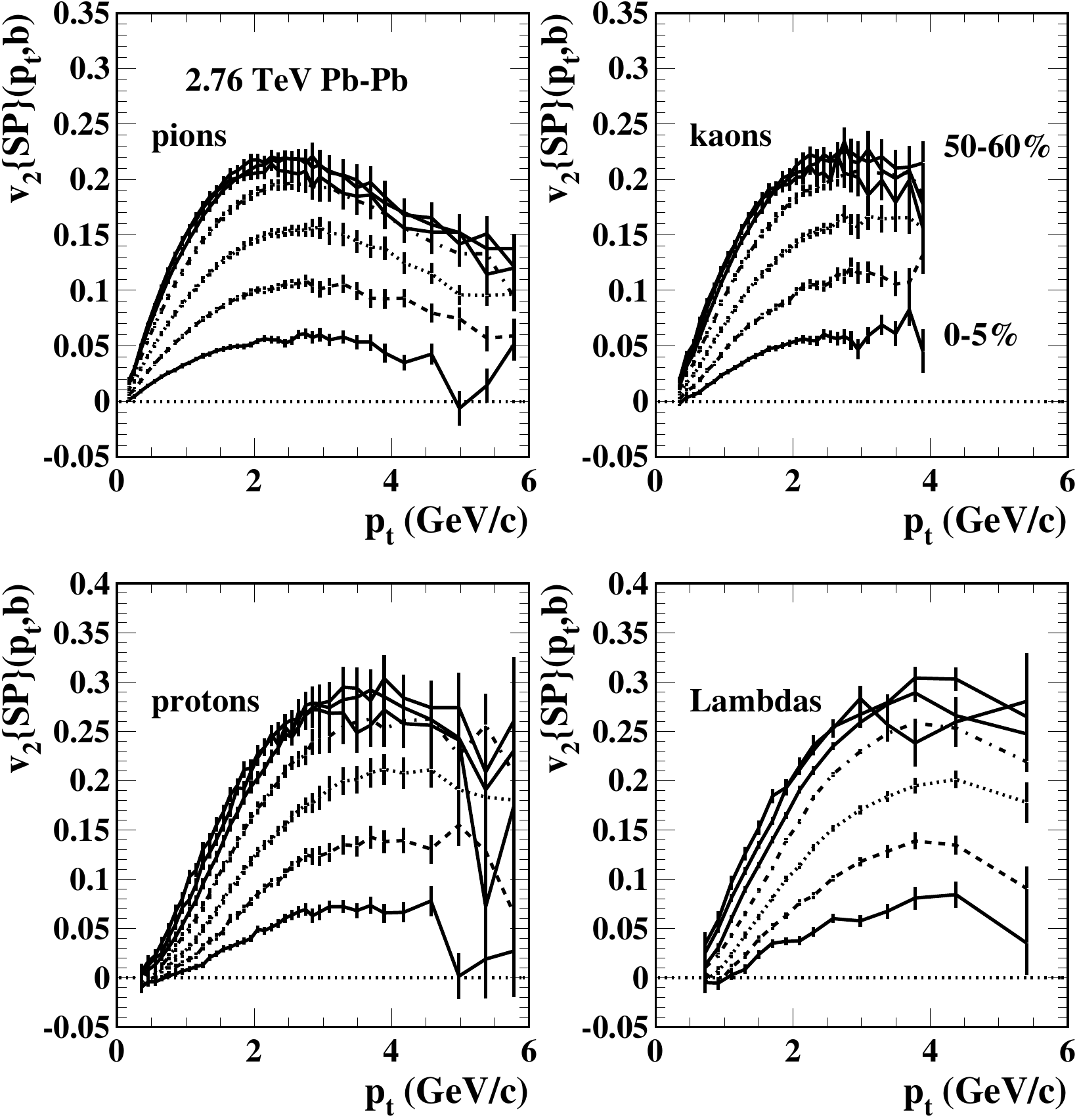}
\put(-140,233) {\bf (a)}
\put(-22,233) {\bf (b)}
\put(-140,108) {\bf (c)}
\put(-22,108) {\bf (d)}
\caption{\label{alice1}
$v_2(p_t,b)$ data for four hadron species from seven centrality classes of 2.76 TeV \pbpb\ collisions~\cite{alicev2ptb}. Bars represent statistical and systematic errors combined quadratically.
} 
\end{figure}
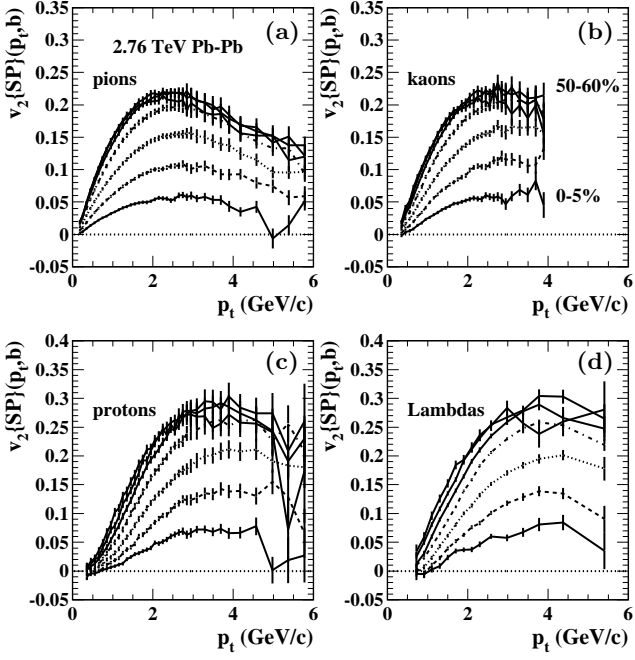

Figure~\ref{alice1} shows $v_2\{SP\}(p_t,b)$ data for four hadron species from 15 million 2.76 TeV \pbpb\ collisions in seven centrality bins from 0-5\% to 50-60\%~\cite{alicev2ptb}. The NGNM employed for that analysis is the scalar-produce or SP method. For each hadron species (charged pions, kaons, protons and neutral Lambdas) particles and antiparticles are combined. Error bars represent statistical plus systematic uncertainties combined in quadrature. As noted, this conventional plotting format conceals essential information carried by $v_2(p_t,b)$ data that is relevant to hydro theory. The first step in deriving quadrupole spectra is to rescale the data with measured \pt-integral $v_2(b)$ values.

\begin{figure}[h]
     \includegraphics[width=3.3in]{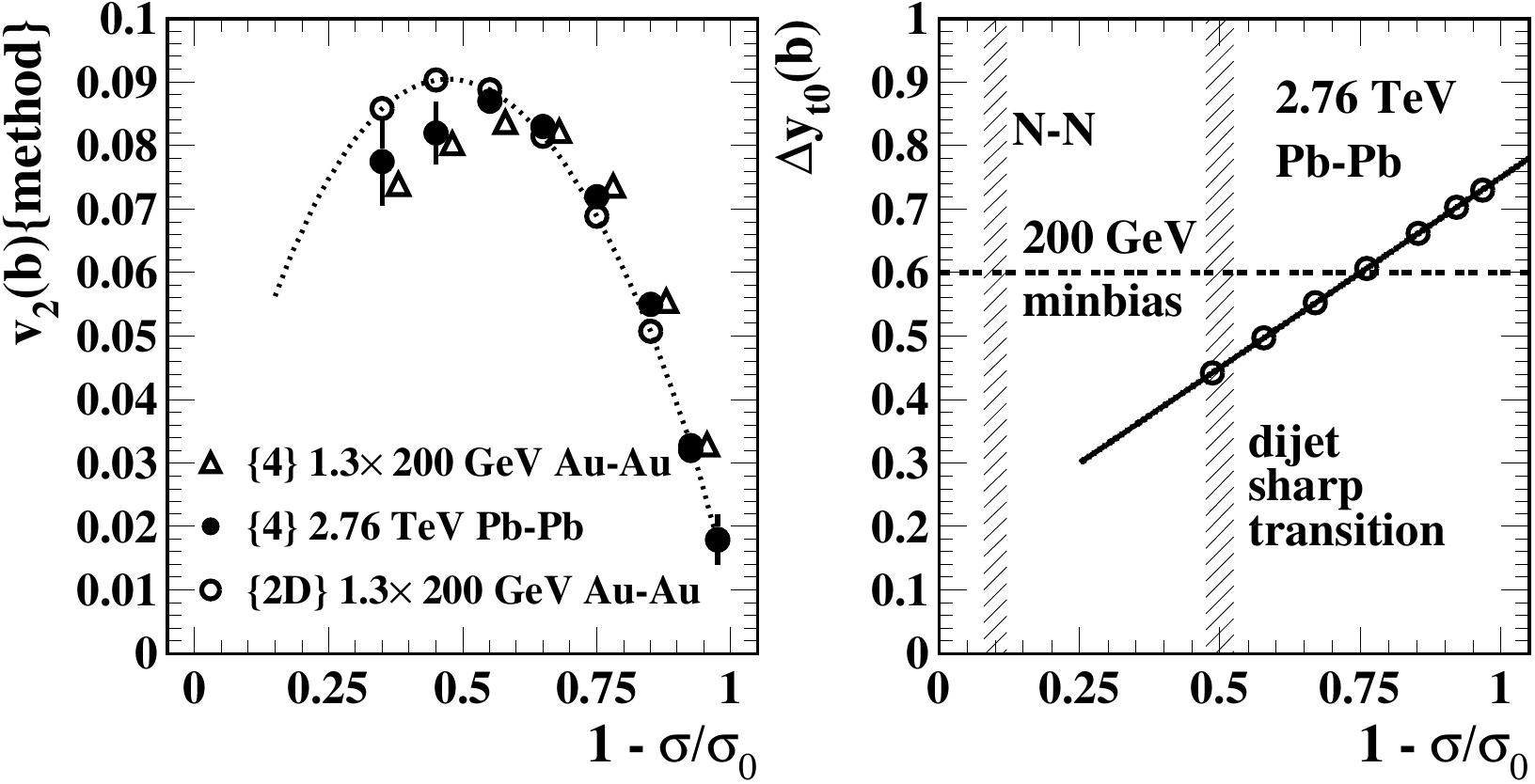}
\caption{\label{alice4}
Left: $v_2\{4\}(b)$ \pt-integral data for eight centralities of 2.76 TeV \pbpb\ collisions (solid points)~\cite{alicev2b} compared to the 200 GeV $v_2\{2D\}(b)$ data trend from Eq.~(\ref{magic}) multiplied by 1.3 (dotted curve)~\cite{davidhq,noelliptic}. The open circles on that curve are values applied in the present analysis.
Right: Quadrupole source boosts $\Delta y_{t0}(b)$ for seven centralities of 2.76 TeV \pbpb\ collisions (points) inferred from $v_2(p_t,b)$ data in this study.
} 
\end{figure}

Figure~\ref{alice4} (left) shows $v_2\{4\}(b)$ data from Ref.~\cite{alicev2b} (solid) vs centrality measured by fractional cross section $\sigma / \sigma_0$.
The dotted curve is derived from Eq.~(\ref{magic}) describing 200 GeV $v_2\{2D\}(b)$ data multiplied by factor 1.3. The open circles are resulting 2.76 TeV $v_2(b)$ values used to rescale  $v_2(p_t,b)$ data below. The open triangles are the 200 GeV $v_2\{4\}(b)$ data in Fig.~\ref{v2ep} (right) multiplied by 1.3 and shifted slightly to the right. See Sec.~\ref{quadenergy} for discussion of $v_2$\{4\} jet bias in more-central \aa\ collisions.
Figure~\ref{alice4} (right) is discussed below. 

\begin{figure}[h]
     \includegraphics[width=3.3in]{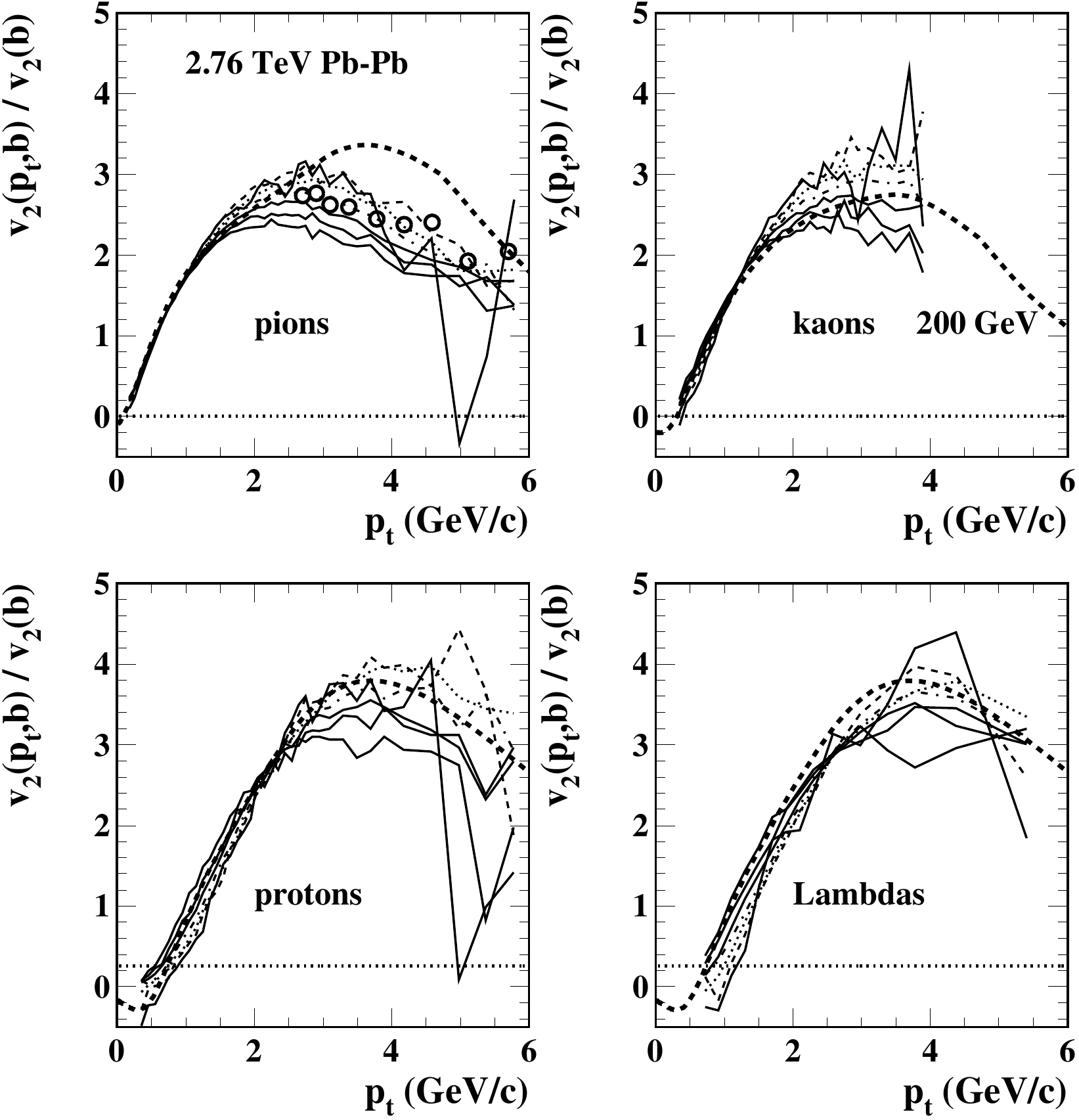}
\put(-140,233) {\bf (a)}
\put(-22,233) {\bf (b)}
\put(-140,108) {\bf (c)}
\put(-22,108) {\bf (d)}
\caption{\label{alice2}
Data from Fig.~\ref{alice1} rescaled by factor $1 / (1.3\, v_2(b)\{2D\})$ derived from 200 GeV \auau\ $v_2(b)\{2D\}$ data. The dashed curves are the same that appear in Fig.~\ref{x1} (left) for 200 GeV MB data~\cite{quadspec} scaled by factor 1/0.055. Open points in panel (a) are recent high-statistics 200 GeV pion data from Ref.~\cite{newstarpion} that agree with rescaled LHC data.
} 
\end{figure}
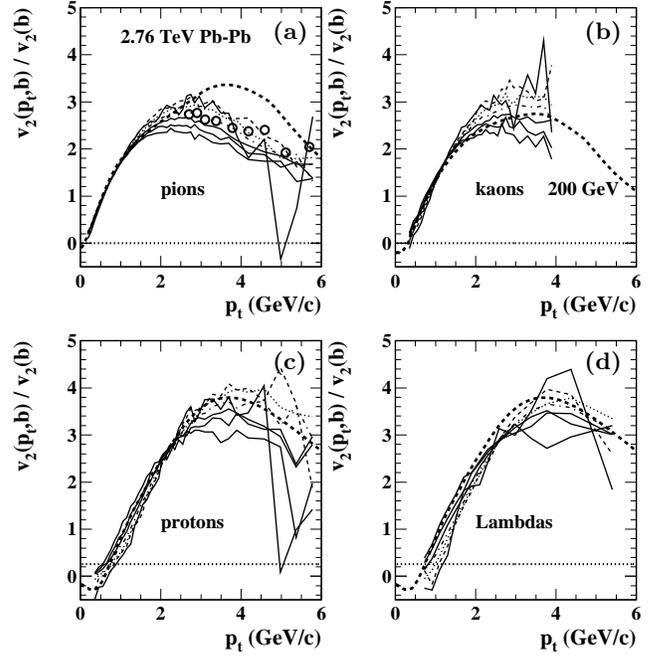

Figure~\ref{alice2} shows data from Fig.~\ref{alice1} rescaled by factor $1/v_2(b)$ where 
$v_2(b)$ is the open circles in Fig.~\ref{alice4} (left) as proxy for $v_2\{4\}$ data for unidentified hadrons ($\approx 80$\% pions) at 2.76 TeV from Ref.~\cite{alicev2b}.
The data for pions in particular fall on a single locus below 1 GeV/c. The bold dashed curves are the curves passing through 200 GeV MB $v_2(p_t)$ data in Fig.~\ref{x1} (left) derived in Ref.~\cite{quadspec} divided by 200 GeV $v_2(\text{MB}) \approx 0.055$. The open points in panel (a) are more-recent 200 GeV pion data with higher statistics~\cite{newstarpion} that agree well with the rescaled LHC data.

\begin{figure}[h]
     \includegraphics[width=1.65in]{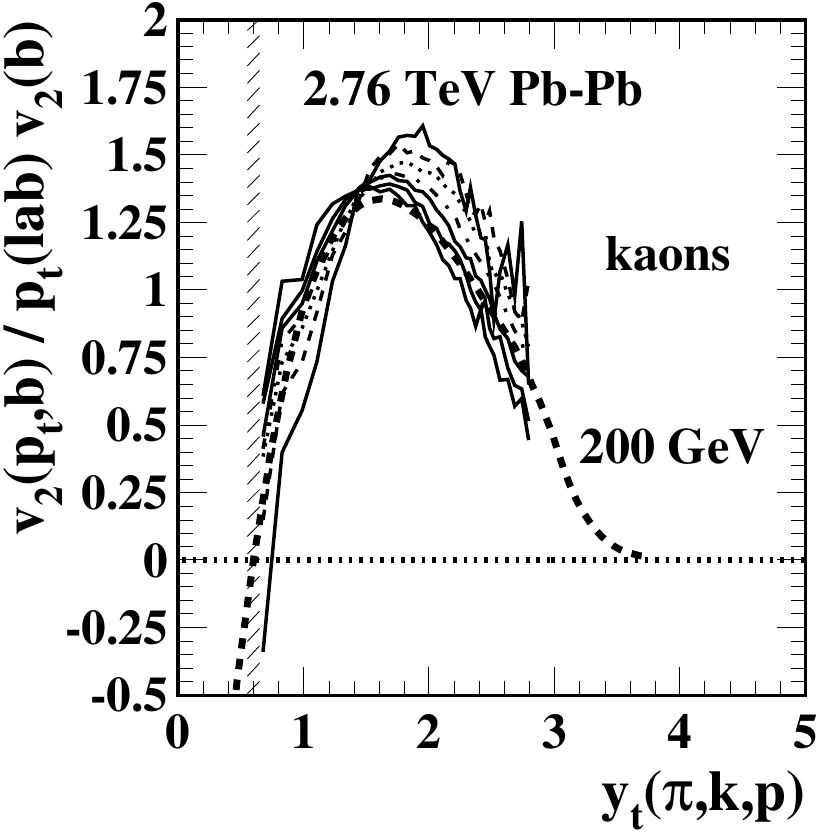}
     \includegraphics[width=1.65in]{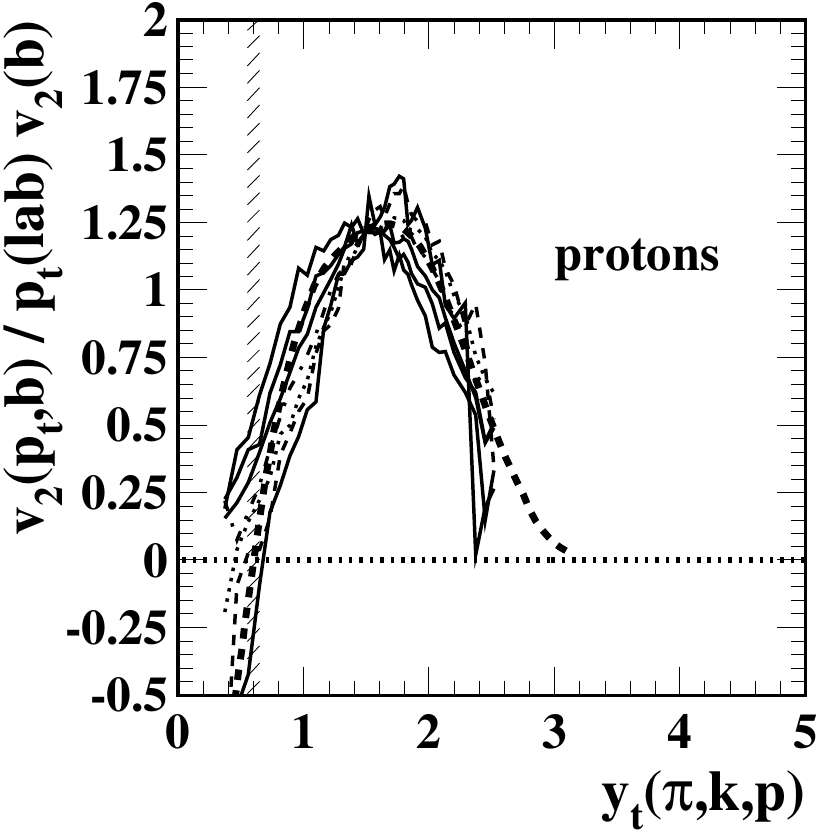}
 \caption{\label{alice3shift}
Rescaled $v_2(p_t,b)$ data from Fig.~\ref{alice2} divided by $p_t$ in the lab frame for two hadron species. A systematic variation in the apparent source boost (zero intercept) is apparent. The hatched bands indicate the nominal source boost $\Delta y_{t0} = 0.6$ inferred in Ref.~\cite{quadspec} from 200 GeV \auau\ MB $v_2$ data.
} 
\end{figure}

Figure~\ref{alice3shift} shows data from Fig.~\ref{alice2} divided by \pt\ in the lab frame and plotted on proper \yt\ for each hadron species as in Fig.~\ref{x1} (right). The bold dashed curves in this figure are derived from those in Fig.~\ref{x1} (right) again divided by $v_2(\text{MB}) = 0.055$. The general trend for kaons and protons is a zero intercept near $y_t = 0.6$ as in the 200 GeV MB study, interpreted there as a source boost common to several hadron species. However, close examination of the data reveals systematic variation of source boost $\Delta y_{t0}$ (\yt\ intercept) with collision centrality.

Figure~\ref{alice4} (right) shows boost deviations (from 0.6) required to bring data as in Fig.~\ref{alice3shift} onto a common locus corresponding to source boost $\Delta y_{t0} = 0.6$. The hatched band indicates the location on centrality of the sharp transition in jet systematics (onset of ``jet quenching'') noted in Ref.~\cite{anomalous}. There is no apparent correspondence.

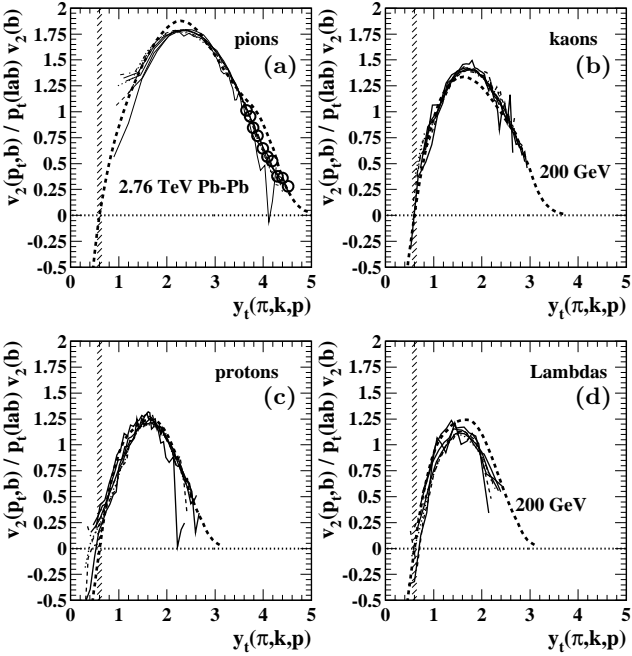
\begin{figure}[h]
     \includegraphics[width=3.3in]{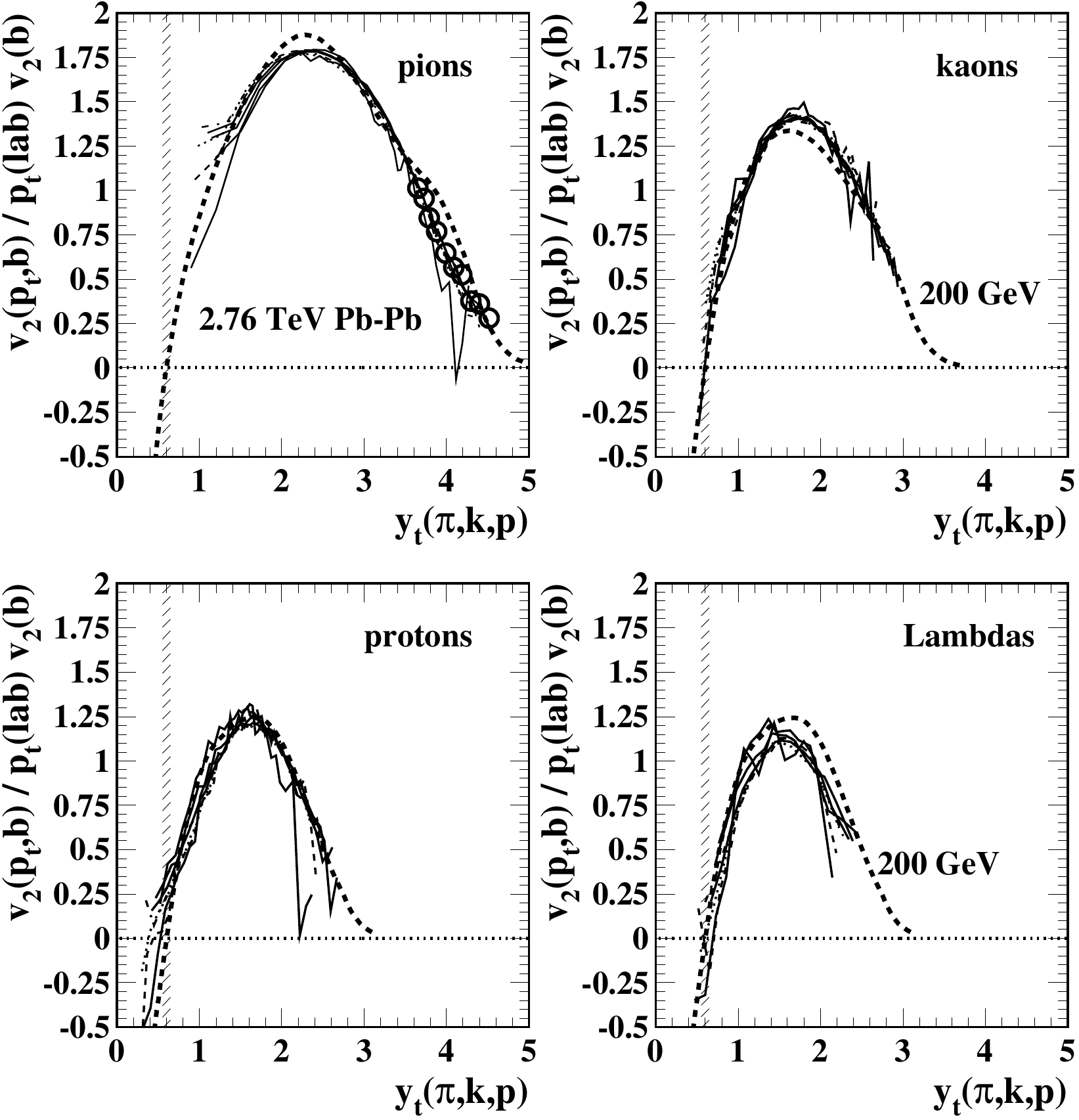}
\put(-140,220) {\bf (a)}
\put(-22,220) {\bf (b)}
\put(-140,95) {\bf (c)}
\put(-22,95) {\bf (d)}
\caption{\label{alice3}
Data prepared as in Fig.~\ref{alice3shift} but shifted on \yt\ to a common source boost $\Delta y_{t0} = 0.6$ based on the boost trend in Fig.~\ref{alice4} (right). The dashed curves are  from Fig.~\ref{x1} (right).  The open points in panel (a) are more-recent high-statistics 200 GeV pion data from Ref.~\cite{newstarpion}.
}  
\end{figure}

Figure~\ref{alice3} shows data from Fig.~\ref{alice2} divided by \pt\ in the lab frame and plotted on proper \yt\ for each hadron species as in Fig.~\ref{x1} (right). In this case the 2.76 TeV data for several centralities are shifted on \yt\ according to Fig.~\ref{alice4} (right) corresponding to a common source boost $\Delta y_{t0} = 0.6$. The data for each of four hadron species coincide for seven centralities within their uncertainties and with equivalently-scaled 200 GeV MB (dashed) trends. 

This result confirms that \pt-differential and \pt-integral $v_2$ data at 2.76 TeV are quantitatively consistent and correspond with 200 GeV $v_2(p_t,b)$ data scaled up by common factor 1.3, contradicting a claim in Ref.~\cite{alicev2b} (see Sec.~\ref{conflict}). The result also suggests that $v_2(p_t',b)$ transformed to the boost frame factorizes (see Sec.~\ref{quadcent}).
\subsection{SP spectra vs quadrupole spectra}

The next step in deriving quadrupole spectra requires SP spectra for identified hadrons corresponding to these $v_2(p_t,b)$ data. Ideally, SP spectra for each centrality and hadron species would be available. Included in Ref.~\cite{alicespec2} are SP spectra only for pions, kaons and protons, and only for 0-5\% and 60-80\% central 2.76 TeV \pbpb\ collisions and \pp\ collisions as in App.~\ref{spspec}. Since the data trends in Fig.~\ref{alice3} do not vary significantly with centrality the data for 30-40\% central are adopted as representative. Corresponding SP spectra are averages of 0-5\% and 60-80\% per-participant-scaled spectra as shown in Fig.~\ref{alice5}.

\begin{figure}[h]
     \includegraphics[width=3.3in]{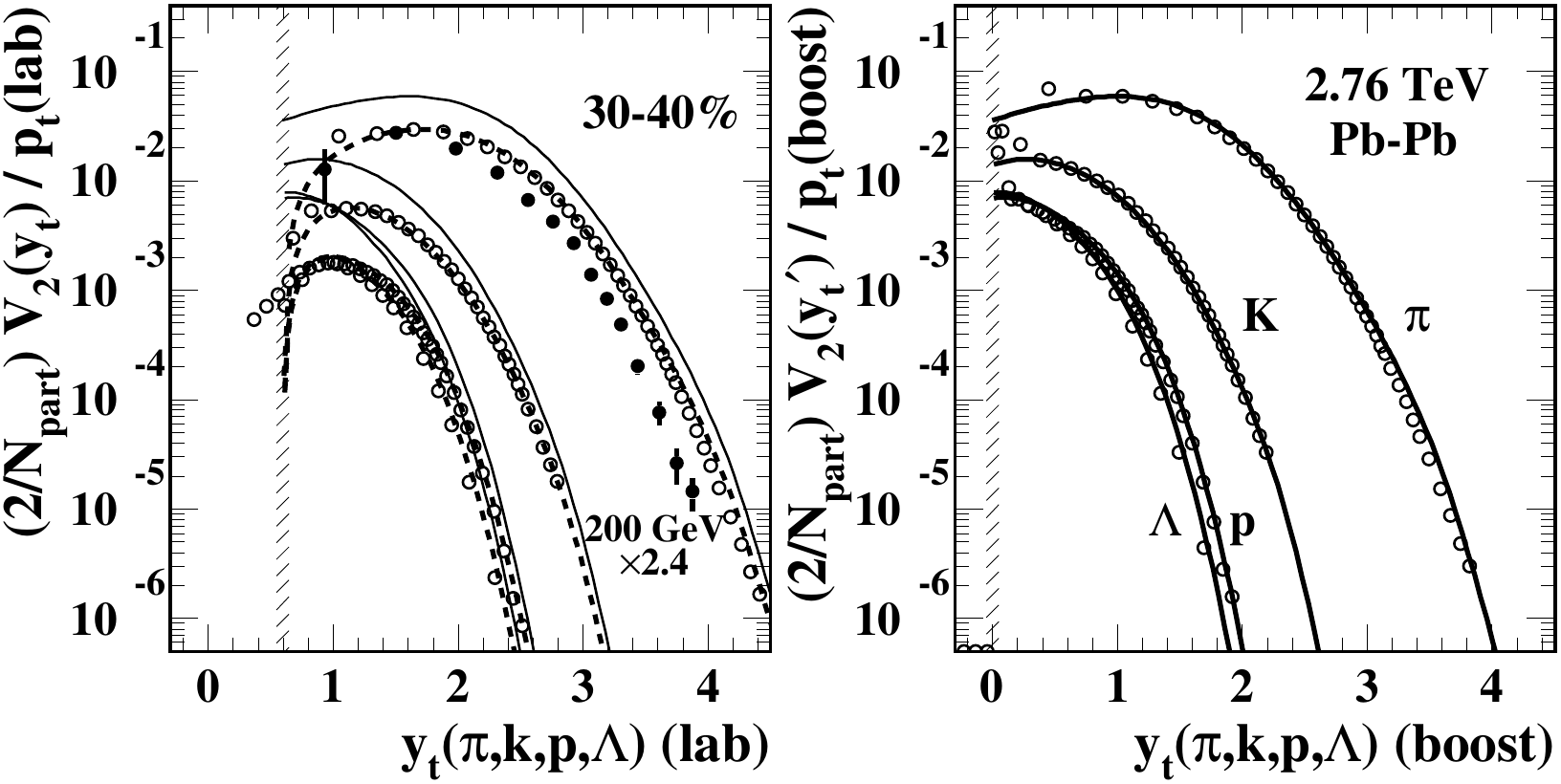}
\caption{\label{alice8ab}
Left: 2.76 TeV data and 200 GeV dashed curves from Fig.~\ref{alice3} multiplied by corresponding SP spectra in the form $(2/N_{part}) \bar \rho_0(y_t,b)$ for the 30-40\% centrality class of 2.76 TeV \pbpb\ collisions (rescaled as in App.~\ref{spspec}). The solid points are 200 GeV pion data from Fig.~\ref{x2} (right) scaled up by collision-energy factor 2.4 defined in Sec.~\ref{edep}.
Right: Data and dashed curves from the left panel divided by ratio $p_t' / p_t$ derived from Eq.~(\ref{kine}) and shifted by $\Delta y_{t0} = 0.6$ to the boost frame. The transformed dashed curves become the solid curves which are repeated also in the left panel (but in the lab frame).
}  
\end{figure}

Figure~\ref{alice8ab} (left) shows data for four hadron species from Fig.~\ref{alice3} multiplied by $v_2(b)$ and corresponding SP spectra in the form $(2/N_{part}) \bar \rho_0(y_t,b)$ (except Lambda $v_2$ data are multiplied by the proton SP spectrum). All \pbpb\ spectra have been rescaled by factor 1/1.65 as discussed in App.~\ref{spspec} and shown in Fig.~\ref{alice5aa}. This 2.76 TeV result can be compared with 200 GeV data in Fig.~\ref{x2} (right). The dashed curves from Fig.~\ref{x1} (right) are processed identically to obtain the various dashed curves through 2.76 TeV data in this panel. The solid curves are reproduced from the right panel for comparison. The solid points are 200 GeV pion data from Fig.~\ref{x2} (right) scaled up by collision-energy factor 2.4 (explained in Sec.~\ref{edep}).

Figure~\ref{alice8ab} (right) shows data from the left panel (lab frame) multiplied by kinematic factor $p_t / p_t'$ from Eq.~(\ref{kine}) [without factor $\gamma_t(1-\beta_t) \approx 0.55$] and transformed  to the boost frame (shifted left by $\Delta y_{t0}$ = 0.6) to obtain data  proportional to quadrupole spectra $\bar \rho_2(y_t',b)$ for four hadron species defined in Eq.~(\ref{stuff}). The solid curves are the dashed curves from the left panel treated the same. The last step is transformation to $m_t' - m_h$.

\begin{figure}[h]
     \includegraphics[width=3.3in]{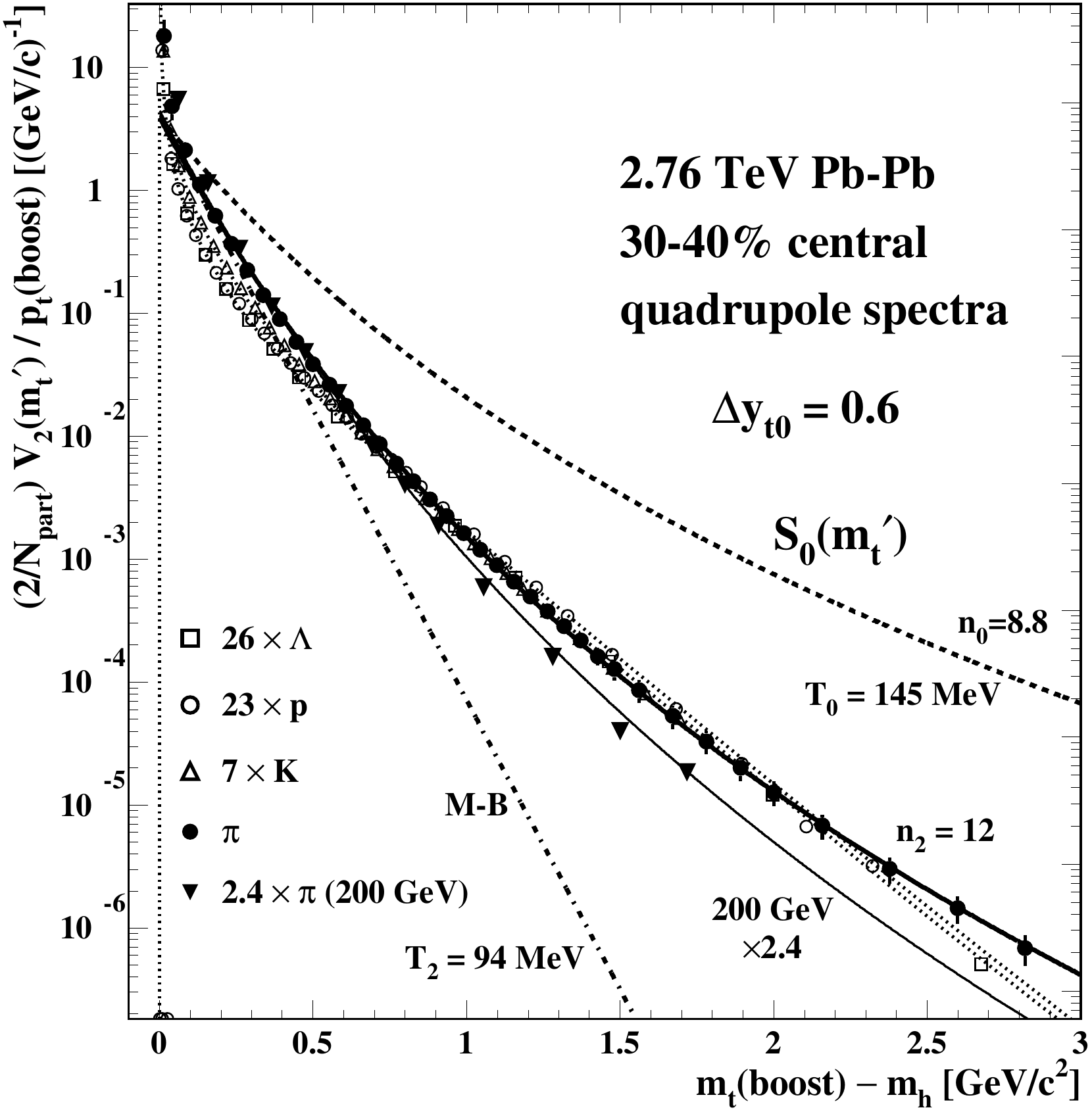}
\caption{\label{alice8c}
Data from Fig.~\ref{alice8ab} (right) transformed  to $m_t' - m_h$ and rescaled (relative to pions) as indicated in the plot. The error bars for pion data are increased 3-fold for visibility. The bold solid curve is a L\'evy distribution that describes rescaled 2.76 TeV data. The dash-dotted curve is the M-B limit for $T_2 = 94$ MeV. The dashed curve is proportional to the SP spectrum soft component for unidentified hadrons from 2.76 TeV \pp\ collisions derived in Ref.~\cite{alicetomspec}. The thin solid curve is the 200 GeV quadrupole spectrum from Fig.~\ref{xbig} rescaled by energy factor 2.4. The dotted curves are explained in the text.
}  
\end{figure}

Figure~\ref{alice8c} shows quadrupole spectra for four hadron species transformed to  $m_t' - m_h$ and rescaled as noted on the plot (corresponding to the 200 GeV result in Fig.~\ref{xbig} from Ref.~\cite{quadspec}). Above 0.7 GeV/$c^2$ the spectra coincide as for 200 GeV data, but below that point there are significant deviations. As discussed in App.~\ref{spspec} the deviations may arise from bias in the low-\pt\ parts of some SP spectra. The two dotted curves are the 200 GeV dashed curves from Fig.~\ref{alice3} (b) and (c) for kaons and protons transformed in the same manner as the 2.76 TeV data. The origin of those curves is a single universal quadrupole spectrum back-transformed via 200 GeV SP spectra to describe 200 GeV $v_2$ data in Ref.~\cite{quadspec}. The deviations in Fig.~\ref{alice8c} appear to be due to the 2.76 TeV SP spectra. Otherwise, quadrupole spectra for four hadron species at 2.76 TeV are well-described by a single L\'evy distribution (bold solid curve) with $T_2 \approx 94$ MeV and $n_2 \approx 12$ compared to  $T_2 \approx 92$ MeV and $n_2 \approx 14$ for 200 GeV pion data from Fig.~\ref{xbig} ( inverted solid triangles and thin solid curve, both rescaled by energy factor 2.4). The dash-dotted curve is the \mbox{M-B} equivalent with $1/n_2 \rightarrow 0$. The dashed curve is proportional to hadron SP spectrum soft component $\hat S_0(m_t')$ for 2.76 TeV \pp\ collisions~\cite{alicetomspec} plotted in the boost frame for comparison.

The data-derived quantity in Fig.~\ref{alice8c} is 
\bea \label{stuff2}
\frac{2}{N_{part}}\, \frac{V_2(m_t')}{p'_t} &=&  f(m_t';\Delta y_{t0},\Delta y_{t2})
\\ \nonumber 
&\times&   \frac{2}{N_{part}}\,\left\{\frac{\Delta y_{t2}}{2T_2}\right\}\, \bar \rho_2(m_t';T_2,n_2)
\\ \nonumber 
&\approx&   \frac{2}{N_{part}}\, \left\{\frac{\Delta y_{t2}\bar \rho_2}{2T_2}\right\} \hat S_2(m_t';T_2,n_2)
\eea
plotted as points for four hadron species. The function $f(m_t';\Delta y_{t0},\Delta y_{t2})$ is unity at lower \mt\ but increases monotonically with increasing \mt\ at a rate determined by the unknown ratio $\Delta y_{t2}/\Delta y_{t0}$. Exponent $n_2 = 12$ is then a lower limit for the actual quadrupole spectrum. Unit-normal $\hat S_2(m_t';T_2,n_2)$ estimates the functional form of a universal quadrupole spectrum shape for 2.76 TeV. 

The product $\Delta y_{t2}(b)\bar \rho_2(b)$ represents the ``amplitude'' of the NJ quadrupole. At present there is no way to determine the two factors separately but some limiting cases can be considered. The condition $\Delta y_{t2} < \Delta y_{t0}$ (positive-definite boost) determines a lower limit on quadrupole SP density $\bar \rho_2$. Comparison of the distinctive shape of the NJ quadrupole spectrum (cutoff at $\Delta y_{t0}$ and very soft spectrum) with SP spectra may establish an upper limit on $\bar \rho_2$. In Ref.~\cite{quadspec} an upper limit on pion $\bar \rho_2$ of 5\% of the total hadron density $\bar \rho_0$ was estimated by such a spectrum comparison.

\section{Centrality and energy trends} \label{edep}

Reference~\cite{davidhq} demonstrated that $v_2\{2D\}$ data for 62 and 200 GeV \auau\ collisions in the per-particle form $A_Q\{2D\}(b,\sqrt{s_{NN}})$ factorize approximately as
\bea
A_Q\{2D\}(b,\sqrt{s_{NN}}) \propto N_{bin} \epsilon_{opt}^2(b) \log(\sqrt{s_{NN}} / \text{13 GeV})~
\eea 
as in Fig.~\ref{v2ep} (right). It was further demonstrated that 200 GeV $v_2\{2D\}(y_t,b)$ data for unidentified hadrons in the form $V_2\{2D\}(y_t,b) = \bar \rho_0(y_t,b) v_2\{2D\}(y_t,b)$ factorize approximately as~\cite{davidhq2,v2ptb}
\bea \label{davidhq}
V_2\{2D\}(y_t,b) &\approx& \langle 1/p_t \rangle V_2\{2D\}(b) \, p_t \, Q_0(y_t)
\eea
where a common factor $V_2\{2D\}(b)$ has been canceled on both sides relative to Eq.~(14) of Ref.~\cite{v2ptb}, and $Q_0(y_t)$ is a L\'evy distribution on $m_t'$, as defined in Fig.~\ref{xbig}, transformed to $y_t'$ in the boost frame and boosted by $\Delta y_{t0} \approx 0.6$ to \yt\ in  the lab frame. In this study procedures established with RHIC data are extended to address LHC $v_2$ data at 2.76 TeV. Is factorization a good approximation over a large collision-energy interval, and if so what are the implications for the hydro narrative?

For a given \aa\ collision system  the complete argument dependence is $(p_t,b,\sqrt{s_{NN}};h)$ for identified hadrons $h$. Ideally, from Eq.~(\ref{shat}) reexpressed in the boost frame one may conjecture that
\bea
 V_2(m_t',b,\sqrt{s};h)/p_t' \hspace{-.03in} &\approx& \hspace{-.03in}\langle 1/p_t' \rangle V_2(b,\sqrt{s};h)  \hat S_2(m_t',\sqrt{s}),~~~~
\eea
where $s_{NN} \rightarrow s$, $\hat S_2(m_t',\sqrt{s_{NN}})$ is defined by the combination $(n_2,T_2)$, $T_2 \approx 93$ MeV appears to be universal and $n_2$ may depend only weakly on $\sqrt{s_{NN}}$. The principal (lab-boost) difference from the approximation in Eq.~(\ref{davidhq}) occurs near the lab-spectrum lower bound at $y_t = \Delta y_{t0}$ per Fig.~\ref{boost3} (right). The discussion below considers evidence for individual pair-wise factorizations $(b,\sqrt{s_{NN}})$, $(p_t,b)$ and $(p_t,\sqrt{s_{NN}})$ in that order. Note that $V_2(p_t,b,\sqrt{s_{NN}})$ may have different properties from $v_2(p_t,b,\sqrt{s_{NN}})$, since ratio $v_2$ incorporates properties of denominator $\bar \rho_0(p_t,b,\sqrt{s_{NN}})$ per Eq.~(\ref{v2struct}).

\subsection{Quadrupole A-A centrality-energy factorization} \label{quadenergy}

The first issue is \pt-integral $(b,\sqrt{s_{NN}})$ factorization. Figure~\ref{quadx} (right) indicates that pair data for 62 and 200 GeV \auau\ collisions follow the same trend
\bea \label{fig2trends}
V_2^2(b) &=& \bar \rho_0(b) A_Q(b) 
\\ \nonumber
&\propto& N_{part} N_{bin} \epsilon^2(b)
\propto N_{part}^2 \epsilon_{opt}^2(b) \, \nu
\eea
represented by the solid curve for both energies, or
\bea \label{bigv2b}
\frac{2}{N_{part}} V_2(b) &=&\frac{2}{N_{part}} \bar \rho_0(b) v_2(b) \propto  \epsilon_{opt}(b) \sqrt{\nu}
\eea
for single particles, with $\nu \approx (N_{part}/2)^{1/3}$ and $\sqrt{\nu} \in [1,2.4]$ for \auau\ collisions. The same argument may hold for the centrality dependence of $V_2(p_t,b)$ modulo the centrality dependence of source boost $\Delta y_{t0}(b)$.

Figure~\ref{alice4} (left) compares $v_2(b,\sqrt{s_{NN}})$ data for 200 GeV [dotted curve, Eq.~(\ref{magic})] and 2.76 TeV (solid points), and the shapes seem to be compatible modulo a factor 1.3. But such compatibility would be in conflict with Eq.~(\ref{fig2trends}), since the shape of SP yield trend $\bar \rho_0(b)$ changes significantly from 200 GeV to 2.76 TeV due to increased dijet production (see Fig.~\ref{alice5aa} and Ref.~\cite{nominijets}, App.~B). Thus $V_2(b,\sqrt{s_{NN}})$ and $v_2(b,\sqrt{s_{NN}})$ should not both factorize. The apparent contradiction may be due to different $v_2$ analysis methods. The 200 GeV $v_2\{2D\}$ data trend in Fig.~\ref{alice4} (left) accurately distinguishes jet structure from the NJ quadrupole based on model fits, whereas the 2.76 TeV data are $v_2\{4\}$. Although it is claimed (based on a toy-model simulation {\em emphasizing hadronic resonances}~\cite{v24nojets}) that the latter method is resistant to ``nonflow''~\cite{alicev2b},  simulations based on real jet properties or \auau\ data analysis each indicate that $v_2\{4\}$ may acquire a substantial MB jet contribution in more-central \aa\ collisions~\cite{davidhq}. Based on available data one may then conjecture that $V_2\{2D\}(b,\sqrt{s_{NN}})$ actually factorizes as
\bea \label{vv2b}
V_2\{2D\}(b,\sqrt{s_{NN}}) 
 &\propto&  N_{part} \, \epsilon_{opt}(b) \sqrt{\nu} \log(s_{NN}/s_0)~~~
\eea
with $\sqrt{s_0} \approx 10$ GeV.

\begin{figure}[h]
     \includegraphics[width=3.3in]{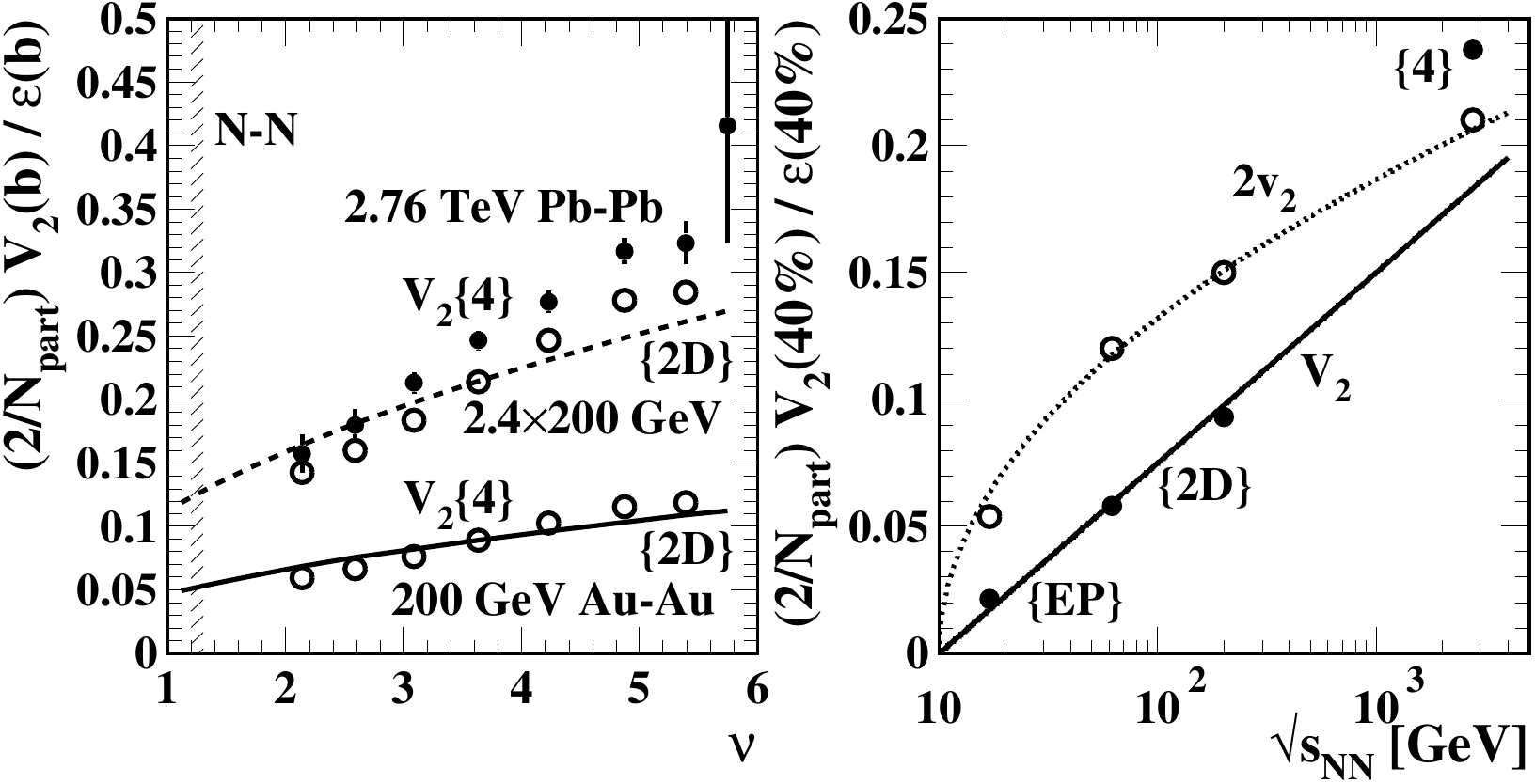}
\caption{\label{aliceedep}
Left: $V_2\{4\}(b)$ data (solid points) reconstructed from 2.76 TeV $v_2\{4\}(b)$ data in Fig.~\ref{alice4} (left) and corresponding $\bar \rho_0(b)$ yield trend from Ref.~\cite{nominijets}. The solid curve and lower open points are derived from Eq.~(\ref{vv2b}) $\propto \sqrt{\nu}$ and 200 GeV \auau\ $v_2\{4\}(b)$ data from  Fig.~\ref{v2ep}. The dashed curve and upper open points are the 200 GeV trends scaled up by energy factor 2.4 explained in the  text.
Right: $V_2(b)$ data (solid points) vs $\sqrt{s_{NN}}$ evaluated at $1-\sigma/\sigma_0 \approx 40$\% ($\nu \approx 3.5$). The solid line is $0.075 \log_{10}(\sqrt{s_{NN}} / \text{10 GeV})$. The dotted curve through $v_2$ data $\propto \sqrt{\log(\sqrt{s_{NN}} / \text{10 GeV})}$ guides the eye.
}  
\end{figure}

Figure~\ref{aliceedep} (left) shows $V_2\{4\}(b)$ data (solid points) reconstructed from the 2.76 TeV $v_2\{4\}(b)$ data in Fig.~\ref{alice4} (left) and the corresponding 2.76 TeV \pbpb\ $\bar \rho_0(b)$ yield trend (from Ref.~\cite{nominijets}, App.~B). The 200 GeV solid curve is derived from Eq.~(\ref{magic}) consistent with Eq.~(\ref{vv2b}). The dashed curve is the solid curve multiplied by factor $1.3 \times 1.87 \approx 2.4$, where 1.3 is the $v_2(b)$ scaling in Fig.~\ref{alice4} (left) and 1.87 represents the $\log(s_{NN}/s_0)$ scaling of $\bar \rho_0(b)$ (soft component) as reported in Ref.~\cite{alicetomspec}, Sec.~V-B between 200 GeV and 2.76 TeV.  The lower open circles are $V_2\{4\}$ values derived from the $A_Q\{4\}(b)$ data shown in Fig.~\ref{v2ep} (right). The upper open circles are those data multiplied by factor 2.4. The correspondence is remarkable in that the upper two sets of points are derived from four independent measurements (of $v_2$ and hadron yields at each of two energies).
Larger deviations of data from the dashed curve for more-central collisions are consistent with jet bias in $v_2\{4\}(b)$ observed for RHIC 200 GeV \auau\ data~\cite{davidhq} which should increase  with stronger jet production at LHC energies. The apparent agreement between $v_2(b)$ data for two energies in Fig.~\ref{alice4} (left) may then be misleading (at least for more-central collisions).

Figure~\ref{aliceedep} (right) shows the energy trend of $V_2(b,\sqrt{s_{NN}}) = \bar \rho_0(b,\sqrt{s_{NN}}) v_2(b,\sqrt{s_{NN}})$ (solid points) with $b$ corresponding to fractional cross section $1 - \sigma/\sigma_0 = 0.4$ ($\nu \approx 3.5$), and with analysis methods for different $v_2$ data noted. The general trend is $V_2(b,\sqrt{s_{NN}}) \propto  \log(s_{NN}/s_0)$ (solid line) as in Eq.~(\ref{vv2b}), modulo the high $V_2\{4\}$ point consistent with the left panel. The 62 and 200 GeV $V_2\{2D\}$ points are consistent with the ratio $1.56 = \sqrt{2.45}$ derived from Fig.~\ref{quadx} (right).
Also shown are corresponding $v_2(b,\sqrt{s_{NN}})$ data (open points). 

Given the structure of Eq.~(\ref{v2struct}) the energy dependence of $v_2(b,\sqrt{s_{NN}})$ is difficult to interpret. The dotted curve $\propto \sqrt{\log(s_{NN}/s_0)}$ is intended only to guide the eye. With no jet or valence-quark contributions $v_2(b,\sqrt{s_{NN}})$ might be nearly independent of energy, the $\log(s_{NN}/s_0)$ factors in $V_2(b,\sqrt{s_{NN}})$ and $\bar \rho_0(b,\sqrt{s_{NN}})$ nearly canceling in the ratio. At lower energies the valence-quark contribution from projectile nucleons should return to midrapidity and reduce the $v_2$ ratio. At higher energies the increasing jet contribution to $\bar \rho_0(b,\sqrt{s_{NN}})$ in the denominator might reduce the $v_2$ ratio, but only if jet-related (nonflow) contributions to the numerator in Eq.~(\ref{v2struct}) are excluded.

\subsection{Quadrupole-spectrum centrality dependence} \label{quadcent}

The second issue is $(p_t,b)$ factorization for given collision energy. Reference~\cite{v2ptb} presented $v_2(p_t,b)$ data for unidentified hadrons from 62 and 200 GeV \auau\ collisions. A general result was the approximate factorization in Eq.~(\ref{davidhq}), $Q_0(y_t)$ being a universal function (boosted L\'evy distribution) approximately independent of centrality and corresponding to fixed source boost $\Delta y_{t0} = 0.6$. That result is consistent with the analysis of MB data in Sec.~\ref{200gevquad} but unidentified hadrons ($\approx$ pions) are relatively insensitive to source boost (see Fig.~\ref{x1}, right). 

Figure~\ref{alice3} of this study suggests that in the lab frame
\bea
v_2(p_t,b;h)/ p_t \, v_2(b;h)  &\approx&F[p_t;\Delta y_{t0}(b),h],
\eea
common to 200 GeV and 2.76 TeV but specific to each hadron species $h$. Source boost $\Delta y_{t0}(b)$  varies with centrality as in Fig.~\ref{alice4} (right), but in the boost frame
\bea
v_2(p_t',b;h) /p_t'\, v_2(b;h) &\approx& F(p_t';h)
\eea
and $v_2(p_t',b;h)$ factorizes, with hadron-specific functions $F(p_t';h)$. Alternatively, Fig.~\ref{alice8c} demonstrates that for a given energy and centrality the quadrupole spectra $\propto V_2(m_t',b;h)$ for several hadrons species are identical in the boost frame, as shown  also in Fig.~\ref{xbig} for 200 GeV. In principal $v_2(p_t,b)$ and $V_2(p_t,b)$ cannot both factorize because of the complex behavior of SP spectrum $\bar \rho_0(p_t,b)$ that relates them. However, one can conjecture that
\bea
V_2(m_t',b;h) /p_t'\, V_2(b;h) &\approx& G(m_t')
\eea
is the more fundamental relation, and that the $v_2(p_t,b;h)$ factorizations suggested by Figure~\ref{alice3} arise only because of the relatively small changes in $\bar \rho_0(p_t,b;h)$ with centrality (e.g.\ Fig.~\ref{alice5}). Tests of that hypothesis would require more complete and accurate correlation {\em and} spectrum data for identified hadrons than are currently available.

\subsection{Quadrupole-spectrum energy dependence}

The third issue is $(p_t,\sqrt{s_{NN}})$ factorization for given centrality $b$. Figure~\ref{alice3} of the present study suggests that $v_2(p_t,b,\sqrt{s_{NN}})/p_t\,v_2(b,\sqrt{s_{NN}}) $ has the same functional form in the boost frame for 200 GeV MB $v_2$ data (bold dashed curves) and for all centralities of 2.76 TeV data (thin curves of several line styles), {\em separately} for three hadron species (pions, kaons, Lambdas). But comparison of Fig.~\ref{xbig} and Fig.~\ref{alice8c} reveals that $V_2(m_t',\sqrt{s_{NN}})$ in the boost frame has the same form for several hadron species at each energy but does change significantly with energy: quadrupole L\'evy exponent $n_2$ becomes significantly smaller with increasing energy, matching a similar trend for SP spectrum soft-component exponent $n_0$ (attributed to Gribov diffusion)~\cite{alicetomspec}.
Again, $v_2(p_t,\sqrt{s_{NN}})$ and $V_2(p_t,\sqrt{s_{NN}})$ should not both factorize.

However, quadrupole spectrum $V_2(p_t,\sqrt{s_{NN}})$ and SP spectrum $\bar \rho_0(p_t,\sqrt{s_{NN}})$ (Fig.~\ref{alice5aa}) both experience similar shape changes with increasing energy (measured by respective L\'evy exponents $n_x$) due to a common origin: low-$x$ gluons. With increasing collision energy projectile PDFs extend to lower momentum fraction resulting in increased transverse-momentum dispersion. The distribution tails rise as a consequence (L\'evy exponents decrease). The changes may then nearly cancel in the $v_2(p_t,\sqrt{s_{NN}})$ {\em ratio} which gives a misleading impression.

\subsection{Quadrupole-spectrum factorization summary}

Those several results can be summarized by ($s_{NN} \rightarrow s$)
\bea \label{fac01}
 V_2(m_t',b,\sqrt{s};h) \hspace{-.04in}  
&=& \hspace{-.04in}\langle 1/p_t' \rangle V_2(b,\sqrt{s};h) \, p_t' \, \hat S_2(m_t',\sqrt{s})~~
\\ \nonumber
&\approx& p_t' \frac{\Delta y_{t2}(b) \bar \rho_2(b,\sqrt{s};h)}{2 T_2} \hat S_2(m_t',\sqrt{s}),
\eea
where the first line is empirical, inferred from data, and the second line is based on Eq.~(\ref{stuff2}).
The quadrupole source-boost MB value is $\Delta y_{t0} \approx 0.6$ at both 200 GeV and 2.76 TeV. The variation with centrality at 200 GeV is unknown, but the variation at 2.76 TeV (Fig.~\ref{alice4}, right) is modest over the measured centrality interval. The quadrupole spectrum shape $\hat S_2(m_t',\sqrt{s_{NN}})$ (common to several hadron species) is  defined by the combination $(n_2,T_2)$, where $T_2$ is apparently independent of collision system and and L\'evy exponent $n_2$ depends only weakly on $\sqrt{s_{NN}}$ (a trend similar to the SP spectrum but with different values). Mean value $\langle 1/p_t' \rangle$ follows accordingly. $\Delta y_{t2}(b)\, \bar \rho_2(b,\sqrt{s_{NN}};h)$ centrality and energy trends factorize, and hadron species dependence is consistent with the statistical model at both energies. 
Quadrupole slope parameter $T_2 \approx 93$ MeV is universal like SP spectrum slope parameter $T_0 \approx 145$ MeV but the two values are very different: quadrupole and SP spectra are distinct.

The universality of $v_2$ data manifested by quadrupole spectrum $V_2(m_t',b,\sqrt{s_{NN}};h)$ in the boost frame is consistent with NJ quadrupole production based on low-$x$ gluons~\cite{gluequad}. $V_2(m_t',b,\sqrt{s_{NN}})$ does not rely on \pp\ or \aa\ collision energy except for the $\log(s_{NN}/s_0)$ low-$x$ gluon density trend. The same energy trend is the basis for MB dijet production but the centrality trend is very different: the two production mechanisms are {\em related but distinct}.

\section{Systematic uncertainties} \label{syserr}

\subsection{200 GeV quadrupole spectra}

Systematic uncertainties for the analysis in Ref.~\cite{quadspec} were presented in that article. However, two further comments are appropriate here: (a) The  Lambda data from 0-10\% central 200 GeV \auau\ collisions in Fig.~\ref{x2} (left) (solid points) were released after the analysis in Ref.~\cite{quadspec} was completed. The significant negative values below the zero intercept near $y_t = 0.6$ confirm the prediction of the quadrupole-spectrum analysis represented by the dash-dotted curve. (b) It is instructive to compare the statistical uncertainties in Fig.~\ref{x1} (left) of this article with those in Fig.~\ref{xbig}. In the former case the errors for larger \pt\ are comparable to the panel range whereas the errors at smaller \pt\ are tiny. In the latter case relative errors (on a semilog scale) are comparable for all \pt\ values except the last few points. The difference is a consequence of the structure of Eq.~(\ref{v2ep}). Relative to the errors of Fourier amplitude $V_2(p_t)$ in Fig.~\ref{xbig} the errors of $v_2(p_t)$ include an extra factor $1/\sqrt{\bar \rho_0(p_t)}$ that increases rapidly with increasing \pt\ implying that statistically-significant information at lower \pt\ may be visually suppressed.

\subsection{$\bf v_2$ data: 200 GeV vs 2.76 TeV }

Figure~\ref{alice1} shows $v_2\{SP\}(p_t,b)$ data for 15 million 2.76 TeV \pbpb\ events with statistical and systematic errors combined in quadrature. The error bars are much reduced from the 200 GeV data in Fig.~\ref{x1} (left) (e.g.\ 200 GeV kaon and Lambda data were based on $\approx 1.5$ million minimum-bias \auau\ collisions). However, the trend of errors is the same: $v_2$ errors at lower \pt\ are tiny suggesting that important information in that interval is visually suppressed, whereas \v2\ data transformed to quadrupole spectra make the same information visually accessible.

Significant systematic differences among $v_2\{2D\}$, $v_2\{4\}$ and $v_2\{SP\}$ indicate continuing issues with NGNM $v_2$ data arising from jet (nonflow) bias. Fig.~\ref{alice4} (left) compares $v_2\{4\}$ data from 200 GeV and 2.76 TeV collisions with 200 GeV $v_2\{2D\}$ data [represented by Eq.~(\ref{magic}), dotted curve] scaled up by factor 1.3. In this plot format the $v_2\{4\}$ data for two energies appear compatible but differ systematically from the $v_2\{2D\}$ trend. 

Fig.~\ref{alice3} illustrates the combined systematic consistency of 200 GeV $v_2\{2D\}(b)$ data, the simple $\Delta y_{t0}(b)$ trend in Fig.~\ref{alice4} (right) and 2.76 TeV $v_2\{SP\}(p_t,b)$ data. If the $v_2\{SP\}(p_t,b)$ data were rescaled by $v_2\{4\}(b)$ data instead significantly more scatter would be introduced. Fig.~~\ref{aliceedep} (left) shows a $v_2\{2D\}$ vs $v_2\{4\}$ comparison in more detail in terms of Fourier amplitudes $V_2(b)$ that eliminate the $1/\bar \rho_0(b)$ factor in ratio measure $v_2(b)$.  Fig.~\ref{aliceprl} also reveals issues with $v_2\{4\}(p_t,b)$ vs $v_2\{SP\}(p_t,b)$  in the context of hydro predictions relating to \v2\ energy dependence.

\subsection{2.76 TeV SP spectrum data}

Relative uncertainties in the \pt\ (or \mt) structure of reconstructed quadrupole spectra $\bar \rho_2(m_t',b)$, or equivalently Fourier amplitudes $V_2(m_t',b)$, depend on $v_2(p_t,b)$ data and SP spectrum data $\bar \rho_0(p_t,b)$. SP spectra for four hadron species required for the present study are described in App.~\ref{spspec} where two contributions to systematic error are discussed: (a) an apparently extraneous factor 1.65 for 2.76 TeV \pbpb\ spectra relative to \pp\ spectra and (b) a possible uncorrected inefficiency for proton and kaon spectra at lower \pt. Issue (a) is accommodated in the present study by rescaling the \pbpb\ spectra with factor 1/1.65. Issue (b) relates to distortion of the reconstructed quadrupole spectra for protons and  kaons.

Figure~\ref{alice5aa} (c) and (d) illustrate the relation of 200 GeV \auau\ and \pp\ spectra.
The per-participant-pair (normalization by $N_{part}/2$) spectrum format provides an accurate quality-assurance test for \aa\ vs \pp\ data. The very-peripheral \aa\ data should approach \pp\ data smoothly as a limit, as is observed for those data. The spectrum values at low \pt\ also coincide as expected.

Figure~\ref{alice5aa} (a) and (b) show the 2.76 TeV SP spectrum data from Ref.~\cite{alicespec2} used in the present study. Given rescaling of the \pbpb\ data by factor 1/1.65 the pion spectra vary as expected from the 200 GeV case. However, the proton data show large deviations from expected behavior at lower \pt\ as noted elsewhere in the text. Those systematic deviations far exceed what might be expected for these high-statistics data and dominate the uncertainty in inferred quadrupole spectra as in Fig.~\ref{alice8c}.

\subsection{2.76 TeV quadrupole spectra and source boosts}

The plotted error bars for the quadrupole-spectrum data in Fig.~\ref{alice8c} are simply the published $v_2\{SP\}(p_t,b)$ data uncertainties in Fig.~\ref{alice1} transformed the same as the data values. The resulting error bars are typically smaller than the points. For illustration the pion error bars plotted  in Fig.~\ref{alice8c} have been increased by factor 3 to insure visibility at least for the points at largest $m_t'$. For the pion data there appears to be  excellent systematic control, especially in relation to the 200 GeV quadrupole spectrum data (inverted solid triangles).
However, as noted elsewhere there are substantial systematic deviations for kaon and especially proton data that appear to be directly related to SP spectrum issues noted in App.~\ref{spspec}; e.g.\ a factor-3 proton deviation at lower \pt\ noted in Fig.~\ref{alice5aa} (b) matches the similar deviation in Fig.~\ref{alice8c}.

The centrality and energy dependence of unit-integral $\hat S_0(m_t';T_2,n_2)$  defined in Eqs.~(\ref{shat}) and (\ref{stuff2}) depends on parameters $T_2$ and $n_2$, determined at 2.76 TeV by pion data alone as in Fig.~\ref{alice8c}. Presently-available data do not require any significant change in $T_2 \approx 93\pm 1$ MeV with either centrality or energy. L\'evy exponent $n_2$ decreases significantly with energy from $n_2 = 14\pm 1$ at 200 GeV to $n_2 = 12\pm 1$ at 2.76 TeV as is evident in the same figure.

The centrality and energy dependence of \pt-integral $V_2(b,\sqrt{s}) \propto \Delta y_{t2}(b,\sqrt{s}) \bar \rho_2(b,\sqrt{s})$ is shown in Fig.~\ref{aliceedep}. The centrality trend of the plotted ratio in the left panel inferred from 200 GeV $v_2\{2D\}$ data is $\propto \sqrt{\nu}$ (solid and dashed curves). The trends indicated by $v_2\{4\}$ data deviate significantly from the $\{2D\}$ trends in ways expected for jet-related bias (nonflow), including deviations increasing with collision energy. The $V_2(\sqrt{s})$ energy dependence (right panel) appears to be close to $\propto \log(s/s_0)$ with $\sqrt{s_0} \approx$ 10 GeV but the uncertainty at 2.76 TeV is large.
There is currently no evidence for a varying (or any) thermodynamic EoS or QCD phase transition from observed \v2\ data trends that are simple and consistent from low-multiplicity \pp\ collisions to central \aa\ collisions (Fig.~\ref{quadx}) and over a large energy interval (Fig.~\ref{aliceedep}).

That quadrupole source boost $\Delta y_{t0}(b)$ varies significantly with \aa\ centrality at 2.76 TeV is demonstrated by comparison of Figs.~\ref{alice3shift} and \ref{alice3}. An inferred centrality variation is sketched as the linear trend in Fig.~\ref{alice4} (right). A 20\% change in the slope of Fig.~\ref{alice4} (right) cannot be excluded by data, and the  trend could be significantly nonlinear on fractional cross section $\sigma/\sigma_0$. The \v2\ data are consistent with no significant energy dependence of $\Delta y_{t0}$ between 200 GeV and 2.76 TeV at the current level of uncertainty in inferred boost values. Presently-available \v2\ data do not require significant dispersion in the source boost for a given collision system (no evidence from \v2\ data for Hubble expansion of a bulk medium).

\section{discussion} \label{disc}

\subsection{Conflicting reports of $\bf v_2$ energy dependence} \label{conflict}

Reference~\cite{alicev2b} reported the first LHC measurements of \pt-integral and \pt-differential $v_2$ for unidentified charged hadrons from \pbpb\ collisions at 2.76 TeV. The analysis method used is denoted by $v_2\{4\}$. While \pt-integral $v_2\{4\}$ was observed to increase by factor 1.3 compared to 200 GeV data, as in  Fig.~\ref{alice4} of this study, the \pt-differential data were said to be equivalent (within uncertainties) to comparable 200 GeV data from the STAR collaboration. It was further reported that those results confirm certain hydro model predictions~\cite{kestin,niemi} and that factor 1.3 corresponds to increase of ensemble-mean \pt\ due to increased radial flow. Those \pt-differential data are in conflict with Ref.~\cite{alicev2ptb} data presented in this study.

\begin{figure}[h]
     \includegraphics[width=3.3in]{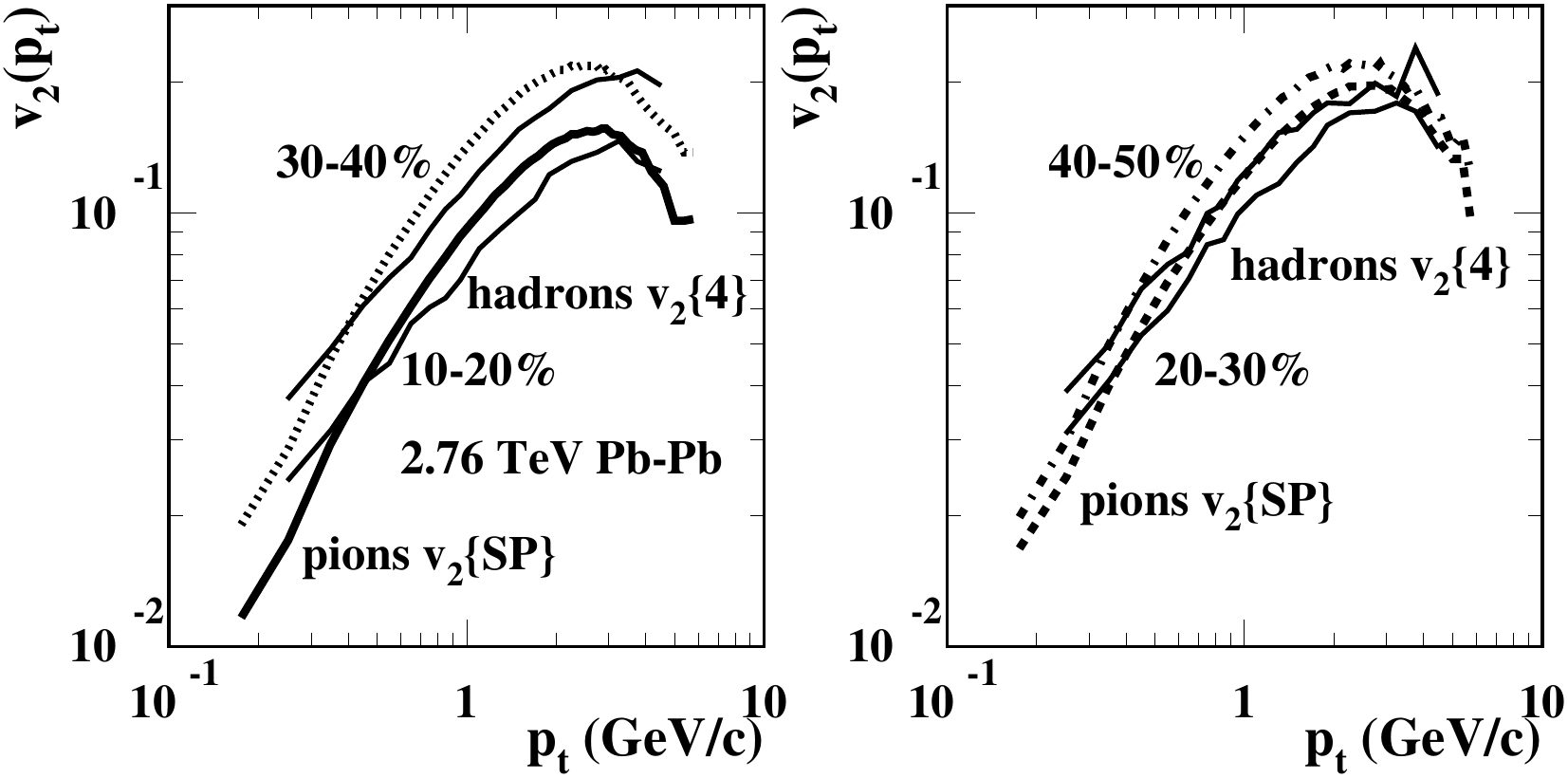}
\caption{\label{aliceprl}
\pt-differential $v_2\{4\}$ data for unidentified hadrons (thin solid curves~\cite{alicev2b}) and  $v_2\{SP\}$ data for identified pions (bold curves of several line styles~\cite{alicev2ptb}) from four centralities of 2.76 TeV \pbpb\ collisions illustrating systematic differences.
}  
\end{figure}

Figure~\ref{aliceprl} shows $v_2\{4\}$ data for unidentified charged hadrons from Ref.~\cite{alicev2b} for four centralities of 2.76 TeV \pbpb\ collisions (thin solid curves) compared to corresponding $v_2\{SP\}$ data for identified pions from Ref.~\cite{alicev2ptb} as presented in Fig.~\ref{alice1} (bold curves of several line styles). The log-log format provides the best visual access to differential structure. Because $v_2$ for more-massive hadrons is typically larger in magnitude (see Fig.~\ref{alice1}) one expects the data for unidentified hadrons to exceed significantly that for identified pions over a relevant \pt\ interval but to have a similar shape on \pt. Both those expectations are contradicted by the unidentified-hadron $v_2\{4\}$ data from Ref.~\cite{alicev2b} in Fig.~\ref{aliceprl} (thin solid curves). For the \pt\ interval most apparent in the conventional linear format the hadron $v_2\{4\}$ data are about 20\% low compared to the pion $v_2\{SP\}$ data (and perhaps 30\% low compared to unbiased hadron data), and thus {\em seemingly} compatible with 200 GeV measurements.

The relevant theory predictions include ``...for heavier particles like protons $v_2(p_T)$ will be below the values measured at RHIC, even if the $p_T$-integrated $v_2$ is larger''~\cite{niemi}, and ``...while $p_t$-integrated elliptic flow {\em increase}[s] from RHIC to LHC the differential elliptic flow...{\em decreases} in the same...energy range''~\cite{kestin} (both attributed to effects of radial flow). While the first $v_2\{4\}$ results for unidentified hadrons from Ref.~\cite{alicev2b} seemed to support those hydro predictions the later $v_2\{SP\}$ data for identified hadrons from Ref.~\cite{alicev2ptb} (the same collaboration) strongly contradict the theory predictions. The later result is also consistent with a previous study revealing that evidence for radial flow in \auau\ \pt\ spectra from the RHIC is negligible~\cite{hardspec}. Evolution of \pt\ spectra is dominated by a MB dijet contribution predicted by pQCD~\cite{fragevo} and consistent with jet-related 2D angular correlations~\cite{jetspec}. The present study confirms that \pt-integral and -differential $v_2$ data are precisely compatible, one being the simple integral of the other as in Eq.~(\ref{v2int}). Increase of ensemble-mean \pt\ from RHIC to LHC energies responds to increased dijet production as demonstrated in Ref.~\cite{tomalicempt}, not radial flow.

\subsection{IS parton and FS hadron production models}

According to the conventional flow narrative copious particle (parton and/or hadron) rescattering is required to convert any IS \aa\ configuration-space asymmetry to a FS momentum-space asymmetry measured by $v_2$. In that context observation of substantial $v_2$ interpreted as elliptic flow is seen as confirming formation of a dense flowing medium by rescattering. Estimates of  copious IS parton scattering seem to provide conditions for the required {\em re}\,scattering, but such estimates can be questioned based on differential spectrum analysis~\cite{fragevo}. If most FS hadrons belong to a TCM soft component (as observed) formed outside the collision volume and therefore do not rescatter (as established by fixed-target \ha\ experiments), and the NJ quadrupole is an independent phenomenon unrelated to the TCM soft component, the bases for claiming a dense flowing medium are negated.

There are thus two competing scenarios for the IS: (a) projectile-nucleon dissociation leading to isolated gluons fragmenting to charge-neutral hadron pairs (that do not rescatter) as the great majority of FS hadrons or (b) copious IS large-angle parton scattering as the dominant mechanism for FS hadron production. The phenomenology of low-energy jets in yields, spectra and correlations tracked from NSD \pp\ collisions continuously on centrality to central \aa\ collisions overwhelmingly prefers scenario (a). All jets predicted by measured cross sections down to 3 GeV survive to the FS~\cite{anomalous,hardspec,fragevo,jetspec}. Jets are {\em unmodified} over the more-peripheral half of the total cross section (for 200 GeV \auau\ collisions). In the more-central half jets are indeed substantially modified~\cite{anomalous} but are still described quantitatively by pQCD (modified DGLAP equations)~\cite{fragevo}. Low-energy jets do serve as sensitive probes of the collision system but fail to demonstrate a dense bulk medium. The NJ quadrupole represents a small fraction of the FS, with manifestations in angular correlations but not in SP spectra or yields. Quadrupole trends are also inconsistent with a flowing bulk medium as discussed further below.

\subsection{Hydro vs $\bf V_2^2(b,\sqrt{s_{NN}})$ NJ quadrupole trends}

The characteristics of $v_2(b,\sqrt{s_{NN}})$ data from the RHIC and LHC (Sec.~\ref{edep}) are inconsistent with hydro expectations for $(b,\sqrt{s_{NN}})$ trends in several ways. 
The NJ quadrupole measured by  $V_2^2(n_{ch},\sqrt{s_{NN}})$ in \pp\ collisions and $V_2^2(b,\sqrt{s_{NN}})$ in \aa\ collisions shows a trend $\propto N_{part} N_{bin}$ common to both collision systems. A factor $\epsilon_{opt}^2(b)$ is required by the latter system but not the former, possibly due to quantum effects~\cite{ppquad}. The nominal $(b,\sqrt{s_{NN}})$ density trend relevant to hydro would vary by orders of magnitude from low-multiplicity \pp\ to central \auau, but there is no change in quadrupole systematics throughout that interval, no threshold relating to very large particle densities and copious particle rescattering.


A dramatic change in jet characteristics (sharp transition or ST) observed near 50\% centrality in 62 and 200 GeV Au-Au collisions~\cite{anomalous} could indicate major changes in (or the onset of) a conjectured dense flowing medium or QGP. But the ST induces no corresponding change in quadrupole $v_2$ data for the same collision systems which maintain the same smooth $V_2^2(b) \propto N_{part} N_{bin}\epsilon_{opt}^2(b)$ trend for all \auau\ centralities~\cite{noelliptic}. Similar issues emerge for the TCM.
The SP spectrum soft component shows no change with \aa\ centrality~\cite{hardspec}. The same-side jet peak for MB 2D angular correlations (representing {\em all} FS jets) reveals an azimuth width monotonically {\em decreasing} with \auau\ centrality from peripheral to central collisions, also without correspondence to the ST~\cite{anomalous}. There is thus no indication from those correlation structures associated with MB dijets of copious particle rescattering in a dense flowing medium leading to jet broadening.

\subsection{Hydro vs $\bf V_2^2(p_t,b)$ and quadrupole spectra}

The $p_t$ dependence of ratio measure $v_2(p_t)$ has been considered critical for interpretation of elliptic flow as a hydro phenomenon and for claims of ``perfect liquid'' at the RHIC~\cite{perfect}. However, transformation of $v_2(p_t)$ data to $V_2^2(p_t,b,\sqrt{s_{NN}})$ and inference of quadrupole spectra motivate alternative interpretations of $v_2(p_t)$ data that challenge basic theory assumptions supporting the flow narrative. There are two main issues: (a) quadrupole spectrum shape vs SP spectrum shape and (b) implications from inferred source-boost distributions.


(a) Quadrupole spectra inferred from $v_2(p_t,b)$ data by the method described in Sec.~\ref{quadspecmeth} are very different in shape from SP hadron spectra for the same collision system. The differences falsify the hydro assumption that almost all FS hadrons emerge from a dense, flowing medium. The unique shape of the quadrupole spectrum compared to the SP spectrum may set an upper limit on the fraction of hadrons ``carrying'' the NJ-quadrupole correlation component, suggesting that only a small fraction of FS hadrons participate~\cite{quadspec} and contradicting the fundamental hydro assumption that flows result from large particle densities and copious rescattering. 

The quadrupole spectrum shape in the boost frame, described by a simple L\'evy distribution on $m_t' - m_h$, may be essentially independent of \aa\ centrality or hadron species, and the collision-energy dependence of the spectrum shape is small. The quadrupole spectrum is thus universal,  ruling out constituent-quark (NCQ) models of hadronization from a bulk medium or QGP~\cite{quadspec}. Attempts to ``scale'' LHC $v_2(p_t)$ data to confirm the NCQ hypothesis fail detailed differential data analysis~\cite{alicev2ptb}.


(b) The ``mass ordering'' ascribed to $v_2(p_t)$ vs \pt\ data plots for identified hadrons does imply a source boost common to several hadron species, but that choice of plotting format obscures the boost {\em distribution} that is the {\em primary product of hydro theory}.  As demonstrated in the present study the relevant source-boost distribution is directly accessible in a model-independent way from plots of $v_2(y_t)/ p_t$ vs $y_t$ for several hadron species.

The source-boost distribution inferred directly from $v_2(y_t)/p_t$ data in relation to quadrupole spectra is quite different from hydro expectations. The MB monopole source-boost value $\Delta y_{t0} \approx 0.6$ is the same for two widely-separated collision energies. The boost {\em dispersion} for a given collision system $(b,\sqrt{s_{AA}},A)$ is small and consistent with zero: a single fixed boost value is characteristic of an expanding thin cylindrical shell and inconsistent with assumed Hubble expansion of a flowing bulk medium. Monopole boost $\Delta y_{t0}$ should correspond to radial flow according to the flow narrative, but the  majority of hadrons represented by the SP spectrum exhibit no radial flow: evolution of SP hadron spectra is dominated by the dijet contribution~\cite{hardspec}. And there is no correspondence between NJ-quadrupole and SP-spectrum trends: The NJ quadrupole does not represent azimuth modulation of radial flow carried by a dense, flowing medium.

\section{Summary} \label{summ}

A dominant feature of angular correlations from noncentral nucleus-nucleus (\aa) collisions is the cylindrical-quadrupole component conventionally represented by symbol $v_2$ and interpreted to represent elliptic flow -- azimuth modulation of transverse or radial flow of a conjectured dense bulk medium reflecting the initial-state \aa\ geometry. Elliptic flow is in turn centrally important to the {\em flow narrative} whose main feature is a quark-gluon plasma (QGP) that has been described as a ``perfect liquid'' with minimal viscosity. Recent results from the large hadron collider (LHC) have been interpreted to indicate that ``collectivity'' (flow) is manifested even in small systems (e.g.\ \pp\ collisions). Such strong claims for novelty should be tested rigorously with available data. The present study presents evidence against such claims based on analysis of published \pt-integral $v_2(b)$ and \pt-differential $v_2(p_t,b)$ data, the latter processed to extract {\em nonjet (NJ) quadrupole spectra} for several hadron species.  


The general goals of this study include review of several \v2\ analysis methods and their results, especially identifying and excluding the jet (``nonflow'') contribution to published \v2\ data from several methods, and extension of quadrupole spectrum studies established at the relativistic heavy ion collider (RHIC) to LHC energies and to \aa\ centrality dependence. Based on previous quadrupole-spectrum analysis of RHIC data a NJ quadrupole spectrum shape and hadron source-boost distribution associated with the azimuth quadrupole should be accessible for each collision system. Those results can be compared with predictions from hydrodynamic (hydro) theory.


Direct information about conjectured elliptic flow must come from 2D angular correlations on $(\eta,\phi)$, specifically the quadrupole component of the 1D projection onto azimuth $\phi$. Methods restricted to the 1D projection may include a substantial jet contribution from intrajet correlations (same-side 2D peak or ``jet  cone''). Methods that utilizes full information from 2D angular correlations may successfully eliminate the jet contribution, and simple universal trends are then observed for \pt-integral $v_2(b)$ data from \aa\ {\em and} \pp\ collisions at RHIC energies. In particular, the number of correlated pairs associated with the NJ quadrupole component is simply related in terms of Glauber parameters to the product $N_{part} N_{bin}$ of number of participants and number of binary collisions, participants being projectile nucleons in \aa\ collisions or projectile low-$x$ gluons in \pp\ collisions. The only difference is an additional factor $\epsilon_{opt}^2$ (\aa\ eccentricity squared) in \aa\ collisions but not in \pp\ collisions. The same trend persisting over such a large range of particle densities and collision systems presents a major challenge for the flow narrative and the elliptic-flow interpretation.

More information can be derived from \pt-differential $v_2(p_t,b)$ data, especially in the form of quadrupole spectra and hadron source boosts. Whereas conventional ratio measure $v_2(p_t,b) = V_2(p_t,b) / \bar \rho_0(p_t,b)$ includes single-particle (SP) hadron spectrum $\bar \rho_0(p_t,b)$ in its denominator, Fourier amplitude $V_2(p_t,b)$ provides direct access to quadrupole hadron spectrum $\bar \rho_2(p_t,b)$ associated with the quadrupole correlation component, which may or may not relate to the SP spectrum describing most hadrons. 

In a previous study summarized in this article quadrupole spectra from $v_2(p_t,b)$ data for three hadron species from 0-80\%--central 200 GeV \auau\ collisions revealed interesting trends: (a) the quadrupole components reflect a common fixed source boost $\Delta y_{t0} = 0.6 \approx \beta_t$, (b) quadrupole spectra transformed to $m_t'$ in the boost frame have identical shapes described by a L\'evy distribution with slope parameter $T_2 \approx 93$ MeV, (c) relative hadron abundances follow predictions of the statistical model and (d) the properties of quadrupole spectra are very different from the SP spectra describing most hadrons.

In the present study the same procedures are applied to  $v_2(p_t,b)$ data for four hadron species from seven centralities of 2.76 TeV \pbpb\ collisions, extending the quadrupole-spectrum method to \aa\ centrality dependence and to collision-energy dependence over a large interval. The inferred source boost is again common to all hadron species and does vary significantly with \aa\ centrality. However, the mean source boost at 2.76 TeV is not significantly different from $\Delta y_{t0} = 0.6$ at 200 GeV.


At a more detailed level one may ask to what extent $v_2(p_t,b)$ data {\em factorize}, leading to further simplification that may aid physical interpretation of the data. The relevant control observables are $(p_t,b,\sqrt{s};h)$, where $h$ represents a hadron species. Results from the present study reveal  that the simplest data representation is in terms of the quadrupole Fourier amplitude on transverse mass $m_t'$ {\em in the boost frame}: $V_2(p_t,b,\sqrt{s};h) \rightarrow V_2(m_t',b,\sqrt{s};h)$ with the factorized expression given in Eq.~(\ref{fac01}) of this article, where $p_t'$ is \pt\ in the boost frame, $ \Delta y_{t2}$ is the source-boost quadrupole amplitude, $\bar \rho_2$ is the quadrupole-related hadron density and $\hat S_2(m_t')$ is a unit-integral spectrum shape common to all hadron species and \aa\ centralities. The slight energy dependence of $\hat S_2(m_t')$ is controlled entirely by L\'evy exponent $n_2(\sqrt{s})$. Quadrupole slope parameter $T_2 \approx 93$ MeV appears to be universal.

The \pt-integral quadrupole angular density is given in factorized form by Eqs.~(\ref{fig2trends}) and  (\ref{bigv2b}), where $V_2(b,\sqrt{s};h) \propto \log(s/s_0)$ with $\sqrt{s_0} \approx 10$ GeV and relative hadron abundances follow statistical-model predictions. The universal trend $V_2^2(b) \propto N_{part} N_{bin} \epsilon_{opt}^2(b)$ in terms of the number of quadrupole-related hadron pairs is as described above for 200 GeV \auau\ data.


The simple quadrupole trends may be used to interpret the nonjet (NJ) quadrupole phenomenon. Given the single source-boost value for each collision system and the major differences between quadrupole spectra and SP hadron spectra it is unlikely that the quadrupole component includes most hadrons emerging from a Hubble-expanding bulk medium. It seems more likely that the quadrupole component is ``carried'' by a small minority of final-state hadrons. The quadrupole centrality dependence is very different from the $\propto N_{bin}$ trend observed for dijet (hard) production or the $\propto N_{part}$ trend observed for the soft component arising from projectile-nucleon dissociation. On the other hand, the trend $V_2(\sqrt{s}) \propto \log(s/s_0)$ suggests that the quadrupole is related to low-$x$ gluons in common with dijet production and the soft component but remains a distinct QCD mechanism.


Those observations have significant implications for the flow narrative and related interpretations. Whereas hydro predictions include a broad source-boost distribution corresponding to Hubble expansion of a bulk medium $v_2(p_t,b)$ data  are actually consistent with a very narrow boost distribution, possibly a single value for each collision system. Whereas hydro theory assumes that the great majority of hadrons emerge in common from a flowing bulk medium with azimuth modulation for noncentral \aa\ collisions $v_2(p_t,b)$ data reveal that the quadrupole phenomenon is associated with a small minority of hadrons. Centrality and energy systematics show no indication of a changing (or any) equation of state or QCD phase transition. The same trends are observed from the smallest densities in \pp\ collisions to the largest densities in central \aa\ collisions. Given that $v_2(p_t,b;h)$ data for identified hadrons are accurately represented by a single quadrupole spectrum shape the concept of constituent-quark (NCQ) scaling is ruled out.

This initial study of quadrupole-spectrum centrality and energy dependence is necessarily limited, especially as regards obtaining accurate identified-hadron SP spectra to match $v_2(p_t,b;h)$ data point for point. For improved accuracy future studies should recover $V_2(p_t,b;h)$ Fourier amplitudes directly from corrected $\bar \rho_0(p_t,b;h)$ spectra and $v_2\{2D\}(p_t,b;h)$ data derived from model fits to 2D angular correlations {\em within  the same analysis}.

\begin{appendix}

\section{Boosted hadron sources} \label{boost}

This appendix reviews relativistic kinematics relating to nearly-thermal spectra for hadrons emitted from a moving (boosted) source, essentially a blast-wave model based on the Cooper-Frye description of rapidly-expanding particle sources~\cite{cooperfrye}. I consider only azimuth-monopole and -quadrupole $p_t$ and $y_t$ spectrum components. For algebraic simplicity ``thermal'' spectra are described in the boosted frame by Maxwell-Boltzmann (M-B) exponentials on $m_t$. Relative hadron abundances are assumed to correspond to a statistical model, but not necessarily because of a thermalization process~\cite{statmodel}. The spectrum description may be generalized to L\'evy distributions on $m_t$~\cite{wilk} for more accurate modeling of data. The intent is to provide a general description of hadron production from a source including (but not restricted to) a radially-boosted component with azimuth variation. This material is revised from Ref.~\cite{quadspec}.

\subsection{Radial boost kinematics}

The four-momentum components of a boosted source are first related to transverse rapidity $y_t$. The boost distribution is assumed to be a single value $\Delta y_t$ for simplicity.  The {\em particle} four-momentum components are $m_t = m_h \cosh(y_t)$ and $p_t = m_h\sinh(y_t)$. The {\em source} four-velocity (boost) components are $\gamma_t = \cosh(\Delta y_t)$ and $\gamma_t \, \beta_t= \sinh(\Delta y_t)$, with $\beta_t= \tanh(\Delta y_t)$. Boost-frame variables are defined in terms of lab-frame variables by
\bea \label{boostkine}
m'_t  &\equiv& m_h\, \cosh(y_t - \Delta y_{t}) =\gamma_{t}\, (m_t - \beta_{t}\, p_t)  \\ \nonumber
 &=& m_t\, \gamma_{t} \{ 1 - \tanh(y_t)\, \tanh(\Delta y_{t})  \} \\ \nonumber
p'_t &\equiv&   m_h\, \sinh(y_t - \Delta y_{t}) =\gamma_{t}\, (p_t - \beta_{t}\, m_t) \\ \nonumber
&=& m_t\, \gamma_{t} \{  \tanh(y_t) - \tanh(\Delta y_{t})  \}.
\eea
  
\begin{figure}[h]
   \includegraphics[width=1.65in,height=1.65in]{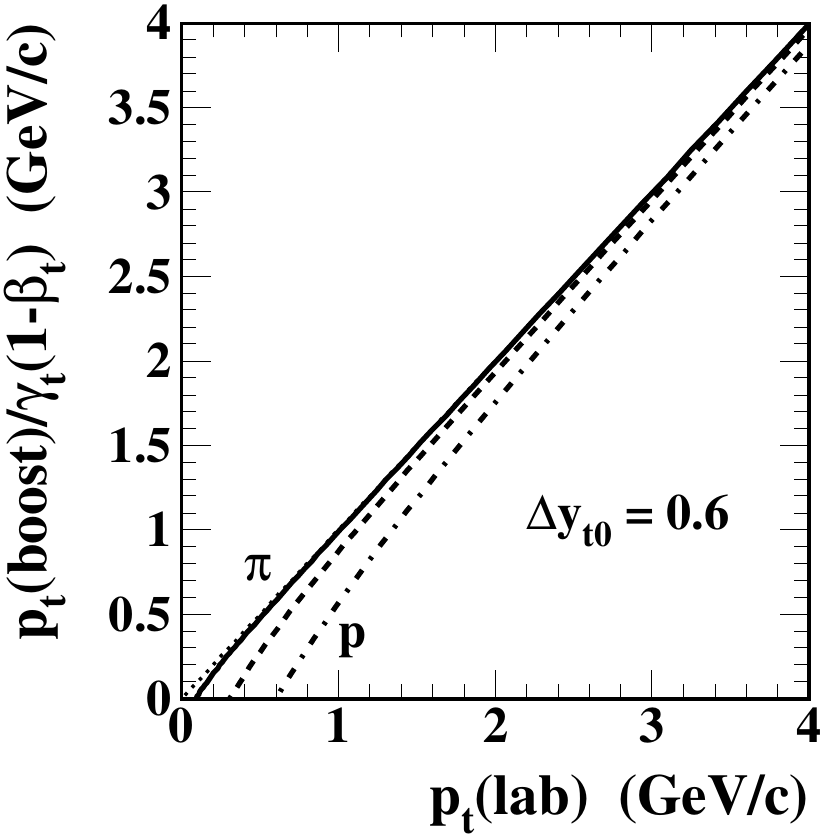}
  \includegraphics[width=1.65in,height=1.63in]{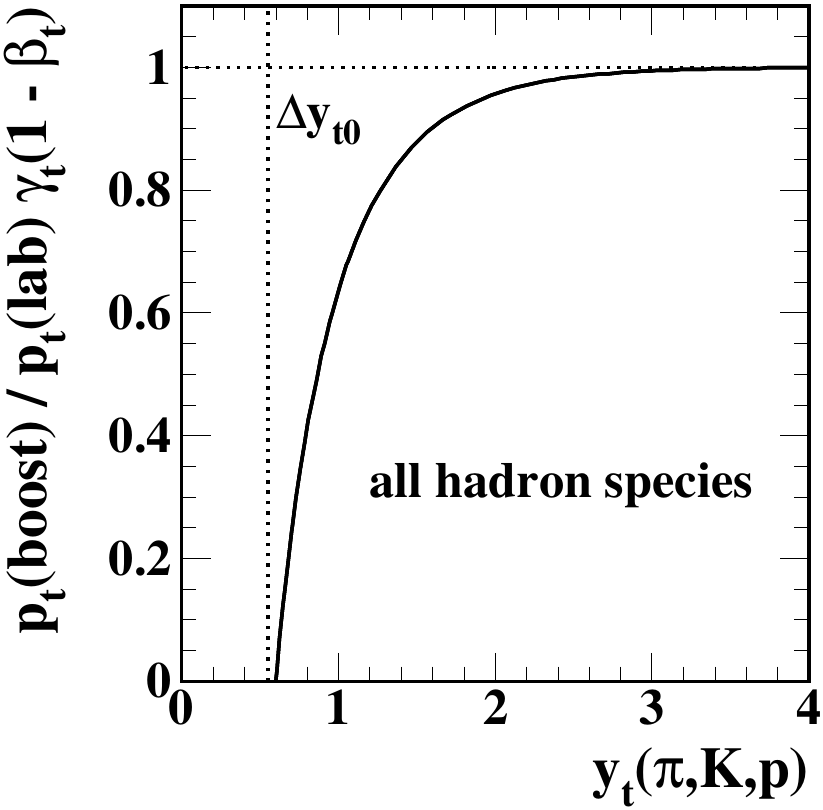}
\caption{\label{boost3}
Left: $p'_t$ ($p_t$ in the boost frame) {\em vs} $p_t$ in the lab frame. The normalizing factor $\gamma_{t}(1 - \beta_{t})$ in the denominator insures that the combination approaches $p_t$ for large $p_t$.
Right:  The quantity in the left panel divided by $p_t$(lab) vs proper $y_t$ for each hadron species, demonstrating Eq.~(\ref{kine}) as a universal trend common to all hadron species.
 }  
 \end{figure}

Fig.~\ref{boost3} (left) relates $p'_t \rightarrow p_t(\text{boost})$ to $p_t \rightarrow p_t(\text{lab})$. The main source of ``mass ordering'' for $v_2(p_t)$ at smaller $p_t$ (lower left), commonly interpreted to indicate ``hydro'' behavior, is a simple kinematic effect. The mass systematics hold for any boosted approximately-thermal hadron source independent of boost mechanism (i.e.\ hydrodynamics is not required). The {\em zero intercepts} ($p'_t = 0$) of the three curves, given by $p_{t0} = m_h \sinh(\Delta y_{t})$, are relevant for discussion of the hydro interpretation of $v_2(p_t)$. 

Fig.~\ref{boost3} (right) relates ratio $p'_t / p_t$ to transverse rapidity $y_t(\pi,K,p)$ and illustrates one reason why plots on $y_t$ are a major improvement over $p_t$ or $m_t$. Normalized $p'_t / p_t$  
\bea \label{kine}
 \frac{p'_t}{p_t\, \gamma_{t}(1 - \beta_{t})} 
    &=& \frac{1 - \beta_t /\tanh(y_t) }{1 - \beta_t}
\eea
increases from zero at monopole boost $\Delta y_{t0} $ and follows a {\em universal curve on $y_t$} to unit value for any hadron species. Thus, normalized $p'_t$ goes asymptotically to $p_t$ for large $p_t$ (or $y_t$) independent of boost. The form in Fig.~\ref{boost3} (right) is important for interpreting $v_2(p_t)$ data in terms of quadrupole spectra for several hadron species. 

The simplified blast-wave model~\cite{cooperfrye} invoked here for illustration assumes longitudinal-boost-invariant normal emission from an expanding thin cylindrical shell, slope parameter $T$ for $m_t$ spectra and thermal parameter $\mu = m_h / T$ for $y_t$ spectra.  Boosted spectra on $y_t$ and $m_t$ are
\bea \label{boostx}
\rho(y_t;\mu,\Delta y_t)  \hspace{-.05in} &=&  \hspace{-.05in} A_{y_t}\,\exp\{ -\mu\, [ \cosh(y_t - \Delta y_t) - 1]\} \\ \nonumber
\rho(m_t;T,\beta_t) \hspace{-.05in} &=&  \hspace{-.05in} A_{m_t}\, \exp\{- [\gamma_t\, (m_t - \beta_t\, p_t) - m_h]/T \},
\eea
providing a simplified description of ``thermal'' radiation from a radially-boosted cylindrical  source. Application of Eq.~(\ref{boostx}) requires a specific radial-boost model $\Delta y_t(r,\phi)$.

\subsection{Radial-boost models} \label{boostc}

In high-energy nuclear collisions there are at least two possibilities for the radial-boost model: (a) a monolithic, thermalized, collectively-flowing hadron source (``bulk medium'') with complex transverse flow (source-boost) distribution dominated by monopole (radial flow or Hubble expansion) and quadrupole (elliptic flow) azimuth components~\cite{starblast}; and (b) several hadron sources, some with azimuth-modulated transverse boost. Hadrons may emerge from a radially-fixed source (soft component), from parton fragmentation (hard component), and possibly from a source with radial boost varying smoothly with azimuth including monopole and quadrupole components. 

An eventwise radial boost distribution with monopole and quadrupole components is represented by
\bea \label{quadboost}
\Delta y_{t}(\phi_r) &=& \Delta y_{t0} + \Delta y_{t2}\, \cos(2 \phi_r ) \\ \nonumber
\beta_t(\phi_r) &=&  \tanh[\Delta y_t(\phi_r)] \\ \nonumber
&\simeq&\beta_{t0} + \beta_{t2}\, \cos(2\phi_r ),
\eea
with $\Delta y_{t2} \leq \Delta y_{t0}$ for {\em positive-definite boost}. The convention $\phi_r \equiv \phi - \Psi_r$ is adopted for more compact notation where $\Psi_r$ is an event-wise reference angle that may relate to an \aa\ reaction plane. Monopole boost component $\Delta y_{t0}$ is easily inferred from $v_2(p_t)$ data, but quadrupole component $\Delta y_{t2}$ is less accessible. Monopole boost $\Delta y_{t0}$ could be associated with ``radial flow'' but may apply to only a small fraction of FS produced hadrons. Quadrupole boost amplitude $\Delta y_{t2}(b)$ may relate to eccentricity $\epsilon(b)$ of the  \aa\ collision geometry.

\section{Single-particle spectra} \label{spspec}

This appendix refers to single-particle (SP) spectra required to infer quadrupole spectra from $v_2(p_t,b)$ data. Conventional differential measure $v_2(p_t,b)$ [e.g.\ as defined in Eq.~(\ref{v2pt2})] includes SP spectrum $\bar \rho_0(p_t,b)$ in its denominator as noted in Eq.~(\ref{v2struct}). The $v_2(p_t,b)$ {\em ratio} may thus introduce a substantial bias from jet contributions to the SP spectrum, aside from possible jet-related contributions to its numerator (``nonflow'') depending on the choice of NGNM. Unique to the NJ quadrupole is the term $V_2\{2D\}(p_t,b)$ which includes the quadrupole spectrum as a factor as noted in Eq.~(\ref{combfac}). To isolate quadrupole spectra from $v_2(p_t,b)$ data for identified hadrons the corresponding SP spectra are required.

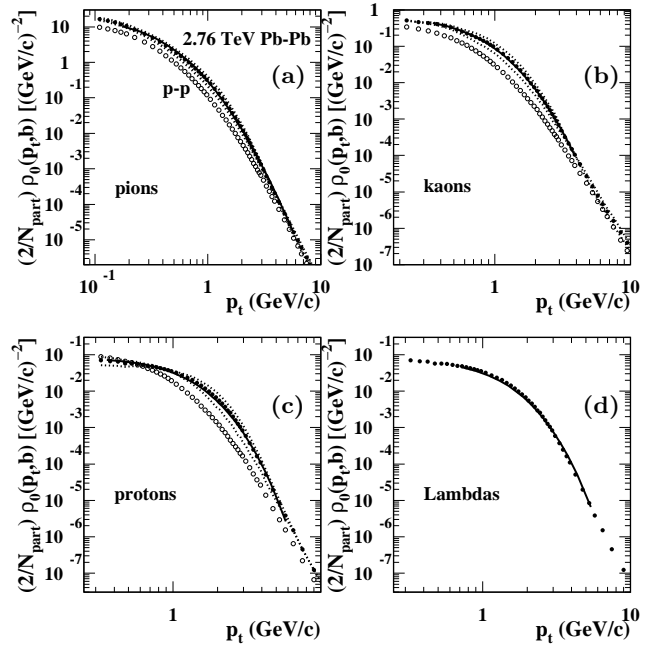
\begin{figure}[h]
     \includegraphics[width=3.3in]{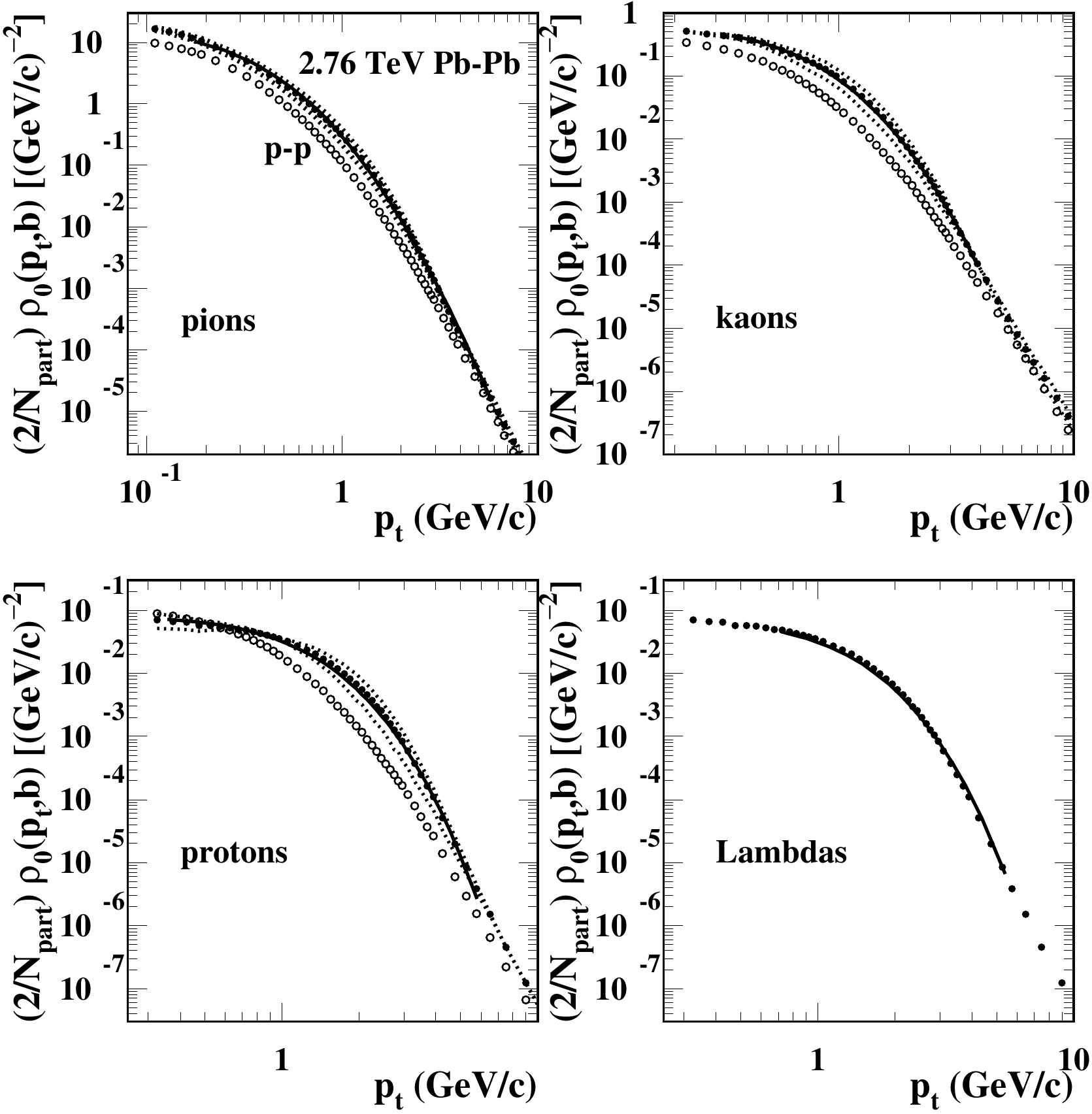}
\put(-140,213) {\bf (a)}
\put(-22,213) {\bf (b)}
\put(-140,88) {\bf (c)}
\put(-22,88) {\bf (d)}
\caption{\label{alice5}
Identified-hadron spectra for four hadron species from 0-5\% and 60-80\% central 2.76 TeV \pbpb\ collisions (dotted curves) and from 2.76 TeV \pp\ collisions (open points)~\cite{alicespec2}. The solid points are averages of the two \pbpb\ centralities to approximate 30-40\% data. The solid curves are parametrizations of those averages on \pt\ values corresponding to $v_2(p_t,b)$ data from Ref.~\cite{alicev2ptb} used for the present study.
}  
\end{figure}

Figure~\ref{alice5} shows SP spectra (densities on \pt) for identified (a) pions, (b) kaons and (c) protons from 0-5\% and 60-80\% central 2.76 TeV \pbpb\ collisions (dotted curves) and from 2.76 TeV \pp\ collisions (open points)~\cite{alicespec2} plotted {\em per participant pair} as in Eq.~(\ref{ppspec}) (second line). Because jet-related contributions to $v_2$ data (nonflow) are relatively largest for peripheral and central collisions (e.g.\ Fig.~\ref{v2ep}, left) representative $v_2(p_t,b)$ data for 30-40\% central are used, as noted in Sec.~\ref{quadspec}. The required SP spectra are then constructed as simple linear averages of the two \pbpb\ spectrum centralities in the form $(2/N_{part}) \bar \rho_0(p_t,b)$ (solid points). The solid curves passing through data are described below. Panel (d) is proton data  (c) with a parametrization on Lambda-$v_2$ \pt\ values.

The solid curves passing through solid points in Fig.~\ref{alice5} are L\'evy parametrizations of the \pbpb\ spectrum data [averages corresponding to 30-40\% $v_2(p_t,b)$ data] defined on the $v_2(p_t,b)$ data \pt\ values. Panel (d) of Fig.~\ref{alice5} shows a ``Lambda'' spectrum (solid curve) on Lambda-$v_2$ \pt\ values derived from  proton spectrum data in panel (c). Those spectra are the $\bar \rho_0(p_t,b)$ distributions used in Sec.~\ref{quadspec}, but while the pion spectrum seems minimally distorted the proton and to lesser extent kaon spectra below 1 GeV/c appear to be substantially reduced.

In Fig.~4 of Ref.~\cite{alicespec2}  \pp\ spectra in the form $N_{bin} \bar \rho_0(p_t,b)$ are compared with \pbpb\ spectra in the form $\bar \rho_0(p_t,b)$, anticipating jet quenching as the principal issue. The plotting format is on linear \pt\ thus visually minimizing the low-\pt\ region, again anticipating jet quenching as the principal issue. The definition of $N_{bin}$ ($N_{coll}$) can be questioned, as to whether the factor $\sigma_{NN}$ in its definition (= 42 mb at 200 GeV) should scale with collision energy (to 65 mb at 2.76 TeV). In contrast there is little uncertainty about the definition of $N_{part}$ since $N_{part,max} \leq 2A$ is within a few percent of maximum value $2A$ (e.g.\ $\approx 382$ for $^{197}$Au at 200 GeV). Thus, estimate $N_{part,max} \approx 404$ for $^{208}$Pb at 2.76 TeV should be correct to about 1\%. There are two significant issues for the spectrum data from Ref.~\cite{alicespec2}. 

(i) According to  Eq.~(\ref{ppspec}) (second line) per-participant-pair spectra should be dominated at lower \pt\ by soft component $\bar \rho_s \hat S_0(p_t)$ (describing MB \pp\ or \nn\ collisions) which was found in previous studies to be approximately independent of  \aa\ centrality~\cite{hardspec}. Thus, all such spectra for given hadron species should {\em coincide at lower \pt} when plotted in the per-participant format. But Fig.~\ref{alice5} reveals that pion and kaon spectra show a substantial difference between \pp\ spectra on the one hand and \pbpb\ spectra on the other, although the \pbpb\ spectra for two centralities do coincide as expected. The common difference between \pp\ and \pbpb\ spectra is a factor 1.65.

(ii) The second issue is best noted in panel (c) where the proton spectra for \pbpb\ collisions descend {\em below} the \pp\ spectrum at lower \pt\ even though the common factor 1.65 applies at higher \pt\ in that case as well. The result suggests a substantial inefficiency for proton detection at lower \pt. Close examination of the kaon spectrum suggests a similar inefficiency although less severe.

Both effects are manifested in corresponding $R_{AA}$ data. The conventional spectrum-ratio measure is $R_{AA}$
\bea \label{raa}
R_{AA} &\equiv& \frac{1}{N_{bin}} \frac{\bar \rho_{0,AA} (p_t)}{\bar \rho_{0,pp}(p_t)}
\nonumber \\
 &\approx&  \frac{1}{N_{bin}} \frac{(N_{part}/2)S_{NN}(p_t) + N_{bin} H_{AA}(p_t)}
{S_{pp}(p_t) + H_{pp}(p_t)}
\nonumber \\
&\rightarrow& r_{AA}(p_t)~~~\text{for larger \pt\ (where $r_{AA} \ll 1$)}
\nonumber \\
&\rightarrow& \frac{1}{\nu}~~~\text{for smaller \pt\ (where $r_{AA} \gg 1$)}
\eea
defined in the first line. The second line is based on TCMs for \aa\ spectra (numerator) and \pp\ spectra (denominator), recalling that $\nu \in [1,6]$ is the mean participant path length in number of encountered participants. Quantity $r_{AA} \equiv H_{AA} / H_{pp}$ is the hard-component ratio that actually represents modified fragmentation to jets over the {\em entire} \pt\ interval~\cite{hardspec}. The last two lines give the $R_{AA}(p_t)$ limiting cases at small and large \pt\ assuming that hard/soft ratio $H_{pp}(p_t) / S_{pp}(p_t)$ is $\ll 1$ at smaller \pt\ and $\gg 1$ at larger \pt, which is typically the case for measured SP spectra~\cite{ppprd,hardspec,alicetomspec}. By definition of $R_{AA}(p_t)$ access to ratio $r_{AA}(p_t)$ and important jet systematics at lower \pt\ is reduced to zero below $p_t \approx 2$-3 GeV/c where $r_{AA} \gg 1$ and the jet fragment yield is {\em strongly enhanced}.

\begin{figure}[h]
     \includegraphics[width=3.3in]{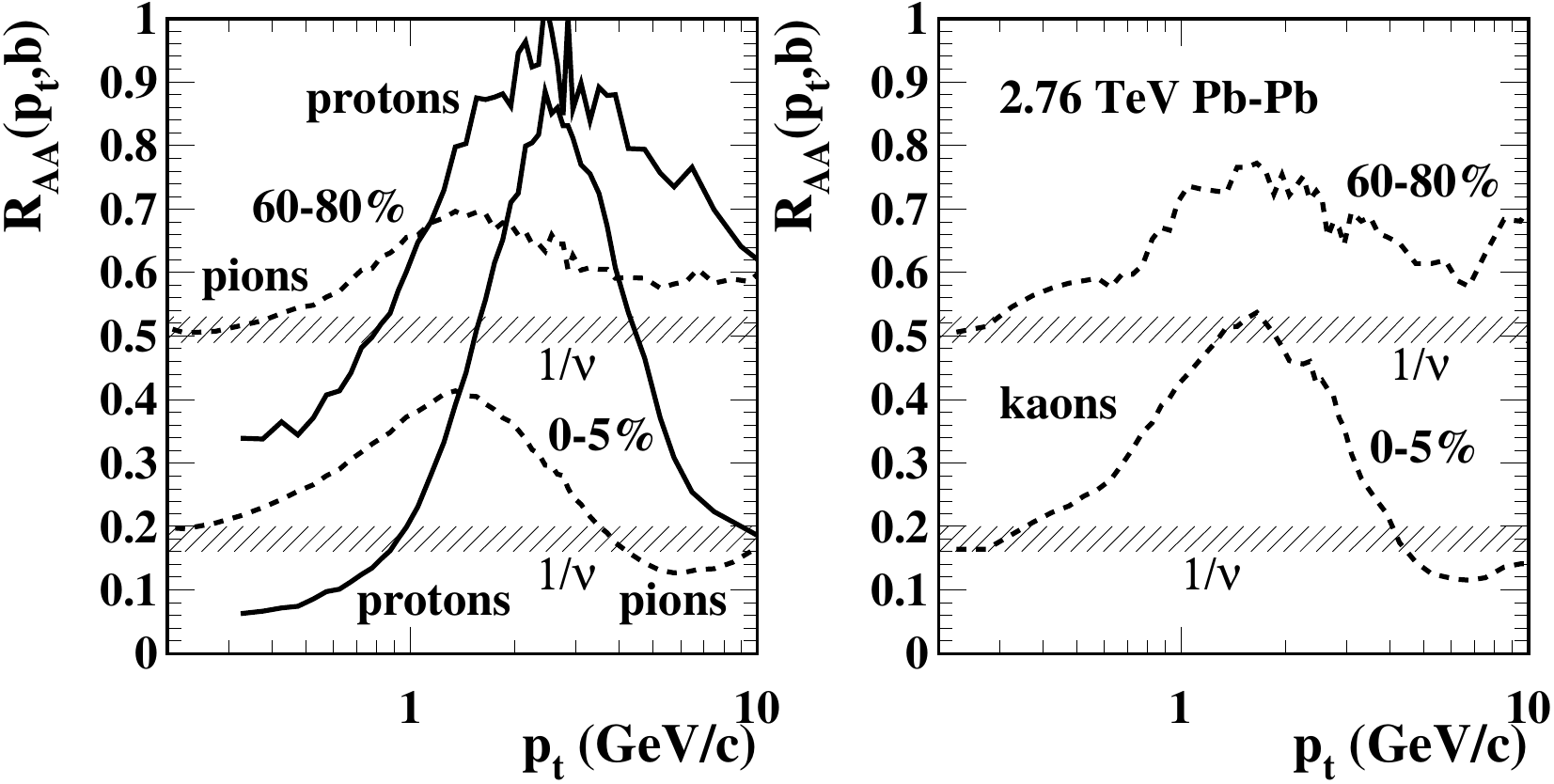}
\caption{\label{alice7x}
Trends of $R_{AA}$ as defined in Eq.~(\ref{raa}) for identified-hadron spectra from 2.76 TeV \pbpb\ and \pp\ collisions as reported in Ref.~\cite{alicespec2}. The \pbpb\ spectra have been rescaled by factor 1/1.65 and the Glauber $N_{part}$ and $N_{bin}$ numbers are those for 200 GeV, both as discussed in the text.
}  
\end{figure}

Figure~\ref{alice7x} shows $R_{AA}(p_t)$ data for three hadron species inferred from published \pbpb\ and \pp\ spectrum data in Fig.~\ref{alice5} according to Eq.~(\ref{raa}) (first line).  In the peripheral \aa\ limit (i.e.\ \nn)  one expects $N_{part}/2,~N_{bin},~r_{AA} \rightarrow 1$, requiring $R_{AA} \rightarrow 1$ as well according to Eq.~(\ref{raa}). But the spectrum data in Fig.~\ref{alice5} correspond to the limit $R_{AA} \rightarrow 1.65$ representing issue (i) above.  That issue is not resolved by rescaling upper limit $N_{bin,max} $, only by rescaling the \pbpb\ spectra down by factor 1/1.65. With that spectrum rescaling Fig.~\ref{alice7x} corresponds to Fig.~6 of Ref.~\cite{alicespec2} (for pions) only when $1/\nu \approx 0.18$ for 0-5\% central, thus defining $N_{bin}$ relative to $N_{part}$ the same as 200 GeV spectra~\cite{hardspec,anomalous}. As noted previously, while $N_{part}/2$ is well defined the definition of $N_{bin}$ is questionable. Figure~\ref{alice7x} is produced by scaling down the \pbpb\ spectra by 1/1.65 and retaining the 200 GeV Glauber parameters modulo few-percent increases corresponding to Au $\rightarrow$ Pb. The resulting pion $R_{AA}$ trends then correspond to 2.76 TeV charged-hadron $R_{aa}$ data reported by CMS for the same \pbpb\ centralities (but $R_{AA}$ data below 1 GeV/c are not shown there)~\cite{cmsraa}.

Issue (ii) is the apparent detection inefficiencies at lower \pt\  revealed by comparing the $R_{AA}$ data trends there  to expected $1/\nu$ limits common to all hadron species. According to Eq.~(\ref{raa}) all $R_{AA}$ data should fall within the hatched bands in the low-\pt\ limit, but only the pion data satisfy that requirement. And the proton data deviate strongly from expectations -- a factor 3 low for the 0-5\% spectrum as in Fig.~\ref{alice5} (c). Note that in Figs.~\ref{alice5} and \ref{alice7x} a logarithmic \pt\ scale is essential for considering these issues productively.

\begin{figure}[h]
     \includegraphics[width=3.3in]{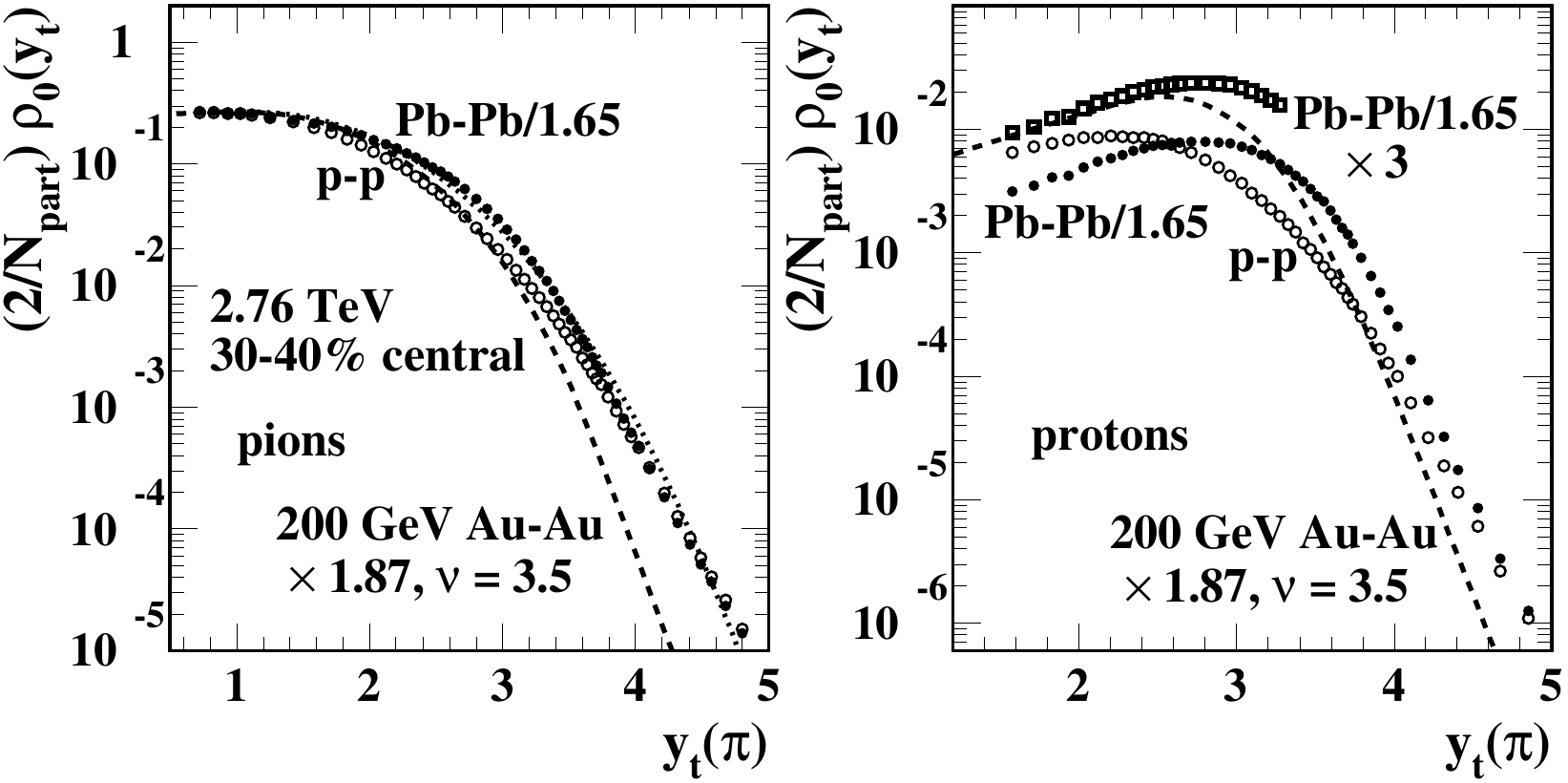}
\put(-143,70) {\bf (a)}
\put(-22,70) {\bf (b)}\\
     \includegraphics[width=1.62in]{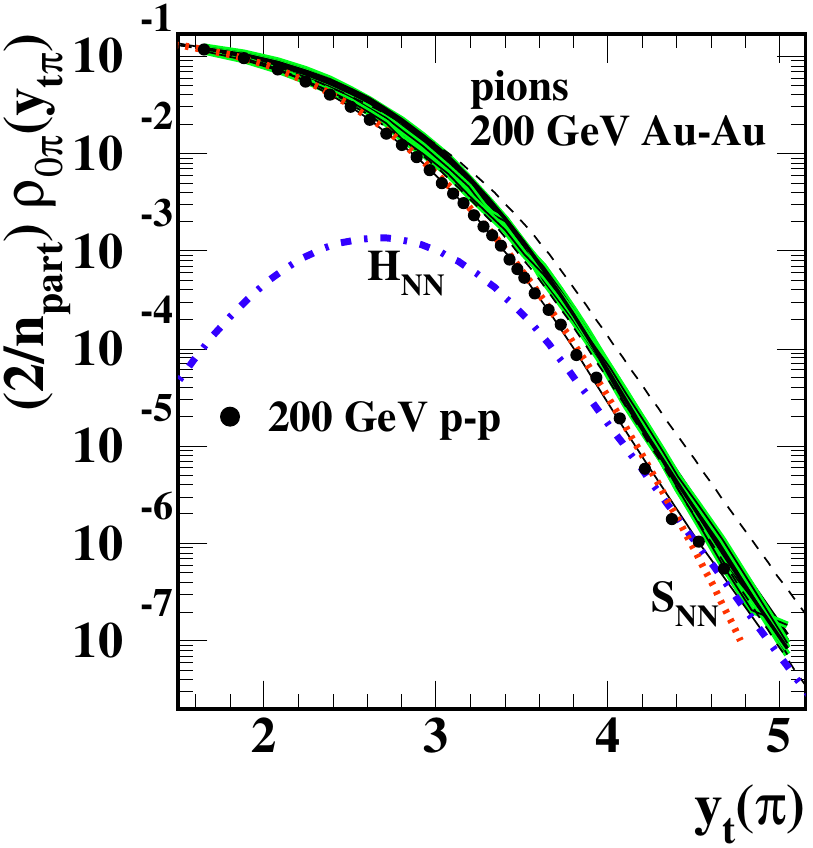}
     \includegraphics[width=1.65in]{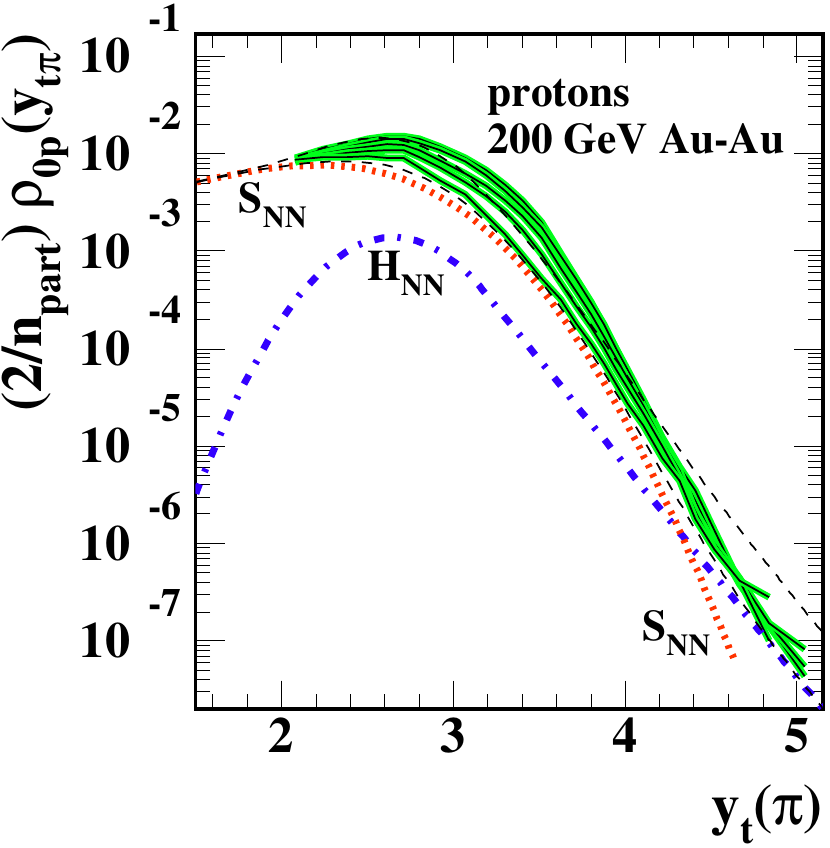}
\put(-143,70) {\bf (c)}
\put(-22,70) {\bf (d)}
\caption{\label{alice5aa}
(a,b) Comparison of pion and proton spectra from 2.76 TeV \pbpb\ collisions (solid points), from 2.76 TeV \pp\ collisions (open points) and from 200 GeV \auau\ collisions scaled up by energy factor 1.87 (dashed curves). (c,d) pion and proton spectra from five centralities of 200 GeV \auau\ collisions (solid curves) compared to unidentified hadrons from 200 GeV \pp\ collisions (points). $S_{NN}$ and $H_{NN}$ denote TCM soft and hard \nn\ models for 200 GeV \auau\ spectra.
}  
\end{figure}

Figure~\ref{alice5aa} (a) shows comparisons of 2.76 TeV pion spectra with lower-energy and unidentified-hadron spectra from other sources as densities on transverse rapidity \yt\ that facilitate precise differential spectrum comparisons. The 2.76 TeV \pp\ pion spectrum (open points) coincides at lower \pt\ with the predicted unidentified-hadron spectrum for that collision system (dotted curve)~\cite{alicetomspec} times factor 0.8 (approximate pion fraction). The deviation centered near $y_t = 3$ is consistent with contributions from kaons and protons to the hadron spectrum. The \pp\ spectrum also coincides at lower \pt\ with a per-participant-pair 200 GeV \auau\ pion spectrum (dashed curve~\cite{hardspec}) scaled up by factor 1.87 representing the expected $\log(s_{NN}/s_0)$ energy trend for a spectrum soft component between 200 GeV and 2.76 TeV~\cite{alicetomspec}. The per-participant-pair 2.76 TeV \pbpb\ spectrum scaled down by factor 1.65 (solid points) also agrees at lower \pt. Thus, pion spectra from several collision systems are quantitatively consistent at lower \pt. The 2.76 TeV \pp\ pion spectrum  appears to provide a proper reference for per-participant-pair \pbpb\ pion spectra rescaled by 1/1.65.

Figure~\ref{alice5aa} (b) shows equivalent comparisons for proton spectra where the picture is less clear. The 2.76 TeV \pp\ proton spectrum (open circles) at lower \pt\ seems to fall significantly below the 200 GeV \auau\ proton spectrum extrapolated to 2.76 TeV (dashed curve). As noted above, the  per-participant-pair 2.76 TeV \pbpb\ proton spectrum scaled down by factor 1.65 (solid points) falls substantially below either of those spectra at lower \pt, whereas the same data scaled up by factor 3 (open squares) coincide at lower \pt\ with the extrapolated 200 GeV \auau\ spectrum. The large deviations between \pp\ and per-participant-pair \aa\ spectra near $y_t = 3$ are a consequence of the much larger {\em relative} contribution of jets to \aa\ proton spectra compared to pion spectra near $p_t = 1.5$ GeV/c [$y_t(\pi) \approx 3$]~\cite{hardspec}. The utility of densities on logarithmic $y_t$ vs linear $p_t$ is further demonstrated by comparison of panel (b) with Fig.~4 of Ref.~\cite{alicespec2}.

Figure~\ref{alice5aa} (c,d) show corresponding pion and proton spectra for five centralities of 200 GeV \auau\ collisions (solid curves) and unidentified hadrons from 200 GeV \pp\ collisions [points in panel (c)] from Ref.~\cite{hardspec}. The pion spectra for \auau\ and \pbpb\ are quite similar modulo the 1/1.65 factor required for \pbpb\ data. However, the \auau\ proton spectra show major differences from the \pbpb\ spectra. Note that for \auau\ proton spectra normalized by $N_{part}/2$ the hard-component contribution at lower \pt\ (below the spectrum mode at $y_t \approx 2.7$ or $p_t \approx 1$ GeV/c) scales as $\nu = 2 N_{bin}/N_{part}$ for all centralities with no evidence of modified fragmentation (``jet quenching'') in  that \pt\ interval whereas above the mode on \pt\ there is strong modification of the proton hard-component shape, but only beyond a {\em sharp transition} on centrality~\cite{hardspec,anomalous}. 

Note that $y_t(\pi) \approx \ln(2p_t / m_\pi)$ is preferred for these plots as a logarithmic representation of \pt\ (with well-defined zero) as opposed to transverse rapidity \yt\ with proper mass for each hadron species, as in Fig.~\ref{x1} (right) for example. In the former case the main issue is the effective endpoint of the underlying scattered-parton or jet energy spectrum near 3 GeV that manifests as modes of the peaked hard-component distributions near $p_t = 1$ GeV/c ($y_t \approx 2.7$) for all hadron species. It is then \pt\ that matters, but the logarithmic representation  $y_t(\pi)$ makes the low-\pt\ region more accessible visually and displays power-law trends on \pt\ as simple straight lines in what is effectively a log-log plot. In the latter case the main issue is hadron emission from a moving or boosted hadron source, which manifests as a zero intercept on \yt\ (with proper hadron mass) common to all hadron species.

\end{appendix}


\end{document}